\documentclass[12pt]{article}
\usepackage{latexsym}
\usepackage{amsmath,amsfonts}
\usepackage{times}
\allowdisplaybreaks[4]

\catcode`\@=11

\newcount\hour
\newcount\minute
\newtoks\amorpm \hour=\time\divide\hour by 60\minute
=\time{\multiply\hour by 60 \global\advance\minute by-\hour}
\edef\standardtime{{\ifnum\hour<12 \global\amorpm={am}%
        \else\global\amorpm={pm}\advance\hour by-12 \fi
        \ifnum\hour=0 \hour=12 \fi
        \number\hour:\ifnum\minute<10
        0\fi\number\minute\the\amorpm}}
\edef\militarytime{\number\hour:\ifnum\minute<10
0\fi\number\minute}

\def\draftlabel#1{{\@bsphack\if@filesw {\let\thepage\relax
   \xdef\@gtempa{\write\@auxout{\string
      \newlabel{#1}{{\@currentlabel}{\thepage}}}}}\@gtempa
   \if@nobreak \ifvmode\nobreak\fi\fi\fi\@esphack}
        \gdef\@eqnlabel{#1}}
\def\@eqnlabel{}
\def\@vacuum{}
\def\marginnote#1{}
\def\draftmarginnote#1{\marginpar{\raggedright\scriptsize\tt#1}}
\def\draft{
        \pagestyle{plain}
        \overfullrule=2pt
        \oddsidemargin -.5truein
        \def\@oddhead{\sl \phantom{\today\quad\militarytime} \hfil
        \smash{\Large\sl DRAFT} \hfil \today\quad\militarytime}
        \let\@evenhead\@oddhead
        \let\label=\draftlabel
        \let\marginnote=\draftmarginnote
        \def\ps@empty{\let\@mkboth\@gobbletwo
        \def\@oddfoot{\hfil \smash{\Large\sl DRAFT} \hfil}
        \let\@evenfoot\@oddhead}
        \def\@eqnnum{(\theequation)\rlap{\kern\marginparsep\tt\@eqnlabel}%
        \global\let\@eqnlabel\@vacuum}  }

\newcommand{\rf}[1]{(\ref{#1})}
\renewcommand{\theequation}{\thesection.\arabic{equation}}
\renewcommand{\thefootnote}{\fnsymbol{footnote}}
\newcommand{\newsection}{    
\setcounter{equation}{0}\section}

\def\appendix#1{\addtocounter{section}{1}\setcounter{equation}{0}
\renewcommand{\thesection}{\Alph{section}}
\section*{Appendix \thesection\protect\indent \parbox[t]{11.15cm}{#1}}
\addcontentsline{toc}{section}{Appendix \thesection\ \ \ #1}}

\def\nline{\,\nabla\kern -0.7em\raise0.2ex\hbox{/}\,\,}
\def\yline{\,y\kern -0.47em /}
\def\aline{\,a\kern -0.49em /}
\def\parline{\,\partial\kern -0.55em /\,\,}
\def\Dline{\,D\kern -0.55em /\,\,}

\def\parline{\,\partial\kern -0.55em /\,\,}
\def\pline{\, p \kern -0.4em /\,}
\def\Poline{\, \Po \kern -0.6em /\,}
\def\half{{\frac{1}{2}}}

\def\alphaline{\, \alpha \kern -0.6em /\,}

\newcommand{\Po}{\mathbb{P}}

\def\be{\begin{equation}}
\def\ee{\end{equation}}
\def\beq{\begin{eqnarray}}
\def\eeq{\end{eqnarray}}

\def\sma{{\scriptscriptstyle (a)}}
\def\smaa{{\scriptscriptstyle (aa)}}
\def\smab{{\scriptscriptstyle (ab)}}
\def\smb{{\scriptscriptstyle (b)}}
\def\smB{{\scriptscriptstyle B}}
\def\smF{{\scriptscriptstyle F}}

\def\oplussm{{\scriptscriptstyle \oplus}}
\def\ominussm{{\scriptscriptstyle \ominus}}

\def\phik{|\phi\rangle}
\def\psik{|\psi\rangle}

\def\smaplusone{{\scriptscriptstyle (a+1)}}
\def\smaplustwo{{\scriptscriptstyle (a+2)}}

\def\smone{{\scriptscriptstyle (1)}}
\def\smtwo{{\scriptscriptstyle (2)}}
\def\smthree{{\scriptscriptstyle (3)}}

\def\smonetwo{{\scriptscriptstyle (12)}}
\def\smtwothree{{\scriptscriptstyle (23)}}
\def\smthreeone{{\scriptscriptstyle (31)}}

\def\smoneone{{\scriptscriptstyle (11)}}
\def\smtwotwo{{\scriptscriptstyle (22)}}
\def\smthreethree{{\scriptscriptstyle (33)}}

\def\smaaplusone{{\scriptscriptstyle (aa+1)}}

\def\smpt{{\scriptscriptstyle [2]}}
\def\smp3{{\scriptscriptstyle [3]}}

\def\smpn{{\scriptscriptstyle [n]}}

\def\sbf{{\bf s}}
\def\Jbf{{\bf J}}
\def\Lbf{{\bf L}}
\def\Mbf{{\bf M}}
\def\Pbf{{\bf P}}

\def\LL{{\cal L}}

\def\Nsf{{\sf N}}
\def\asf{{\sf a}}
\def\bsf{{\sf b}}
\def\nsf{{\sf n}}

\def\nnu{\nu}

\def\i{{\rm i}}
\def\mas{{\rm m}}
\def\bos{{\rm bos}}
\def\fer{{\rm fer}}
\def\kin{{\rm kin}}
\def\dyn{{\rm dyn}}
\def\min{{\rm min}}
\def\max{{\rm max}}
\def\cov{{\rm cov}}
\def\lc{{\rm l.c.}}

\def\LL{{\cal L}}
\def\XX{{\cal X}}

\def\half{\frac{1}{2}}

\def\Krm{{\rm K}}
\def\Frm{{\rm F}}

\begin{document}


\begin{flushright}
FIAN-TD-2007-25 \hspace{1cm} {}~ \\
arXiv: 0712.3526 [hep-th]\\
Modified, December 2011
\end{flushright}

\vspace{1cm}

\begin{center}

{\Large \bf Cubic interaction vertices for fermionic and bosonic

\medskip
arbitrary spin fields}

\vspace{2.5cm}

R.R. Metsaev\footnote{ E-mail: metsaev@lpi.ru }

\vspace{1cm}

{\it Department of Theoretical Physics, P.N. Lebedev Physical
Institute, \\ Leninsky prospect 53,  Moscow 119991, Russia }

\vspace{3cm}

{\bf Abstract}

\end{center}

Using the light-cone gauge approach to relativistic field dynamics, we study
arbitrary spin fermionic and bosonic fields propagating in flat space of
dimension greater than or equal to four. Generating functions of parity
invariant cubic interaction vertices for  totally symmetric and
mixed-symmetry massive and massless fields are obtained. For the case of
totally symmetric fields, we derive restrictions on the allowed values of
spins and the number of derivatives. These restrictions provide a complete
classification of parity invariant cubic interaction vertices for totally
symmetric fermionic and bosonic fields. As an example of application of the
light-cone formalism, we obtain simple expressions for the Yang-Mills and
gravitational interactions of massive arbitrary spin fermionic fields. For
some particular cases, using our light-cone cubic vertices, we discuss the
corresponding  manifestly Lorentz invariant and on-shell gauge invariant
cubic vertices.

\vspace{3cm}


\bigskip
PACS-2011:  11.10.-z;  11.30-j; 11.30.Cp

\newpage
\renewcommand{\thefootnote}{\arabic{footnote}}
\setcounter{footnote}{0}

\section{Introduction}

The light-cone gauge approach to relativistic field and string dynamics
\cite{Dirac:1949cp,Weinberg:1966jm} offers conceptual and technical
simplifications into the study of various problems of quantum field and
string theories. This approach hides the Lorentz symmetries and makes the
notation somewhat cumbersome but eventually turns out to be effective. A
number of important problems of modern quantum field and string theories have
been solved in the framework of this approach. For example, we mention the
solution to the light-cone gauge string field theory
\cite{Kaku:1974zz,Green:1983hw} and the construction of a superfield
formulation for supersymmetric theories (see e.g.
Refs.\cite{Brink:1982pd}-\cite{Ananth:2005vg}.). Sometimes, theories
formulated in the framework of light-cone gauge approach turn out to be a
good starting point for deriving a Lorentz covariant formulation (see e.g.
Refs.\cite{Hata:1986jd}). Use of the light-cone formalism for the building
interaction vertices of massless higher-spin fields may be found in
Refs.\cite{Bengtsson:1983pd}-\cite{Metsaev:1993mj}. Applications of the
light-cone formalism for the studying AdS/CFT correspondence are discussed in
Refs.\cite{Metsaev:1999ui,Brower:2006hf}. Study of super $p$-branes and
string bit models in the light-cone gauge may be found in
Refs.\cite{deWit:1988ig,Bergman:1995wh}.

The problem of gravitational interaction of massless higher-spin fields
attacked in  Ref.\cite{Fradkin:1987ks} was solved in
Refs.\cite{Vasiliev:1990en} (for review, see Refs.\cite{Vasiliev:2004cp}).
This is to say that cubic interaction vertices for totally symmetric massless
higher-spin fields in $AdS_4$ space were found in Ref.\cite{Fradkin:1987ks},
while nonlinear equations of motion to all orders in the coupling constant
for totally symmetric massless higher-spin fields were obtained in
Refs.\cite{Vasiliev:1990en}. At present time it is clear, due to results in
Refs.\cite{Fradkin:1987ks,Vasiliev:1990en}, that self-consistent theories of
massless higher-spin gauge fields require formulating the theories in $AdS$
background. Recent discussion of interacting massless higher-spin AdS fields
theories may be found in Refs.\cite{Alkalaev:2002rq}.

One of interesting problems of modern research in the theories of higher-spin
fields is related to Lagrangian formulation of equations of motions for
massless higher-spin fields found in
Refs.\cite{Vasiliev:1990en}. In this area of research, some
interesting results were recently reported in Ref.\cite{Boulanger:2011dd} (in
earlier literature, see Ref.\cite{Vasiliev:1988sa}). In order to quantize
higher-spin field theories and investigate their ultraviolet behavior, it
would be desirable to find an Lagrangian formulation of higher-spin field
theories. Since the massless higher-spin field theories involve
infinite-dimensional gauge symmetries, one expects that such theories may be
ultraviolet finite. We think that the light-cone approach might be helpful in
developing light-cone gauge Lagrangian formulation of higher-spin fields
theories and understanding their quantum behavior.

In AdS space, the light-cone gauge formulation of free fields dynamics was
developed in Ref.\cite{Metsaev:1999ui} (see also
Refs.\cite{Metsaev:2002vr,Metsaev:2003cu}). Because light-cone gauge
formulations of fields in AdS and flat spaces share many common properties we
believe that methods developed in flat space might be useful and helpful in
studying theories of interacting higher-spin AdS fields. In other words, we
believe that most of our approach to higher-spin fields in flat space, which
we consider in this paper, can straightforwardly be generalized to the case
of higher-spin AdS fields.

In Ref.\cite{Metsaev:2005ar}, we applied light-cone gauge formalism to the
studying cubic interaction vertices of higher-spin massless and massive
bosonic fields in flat space. We found generating functions of parity
invariant cubic interaction vertices for massive and massless fields of
arbitrary symmetry. Also we derive restrictions on the allowed values of
spins and the number of derivatives, which provide the complete
classification of cubic interaction vertices for totally symmetric massless
and massive fields. In earlier literature, application of light-cone
formalism for studying interacting higher-spin fields in flat space may be
found in Refs.\cite{Bengtsson:1983pd}-\cite{Metsaev:1993mj}. In the framework
of Lorentz covariant approaches, discussion of interacting higher-spin fields
in flat space may be found in Refs.\cite{Bekaert:2005jf} (in earlier
literature, see Refs.\cite{Weinberg:1969di}). For interesting review, see
Ref.\cite{Bekaert:2010hw}.

In this paper, we apply the light-cone formalism for studying cubic
interaction vertices for higher-spin bosonic and fermionic fields in flat
space. In Ref.\cite{Metsaev:2005ar}, we studied cubic interaction vertices
involving three bosonic fields, while in the present paper, we study cubic
interaction vertices involving two fermionic fields and one bosonic field. In
view of possible potentially interesting applications of the higher-spin
field theory to superstring theory, it is worthwhile to study cubic
vertices for higher-spin fields in flat space of dimension $d
\geq 4$. We do this in our paper.

This paper is organized as follows.

In Sec. \ref{freesec}, we discuss the light-cone gauge approach to free
fermionic and bosonic fields. We start with the light-cone gauge description
of totally symmetric and mixed-symmetry fields in terms of ket-vectors. After
this, we review realization of the Poincar\'e algebra symmetries on space of the
ket-vectors.

In Sec. \ref{GENVERsec}, we describe restrictions imposed by the Poincar\'e
algebra symmetries on cubic interaction vertices. We present the complete
system of equations for cubic interaction vertices which allows us to
determine the vertices uniquely.

In Sec. \ref{Solcubintversec}, we discuss cubic vertices for massless fields,
i.e., we consider vertices involving two massless fermionic fields and one
massless bosonic field.  We find generating form for the cubic vertices of
massless mixed-symmetry and totally symmetric fields. For the case of totally
symmetric fields, we find restrictions on the allowed number of derivatives and spin
values of the fields entering the cubic vertices. These restrictions provide
the complete classification of the parity invariant cubic vertices for the
totally symmetric fields. We use these results to discuss the no-go theorem
for the gravitational interaction of higher-spin fermionic fields.

Secs. \ref{secMMO}-\ref{Sec1010} are devoted to cubic interaction vertices
involving both massless and massive fields. As before we discuss the
generating form for the cubic vertices of mixed-symmetry and totally
symmetric fields. Also, we find restrictions on the allowed number of derivatives and
spin values of the totally symmetric fields entering the cubic interaction
vertices. We apply our general results to derive the Yang-Mills and
gravitational interactions of massive totally symmetric arbitrary spin
fermionic fields. Our approach allows us to obtain simple expressions
for the cubic vertices of these interactions. We obtain also simple
expressions for cubic interaction vertices involving one massless
arbitrary spin bosonic field and two massive spin-$\half$ fermionic fields.

In Sec. \ref{secMMM}, we discuss cubic vertices for massive fields: two
massive fermionic fields and one massive bosonic field. Generating form of
the vertices is obtained for the mixed-symmetry and totally symmetric fields.
We derive restrictions on the allowed values of spins and the number of
derivatives for cubic vertices of the totally symmetric massive fermionic and
bosonic fields.

In Appendix A, we review the notation and conventions used in this paper.

\newsection{Free light-cone gauge massive and massless fields}\label{freesec}

The method proposed in Ref.\cite{Dirac:1949cp} reduces the problem of finding
a new dynamical system to the problem of finding a new solution to
commutators of defining symmetry algebra. In our case, the defining
symmetries are generated by the Poincar\'e algebra. We begin therefore with a
discussion of the realization of the Poincar\'e algebra on the space of
massive and massless fields. In this section, we consider free fields.

The Poincar\'e algebra of $d$-dimensional Minkowski space contains
translation generators $P^A$ and rotation generators $J^{AB}$ of the Lorentz
algebra $so(d-1,1)$. Non-trivial Poincar\'e algebra commutators are given by
\be \label{pj1} {} [P^A,\,J^{BC}]=\eta^{AB}P^C - \eta^{AC} P^B\,, \qquad {}
[J^{AB},\,J^{CE}] =\eta^{BC}J^{AE} + 3\hbox{ terms}\,, \ee
where the vector indices of the $so(d-1,1)$ algebra take values $A,B,C,E=0,1,\ldots,d-1$ and
$\eta^{AB}$ stands for the mostly positive flat metric tensor.

To discuss the light-cone approach, in place of the Lorentz frame coordinates
$x^A$ we introduce the light-cone frame coordinates $x^+$, $x^-$, $x^I$, $
I=1,\ldots, d-2$, and treat $x^{+}$ as an evolution parameter (for the detailed
discussion of the notation, see Appendix A). In the light-cone frame, the
Poincar\'e algebra  generators can be separated into two groups:
\beq
&& \label{kinnewgen} P^+,\quad P^I,\quad J^{+I},\quad J^{+-},\quad
J^{IJ}, \qquad \hbox{ kinematical generators}\,;
\\
&& \label{dynnewgen} P^-,\quad J^{-I}\,, \hspace{4.5cm} \hbox{ dynamical
generators}\,, \eeq
where the vector indices of the $so(d-2)$ algebra take values $I,J=1,\ldots,d-2$.
For $x^+=0$, the kinematical generators in the field realization are
quadratic in the physical fields, while the dynamical generators receive higher-order
interaction-dependent corrections.%
\footnote{ For $x^+ \ne  0 $, the kinematical generators take the form $G=
G_1 + x^+ G_2$, where $G_1$ is quadratic in fields, while $G_2$ contains
higher order terms in fields.}

In light-cone frame, the commutators of the Poincar\'e algebra  can be
obtained from \rf{pj1} by using the light-cone metric having the following
non vanishing elements: $\eta^{+-}=\eta^{-+}=1$, $\eta^{IJ}=\delta^{IJ}$.
We adapt the following hermitian conjugation rules for the Poincar\'e algebra generators:
\be P^{\pm \dagger}=P^\pm, \quad P^{I\dagger} = P^I, \quad
J^{IJ\dagger}=-J^{IJ}\,,\quad J^{+-\dagger}=-J^{+-}, \quad J^{\pm
I\dagger} = -J^{\pm I}\,. \ee
To find  a realization of the Poincar\'e algebra on the space of massive and
massless fields we use the light-cone gauge description of those fields.

\subsection{ Light-cone gauge massive and massless field}\label{lcgmasfield}

We now review the light-cone gauge description of free fields
propagating in Minkowski space. Throughout the paper a dimension of Minkowski
space is assumed to be equal to $d$. An arbitrary spin field of the
Poincar\'e algebra is labelled by mass parameter $\mas$ and by spin labels
$s_1$, \ldots, $s_\nu$ for bosonic field and by spin labels $s_1+\half$,
\ldots, $s_\nu+\half$, for fermionic field, where all $s_n$, $n=1,\ldots
\nu$, are integers. For massless fields, $\mas=0$,  $\nu=[\frac{d-2}{2}]$,
while for massive fields, $\mas\ne 0$, $\nu=[\frac{d-1}{2}]$. In order to
obtain a description of mixed-symmetry fields it is sufficient to set $\nnu
>1 $. In what follows, a particular value of $\nnu$ does not matter. We now
discuss bosonic and fermionic fields in turn.

{\bf Bosonic massless and massive mixed-symmetry fields}. In order to
streamline the light-cone gauge description of mixed-symmetry fields we
introduce a finite set of oscillators $\alpha_n^I$, $n=1,\ldots,\nnu$, for
the discussion of massless fields and a finite set of oscillators
$\alpha_n^I$, $\zeta_n$, $n=1,\ldots,\nnu$, for the discussion of massive
fields. Using such set of the oscillators, we introduce the following
ket-vectors to discuss the physical D.o.F of  mixed-symmetry massless and
massive bosonic fields having spin labels $s_1,\ldots,s_\nnu$:
\beq
\label{manold-07122011-01} && \hspace{-1.5cm} |\phi_{s_1 \ldots s_\nnu }
(p,\alpha)\rangle \equiv \prod_{n=1}^\nnu
\alpha_n^{I_1^n} \ldots \alpha_n^{I_{s_n}^n}\,
\phi_{s_1\ldots s_\nnu}^{I_1^1\ldots I_{s_1}^1\ldots I_1^\nnu \ldots
I_{s_\nnu }^\nnu }(p)|0\rangle\,,  \hspace{2.2cm} \hbox{massless field},
\\
\label{manold-07122011-02} && \hspace{-1.5cm} |\phi_{s_1 \ldots s_\nnu
}(p,\alpha)\rangle \equiv
\prod_{n=1}^\nnu  \sum_{t_n=0}^{s_n}
\zeta_n^{s_n-t_n} \alpha_n^{I_1^n}\ldots\alpha_n^{I_{t_n}^n} \,
\phi_{s_1\ldots s_\nnu}^{I_1^1\ldots I_{t_1}^1\ldots I_1^\nnu \ldots
I_{t_\nnu}^\nnu }(p)|0\rangle\,,  \hspace{0.5cm} \hbox{massive field}. \eeq
In \rf{manold-07122011-01},\rf{manold-07122011-02} and below, $\alpha$
occurring in the argument of massless field ket-vector
$|\phi(p,\alpha)\rangle$ denotes a set of the oscillators $\{\alpha_n^I\}$,
while $\alpha$ occurring in the argument of massive field ket-vector
$|\phi(p,\alpha)\rangle$ denotes a set of the oscillators
$\{\alpha_n^I\,,\zeta_n\}$. Momentum $p$ occurring in the argument of
ket-vector $|\phi(p,\alpha)\rangle$ and $\delta$- functions denotes a set of
the momenta $\{p^I\,,\beta\equiv p^+\}$. Note that we do not explicitly show
the dependence of the ket-vector $|\phi(p,\alpha)\rangle$ on the evolution
parameter $x^+$. Also, note that ket-vector \rf{manold-07122011-01} is a
degree-$s_n$ homogeneous polynomial in the oscillators $\alpha_n^I$, while
ket-vector \rf{manold-07122011-02} is a degree-$s_n$ homogeneous polynomial
in the oscillators $\alpha_n^I$, $\zeta_n$.

Physical D.o.F of massless and massive fields are described
by irreps of the $so(d-2)$ and $so(d-1)$ algebras respectively. In order for
ket-vectors \rf{manold-07122011-01} and \rf{manold-07122011-02} to describe
the respective irreps of the $so(d-2)$ and $so(d-1)$ algebras we should
impose certain algebraic constraints on the ket-vectors. But to avoid unnecessary
complications, we impose only the tracelessness constraints,
\beq
\label{manold-26122011-01} && \hspace{-1.5cm}\bar\alpha_m^I \bar\alpha_n^I
|\phi_{s_1\ldots s_\nnu} (p,\alpha) \rangle = 0 \,, \qquad\qquad\qquad m,n
=1,\ldots,\nnu\,, \hspace{1cm} \hbox{for massless field},
\\
\label{manold-26122011-02} && \hspace{-1.5cm} (\bar\alpha_m^I \bar\alpha_n^I
+\bar\zeta_m\bar\zeta_n)|\phi_{s_1\ldots s_\nnu} (p,\alpha) \rangle = 0 \,,
\qquad  m,n =1,\ldots,\nnu\,, \hspace{1cm} \hbox{for massive field}.
\eeq
The tracelessness constraints by themselves are not enough to single
out irreps of the $so(d-2)$ and $so(d-1)$ algebras from the ket-vectors. This
implies that ket-vectors \rf{manold-07122011-01} and
\rf{manold-07122011-02} actually describe finite sets of the respective
massless and massive fields.

In what follows, to develop the light-cone gauge description of
arbitrary spin mixed-symmetry bosonic fields on an equal footing we use the
ket-vector defined by
\be
\label{manold-07122011-10} |\phi(p,\alpha)\rangle \equiv \sum_{s_1,\ldots,
s_\nnu = 0}^{\infty}\,\,|\phi_{s_1 \ldots s_\nnu }(p,\alpha)\rangle\,. \ee

{\bf Totally symmetric massless and massive bosonic fields}. Totally
symmetric fields are popular in various studies because these fields are
considerably simpler than the mixed-symmetry fields and therefore allow to
illustrate many important features of higher-spin fields in a relatively
straightforward way. In order to obtain a description of massless and massive
totally symmetric fields it is sufficient to introduce one sort of
oscillators, i.e., we set $\nnu = 1$ in \rf{manold-07122011-01} and
\rf{manold-07122011-02} respectively. This is to say that we introduce the
following ket-vectors to discuss physical D.o.F. of totally symmetric
spin-$s$ massless and massive fields:
\beq
&& \label{manold-07122011-03} |\phi_s(p,\alpha)\rangle = \alpha^{I_1} \ldots
\alpha^{I_s}\, \phi^{I_1 \ldots I_s}(p) |0\rangle\,, \hspace{3cm}
\hbox{massless field},
\\
&& \label{manold-07122011-04} |\phi_s(p,\alpha)\rangle = \sum_{t=0}^s
\zeta^{s-t} \alpha^{I_1} \ldots \alpha^{I_t} \,  \phi^{I_1\ldots
I_t}(p)|0\rangle\,, \hspace{1.7cm}  \hbox{massive field}.
\eeq
We note that ket-vector of massless field \rf{manold-07122011-03}  is
a degree-$s$ homogeneous polynomial in the oscillator $\alpha^I$, while
ket-vector of massive field \rf{manold-07122011-04} is a degree-$s$ homogeneous
polynomial in the oscillators $\alpha^I$,  $\zeta$,
\beq
\label{manold-07122011-05}  && \left(\alpha^I\bar\alpha^I  - s
\right)|\phi_s(p,\alpha\rangle =0\,,\hspace{4.1cm}  \hbox{ for massless
field},
\\
\label{manold-07122011-06}  && \left(\alpha^I\bar\alpha^I + \zeta\bar\zeta -
s \right)|\phi_s(p,\alpha)\rangle =0\,, \hspace{3cm} \hbox{ for massive
field}.
\eeq
Physical D.o.F of massless and massive fields are described by irreps of the
$so(d-2)$ and $so(d-1)$ algebras respectively. In order for ket-vectors
\rf{manold-07122011-03} and \rf{manold-07122011-04} to realize irreps of the
$so(d-2)$ and $so(d-1)$ algebras respectively we impose the respective
tracelessness constraints
\beq
\label{manold-07122011-07} && \bar\alpha^I \bar\alpha^I
|\phi_s(p,\alpha)\rangle =0 \,, \hspace{4.3cm} \hbox{ for massless field},
\\
\label{manold-07122011-08} && \left(\bar\alpha^I  \bar\alpha^I +
\bar\zeta^2\right) |\phi_s(p,\alpha)\rangle =0\,, \hspace{3cm} \hbox{ for
massive field}.
\eeq
As in the case of mixed-symmetry fields in order to treat the totally
symmetric arbitrary spin bosonic fields on an equal footing we use the ket-vector defined by
\be \label{manold-07122011-09}
|\phi(p,\alpha)\rangle \equiv \sum_{s=0}^{\infty}\,\,|\phi_s(p,\alpha)\rangle\,.
\ee

{\bf Mixed-symmetry massless and massive fermionic fields}. In order to
simplify the presentation of light-cone gauge description of mixed-symmetry
fermionic fields we use, as before, the finite set of bosonic oscillators
$\alpha_n^I$, $n=1,\ldots,\nnu$, for the discussion of massless fields and
the finite set of bosonic oscillators $\alpha_n^I$, $\zeta_n$,
$n=1,\ldots,\nnu$, for the discussion of massive fields. In addition to the
bosonic oscillators, we introduce fermionic oscillators denoted by
$\theta^\asf$, $p_{\theta\asf}$, $\eta^\asf$, $p_{\eta\asf}$, where subscript
and superscript $\asf$ is used to indicate spinor indices (for the detailed
description of the fermionic oscillators, see Appendix A). Using such set of
the oscillators, we introduce the following ket-vectors to discuss the
physical D.o.F of massless and massive mixed-symmetry fermionic fields having
spin labels $s_1+\half,\ldots,s_\nnu+\half$:
\beq
&& \label{fer01nnn} |\psi_{s_1 \ldots s_\nnu }(p,\alpha)\rangle =   (
p_\theta \psi_{s_1 \ldots s_\nnu }(p,\alpha) + \psi_{s_1 \ldots s_\nnu
}^\dagger(p,\alpha)\eta) |0\rangle\,,
\eeq
\beq
&& \hspace{-0.5cm} \psi_{s_1 \ldots s_\nnu } (p,\alpha)\equiv
\prod_{n=1}^\nnu \alpha_n^{I_1^n}\ldots\alpha_n^{I_{s_n}^n}\,
\psi_{s_1\ldots s_\nnu}^{I_1^1\ldots I_{s_1}^1\ldots I_1^\nnu \ldots
I_{s_\nnu }^\nnu }(p)\,,
\nonumber\\[-10pt]
\label{manold-07122011-14} && \hspace{11cm} \hbox{massless field}\qquad
\\[-10pt]
&& \hspace{-0.5cm} \psi_{s_1 \ldots s_\nnu }^\dagger (p,\alpha)\equiv
\prod_{n=1}^\nnu \alpha_n^{I_1^n}\ldots\alpha_n^{I_{s_n}^n}\,
\psi_{s_1\ldots s_\nnu}^{\dagger I_1^1\ldots I_{s_1}^1\ldots I_1^\nnu \ldots
I_{s_\nnu }^\nnu }(p)\,,
\nonumber\\
&& \hspace{-0.5cm} \psi_{s_1 \ldots s_\nnu }(p,\alpha) \equiv
\prod_{n=1}^\nnu  \sum_{t_n=0}^{s_n}
\zeta_n^{s_n -t_n} \alpha_n^{I_1^n}\ldots\alpha_n^{I_{t_n}^n} \,
\psi_{s_1\ldots s_\nnu}^{I_1^1\ldots I_{t_1}^1\ldots I_1^\nnu \ldots
I_{t_\nnu}^\nnu }(p)\,,
\nonumber\\[-10pt]
\label{manold-07122011-15} && \hspace{11cm} \hbox{massive field}\qquad
\\[-10pt]
&& \hspace{-0.5cm} \psi_{s_1 \ldots s_\nnu}^\dagger(p,\alpha) \equiv
\prod_{n=1}^\nnu  \sum_{t_n=0}^{s_n}
\zeta_n^{s_n -t_n} \alpha_n^{I_1^n}\ldots\alpha_n^{I_{t_n}^n} \,
\psi_{s_1\ldots s_\nnu}^{\dagger I_1^1\ldots I_{t_1}^1\ldots I_1^\nnu \ldots
I_{t_\nnu}^\nnu }(p)\,.
\nonumber
\eeq
In \rf{fer01nnn}-\rf{manold-07122011-15} and below, the spinor indices of the
fermionic oscillators and component fermionic fields are implicit. We note
that, for massless field, ket-vector \rf{fer01nnn} is a degree-$s_n$
homogeneous polynomial in the oscillator $\alpha_n^I$, while, for massive
field, ket-vector \rf{fer01nnn} is a degree-$s_n$ homogeneous polynomial in
the oscillators $\alpha_n^I$, $\zeta_n$.

In order for ket-vector \rf{fer01nnn} to describe the irreps of the $so(d-2)$
and $so(d-1)$ algebras we should impose certain algebraic constraints on the
ket-vector. But to avoid unnecessary complications, we impose only
$\gamma$-tracelessness constraints of $so(d-2)$ and $so(d-1)$ algebras,
\beq
\label{fer12mix} && \hspace{-1cm} \gamma^I \bar\alpha_n^I \theta
|\psi_{s_1\ldots s_\nnu}(p,\alpha)\rangle = 0 \,,
\nonumber\\[-11pt]
&& \hspace{6cm} n=1,\ldots,\nnu\,, \qquad  \hbox{ for massless field},\qquad
\\[-11pt]
\label{fer13mix} && \hspace{-1cm} p_\eta  \gamma^I \bar\alpha_n^I
|\psi_{s_1\ldots s_\nnu}(p,\alpha)\rangle = 0\,;
\nonumber\\[7pt]
\label{fer05mix} && \hspace{-1cm} ( \gamma^I \bar\alpha_n^I + \gamma_*
\bar\zeta_n )\theta |\psi_{s_1\ldots s_\nnu}(p,\alpha)\rangle = 0 \,,
\nonumber\\[-11pt]
&& \hspace{6cm}  n=1,\ldots,\nnu\,, \qquad  \hbox{ for massive field}.\qquad
\\[-11pt]
&& \hspace{-1cm} p_\eta ( \gamma^I \bar\alpha_n^I - \gamma_*
\bar\zeta_n ) |\psi_{s_1\ldots s_\nnu}(p,\alpha)\rangle = 0\,,
\nonumber
\eeq
This implies that ket-vector \rf{fer01nnn} actually
describes finite sets of massless and massive fields.

Throughout this paper, to discuss the light-cone gauge description of
arbitrary spin mixed-symmetry fermionic fields an equal footing we use the
ket-vector defined by
\beq \label{manold-08122011-01}
|\psi(p,\alpha)\rangle \equiv \sum_{s_1,\ldots, s_\nnu = 0}^{\infty}\,\,
|\psi_{s_1 \ldots s_\nnu }(p,\alpha)\rangle\,.
\eeq

{\bf Fermionic massless and massive totally symmetric fields}. In order to
obtain the light-cone gauge description of a fermionic totally symmetric
field we use the bosonic oscillators $\alpha^I$ for the discussion of
massless field and  the bosonic oscillators $\alpha^I$, $\zeta$ for the
discussion of massive field. In addition to those bosonic oscillators, we use
the fermionic oscillators $\theta^\asf$, $p_{\theta\asf}$, $\eta^\asf$,
$p_{\eta\asf}$. Using such oscillators, we collect physical D.o.F of massless
and massive totally symmetric spin-$(s+\half)$ fermionic fields into the
ket-vector defined by
\beq
&& \label{fer01} |\psi_s(p,\alpha)\rangle =   ( p_\theta \psi_s(p,\alpha) +
\psi_s^\dagger(p,\alpha)\eta) |0\rangle\,,
\eeq
where we use the notation
\beq
\label{fer09} &&  \psi_s(p,\alpha) \equiv
 \alpha^{I_1} \ldots \alpha^{I_s}  \,  \psi^{I_1\ldots
I_s}(p)\,,
\nonumber\\[-8pt]
&& \hspace{10cm} \hbox{massless field},\qquad
\\[-8pt]
\label{fer10} &&  \psi_s^\dagger(p,\alpha) \equiv \alpha^{I_1} \ldots
\alpha^{I_s} \,  \psi^{\dagger\, I_1\ldots I_s}(p)\,,
\nonumber\\[6pt]
\label{fer02} &&  \psi_s(p,\alpha) \equiv \sum_{t=0}^s \zeta^{s-t}
\alpha^{I_1} \ldots \alpha^{I_t}  \,  \psi^{I_1\ldots I_t}(p)\,,
\nonumber\\[-12pt]
&& \hspace{10cm} \hbox{massive field}.\qquad
\\[-12pt]
\label{fer03} &&  \psi_s^\dagger(p,\alpha) \equiv \sum_{t=0}^s \zeta^{s-t}
\alpha^{I_1} \ldots \alpha^{I_t}  \,  \psi^{\dagger\, I_1\ldots I_t}(p)\,,
\nonumber
\eeq
In \rf{fer01}-\rf{fer03} and the subsequent expressions, we do not show the
spinor indices of the oscillators and component fermionic fields explicitly.  For
massless field, ket-vector \rf{fer01} is a degree-$s$ homogeneous polynomial
in the oscillator $\alpha^I$, while, for massive field,  ket-vector
\rf{fer01} is degree-$s$ homogeneous polynomial in the oscillators
$\alpha^I$, $\zeta$,
\beq
\label{fer11} &&  ( \alpha^I\bar\alpha^I - s)|\psi_s (p,\alpha)\rangle=0\,,
\hspace{5cm} \hbox{for massless field},\qquad
\\
\label{fer04} && ( \alpha^I\bar\alpha^I + \zeta\bar\zeta -
s)|\psi_s(p,\alpha)\rangle=0\,, \hspace{4cm} \hbox{for massive field}.\qquad
\eeq
We recall that physical D.o.F of massless and massive fields in
$d$-dimensional Minkowski space are described by irreps of the respective
$so(d-2)$ and $so(d-1)$ algebras. For the ket-vector \rf{fer01} to be a
carrier of the respective $so(d-2)$ and $so(d-1)$ algebra irreps, we impose
the following $\gamma$-tracelessness constraints:
\beq
\label{fer12} && \gamma^I \bar\alpha^I \theta |\psi_s(p,\alpha)\rangle = 0
\,,
\nonumber\\[-12pt]
&& \hspace{10cm} \hbox{for massless field},\qquad
\\[-12pt]
\label{fer13} && p_\eta  \gamma^I \bar\alpha^I |\psi_s(p,\alpha)\rangle =
0\,,
\nonumber\\[7pt]
\label{fer05} && ( \gamma^I \bar\alpha^I + \gamma_* \bar\zeta )\theta
|\psi_s(p,\alpha)\rangle = 0 \,,
\nonumber\\[-12pt]
&& \hspace{10cm} \hbox{for massive field}.\qquad
\\[-12pt]
\label{fer06} && p_\eta ( \gamma^I \bar\alpha^I - \gamma_* \bar\zeta )
|\psi_s(p,\alpha)\rangle = 0 \,,
\nonumber
\eeq
To develop the light-cone gauge description of arbitrary spin totally
symmetric fermionic fields on an equal footing we use the ket-vector defined by
\be
\label{fer07} |\psi(p,\alpha)\rangle \equiv \sum_{s=
0}^{\infty}\,\,|\psi_s(p,\alpha)\rangle\,. \ee

\subsection{ Realization of Poincar\'e algebra symmetries on light-cone gauge
fields}

We now discuss a realization of the Poincar\'e algebra on the space of
massless and massive light-cone gauge fields. A representation of kinematical
generators \rf{kinnewgen} in terms of differential operators acting on the
ket-vectors
$\phik$ and $\psik$ is given by%
\footnote{ In this paper, without loss of generality, we analyze the
Poincar\'e algebra generators and their commutators for $x^+=0$.}
\beq
\label{intver6}&& P^I=p^I\,, \qquad  \qquad\quad P^+=\beta\,,
\\[5pt]
\label{intver9}&& J^{+I}=\partial_{p^I} \beta\,, \qquad\quad
J^{+-}=\partial_\beta \beta + M^{+-}\,, \qquad
J^{IJ}=p^I\partial_{p^J}-p^J\partial_{p^I}+M^{IJ}\,,\qquad
\\[5pt]
&& \hspace{3.6cm} \beta\equiv p^+\,,\qquad
\partial_\beta\equiv \partial/\partial \beta\,, \quad
\partial_{p^I}\equiv \partial/\partial p^I\,,
\eeq
where $M^{+-}$ is a spin operator of $so(1,1)$ algebra, while $M^{IJ}$ is a
spin operator of the $so(d-2)$ algebra,
\be\label{intver13} [M^{IJ},M^{KL}] = \delta^{JK}M^{IL} + 3\hbox{ terms}.\ee
The representation of dynamical generators \rf{dynnewgen} in terms of
differential operators acting on the ket-vectors $\phik$ and $\psik$ is
given by
\beq
\label{intver8} && P^-= p^-\,,\qquad p^- \equiv -\frac{p^Ip^I +
\mas^2}{2\beta}\,,
\\[3pt]
\label{intver12} && J^{-I}=-\partial_{\beta}p^I + \partial_{p^I}P^-
+\frac{1}{\beta}(M^{IJ}p^J + \mas M^I) - \frac{p^I}{\beta}M^{+-}\,,
\eeq
where $\mas$ is the mass parameter and $M^I$ is a spin operator
transforming in the vector representation of the $so(d-2)$ algebra.
This operator satisfies the commutators
\be\label{intver14} [M^I,M^{JK}] = \delta^{IJ}M^K -\delta^{IK}M^J \,,
\qquad  [M^I,M^J ] = -M^{IJ}\,. \ee

From \rf{intver13},\rf{intver14}, we see that the operators $M^{IJ}$ and
$M^I$ satisfy commutators of the $so(d-1)$ algebra. The particular form of
the operators $M^{+-}$, $M^{IJ}$, and $M^I$ depends on the choice of the
realization of spin D.o.F of physical fields. We now present realization of
these operators on space of the ket-vectors $\phik$ and $\psik$ discussed in
Section \ref{lcgmasfield} (see \rf{manold-07122011-10},
\rf{manold-07122011-09}, \rf{manold-08122011-01}, \rf{fer07}). Realization
of the spin operator $M^{+-}$ on space of the ket-vectors $\phik$ and $\psik$ is
given by
\beq
&& M^{+-} =  0\,, \hspace{5.2cm} \hbox{for bosonic fields},
\\
&& M^{+-} = - \half p_\theta \theta - \half \eta p_\eta\,, \hspace{2.8cm}
\hbox{for fermionic fields}.
\eeq
Realization of the operators $M^{IJ}$ and $M^I$ on space of the ket-vectors
$\phik$ and $\psik$ corresponding to the bosonic and fermionic mixed-symmetry
massive fields is given by
\beq
&&\hspace{-1cm} M^{IJ}  =  \sum_{n=0}^\nnu \alpha_n^I\bar\alpha_n^J -
\alpha_n^J\bar\alpha_n^I \,,
\nonumber\\[-11pt]
&& \hspace{8.5cm} \hbox{for mix.-symm. boson. field},
\\[-11pt]
&& \hspace{-1cm} M^I  = \sum_{n=0}^\nnu \zeta_n\bar\alpha_n^I
-\alpha_n^I\bar\zeta_n\,,
\nonumber\\
&& \hspace{-1cm} M^{IJ} = \sum_{n=0}^\nnu \alpha_n^I\bar\alpha_n^J
-\alpha_n^J\bar\alpha_n^I + \half p_\theta \gamma^{IJ}\theta + \half p_\eta
\gamma^{IJ} \eta\,,
\nonumber\\[-11pt]
&& \hspace{8.5cm} \hbox{for mix.-symm. ferm. field}.\qquad
\\[-11pt]
&& \hspace{-1cm} M^I = \sum_{n=0}^\nnu \zeta_n\bar\alpha_n^I
-\alpha_n^I\bar\zeta_n - \half p_\theta \gamma^I\gamma_* \theta + \half
p_\eta \gamma^I\gamma_* \eta\,,
\nonumber
\eeq
Realization of the operators $M^{IJ}$ and $M^I$ on space of the ket-vectors
$\phik$ and $\psik$ corresponding to the bosonic and fermionic totally
symmetric massive fields takes the form
\beq
&&\hspace{-1cm} M^{IJ}  =  \alpha^I\bar\alpha^J - \alpha^J\bar\alpha^I \,,
\nonumber\\[-11pt]
&& \hspace{8.5cm} \hbox{for mix.-symm. boson. field},
\\[-11pt]
&& \hspace{-1cm} M^I  = \zeta\bar\alpha^I - \alpha^I\bar\zeta\,,
\nonumber\\
&& \hspace{-1cm} M^{IJ} = \alpha^I\bar\alpha^J
-\alpha^J\bar\alpha^I + \half p_\theta \gamma^{IJ}\theta + \half p_\eta
\gamma^{IJ} \eta\,,
\nonumber\\[-11pt]
&& \hspace{8.5cm} \hbox{for mix.-symm. ferm. field}.\qquad
\\[-11pt]
&& \hspace{-1cm} M^I = \zeta\bar\alpha^I
-\alpha^I\bar\zeta - \half p_\theta \gamma^I\gamma_* \theta + \half
p_\eta \gamma^I\gamma_* \eta\,,
\nonumber
\eeq
From \rf{intver12}, we see that in the massless limit, $\mas \rightarrow 0$,
the generators of the Poincar\'e algebra are independent of the operator
$M^I$. This implies that the free light-cone gauge dynamics of massive fields
have a smooth limit to the free light-cone gauge dynamics of massless fields.

The above expressions provide a realization of the Poincar\'e algebra in
terms of  differential operators acting on the physical fields collected into
the ket-vectors $\phik$ and $\psik$. We now write a field theoretical
realization of this algebra in terms of the ket-vectors $|\phi\rangle$,
$|\psi\rangle$.  As mentioned above the kinematical generators $G^\kin$ are
realized quadratically in the ket-vectors $|\phi\rangle$, $|\psi\rangle$,
while the dynamical generators $G^\dyn$ are realized non-linearly. At the
quadratic level, both $G^\kin$ and $G^\dyn$ admit the representation
\beq \label{fierep} G_\smpt &= & G_\smpt^\bos + G_\smpt^\fer\,,
\\
\label{fierepbos} &&  G_\smpt^\bos=\int d^{d-1}p\, \langle\phi(p)| \beta G
|\phi(p)\rangle\,,
\\
\label{fierepfer} && G_\smpt^\fer=\int d^{d-1}p\, \langle\psi(p)| G
|\psi(p)\rangle\,,
\eeq
$d^{d-1}p \equiv d\beta d^{d-2}p$, where $G$ are the differential operators
given in \rf{intver6},\rf{intver9}, \rf{intver8}, \rf{intver12} and the
notation $G_\smpt$ is used for the field theoretical free generators. In
\rf{fierepbos}, \rf{fierepfer} and below, the bra-vectors are defined as
$\langle\phi(p)|\equiv (|\phi(p)\rangle)^\dagger$, $\langle\psi(p)|\equiv
(|\psi(p)\rangle)^\dagger$. The bosonic ket-vector $|\phi\rangle$ and
fermionic ket-vector $|\psi\rangle$ satisfy the Poisson-Dirac commutators
\beq
\label{bascomrel} && {}
[\,|\phi(p,\alpha)\rangle\,,\,|\phi(p^\prime\,,\alpha^\prime)\rangle\,]
\bigr|_{equal\, x^+} =  \frac{1}{2\beta} \delta^{d-1}(p+p^\prime) |\rangle
|\rangle'\,,
\\
\label{bascomrelfer} && {}
[\,|\psi(p,\alpha)\rangle\,,\,|\psi(p^\prime\,,\alpha^\prime)\rangle\,]
\bigr|_{equal\, x^+} = \frac{1}{2} \delta^{d-1}(p+p^\prime) |\rangle
|\rangle'\,,
\eeq
where $|\rangle|\rangle'$ is defined by
\beq
|\rangle |\rangle' & = & \left\{ \begin{array}{ll}
\Pi|0\rangle |0'\rangle
&  \hspace{2.3cm} \hbox{for bosonic fields},
\\[7pt]
(p_\theta \eta' + \eta p_\theta')\Pi|0\rangle |0'\rangle
\qquad & \hspace{2.3cm} \hbox{for fermionic fields}\,,
\end{array}\right.
\eeq
and $\Pi$ is a unity operator on a space of ket-vectors given in
\rf{manold-07122011-01},\rf{manold-07122011-02}. Note that, in terms of
component fermionic fields \rf{manold-07122011-14},\rf{manold-07122011-15},
the commutator in \rf{bascomrelfer} is realized as the anti-commutator. With
these definitions, we have the standard commutators
\beq
\label{phig} && {} [ |\phi\rangle,G_\smpt\,]\bigr|_{equal\, x^+} =
G|\phi\rangle\,,
\\
\label{phigfer} && {}  [ |\psi\rangle,G_\smpt\,]\bigr|_{equal\, x^+} =
G|\psi\rangle\,.
\eeq
In the framework of the Lagrangian approach, the light-cone gauge action
takes the standard form
\be \label{lcact} S = \int dx^+  d^{d-1} p\,\, \Bigl( \langle \phi(p)|{\rm
i}\, \beta
\partial^- |\phi(p)\rangle + \langle \psi(p)|{\rm
i}\,
\partial^- |\psi(p)\rangle\Bigr) +\int dx^+ P^-\,, \ee
where $P^-$ is the light-cone frame Hamiltonian. Representation for the
light-cone action given in \rf{lcact} is  valid for the free and for the
interacting theory. The Hamiltonian of free theory can be obtained by using
relations \rf{intver8}, \rf{fierep}.

The internal symmetry can be incorporated into the theory under consideration
by adopting the Chan--Paton method in string theory \cite{Paton:1969je} (see
e.g. Ref. \cite{Metsaev:1991nb}).

\newsection{Equations for cubic interaction
vertices}\label{GENVERsec}

We now discuss the Poincar\'e algebra dynamical generators
given in \rf{dynnewgen}. In theories of interacting fields, the dynamical
generators of the Poinca\'e algebra receive corrections containing higher
powers of physical fields. The dynamical generators can be expanded as
\be\label{GDYN01} G^\dyn=\sum_{n=2}^\infty G^\dyn_\smpn\,, \ee
where $G_\smpn^\dyn$ stands for the functional that has $n$ powers of
physical fields.%
\footnote{ The dynamical generators of  supersymmetric Yang-Mills
theories do not receive corrections of the order higher than four in fields
(see e.g. Refs.\cite{Brink:1982pd}), while the
dynamical generators of (super)gravity theories are nontrivial for all $n\geq
2$ (see e.g. Refs.\cite{Goroff:1983hc}).
Note also that the generators of the closed string field theories, which involve the
graviton field, terminate at cubic correction $G_\smp3^\dyn$
\cite{Green:1983hw}. This implies that, in string theory, the
general covariance is realized in a somewhat nontrivial way. Interesting
discussion of this theme may be found in Ref. \cite{Tseytlin:1986eq}.}
In what follows, the $G_\smpt^\dyn$ and $G_\smp3^\dyn$ contributions
appearing in expansion \rf{GDYN01} will be referred to as the respective
quadratic and cubic dynamical generators. The quadratic dynamical generators for
massless and massive mixed-symmetry fields were discussed in Sec.2. Our
purpose is to find cubic dynamical generators $G_\smp3^\dyn$ for the massless
and massive mixed-symmetry fields. The cubic dynamical generators can be
found by using\\
{\bf i}) restrictions imposed by kinematical symmetries;\\
{\bf ii}) dynamical light-cone principle.

The detailed treatment of restrictions imposed by the kinematical symmetries and
the dynamical light-cone principle may be found in Sec.3 in
Ref.\cite{Metsaev:2005ar}. Therefore to avoid the repetition we briefly
outline our procedure of the derivation of the cubic dynamical generators.

\subsection{ Restrictions imposed by kinematical symmetries}

In this section, we discuss the restrictions imposed by kinematical
symmetries on the cubic dynamical generators $P_\smp3^-$ and $J_\smp3^{-I}$.
Namely, we describe those properties of the cubic dynamical generators that
can be obtained from commutators between $P^-$, $J^{-I}$ and the kinematical
generators given in \rf{kinnewgen}. This is to say that, using commutators
between kinematical generators \rf{kinnewgen} and the dynamical generators
$P^-$, $J^{-I}$, we find the following representation for the cubic dynamical
generators:
\beq
\label{pm1} && P_\smp3^-  = \int d\Gamma_3  \langle \Phi_\smp3|
p_\smp3^-\rangle\,,
\\
\label{npoi4} && J_\smp3^{-I} =\int d\Gamma_3\,\Bigl(\langle \Phi_\smp3 |
j_\smp3^{-I}\rangle - \frac{1}{3} \Bigl(\sum_{a=1}^3
\partial_{p_a^I} \langle\Phi_\smp3|\Bigr)|p_\smp3^-\rangle\Bigr)\,,
\eeq
where we use the notation
\beq
&& |p_\smp3^-\rangle \equiv p_\smp3^- \prod_{a=1}^3 |0\rangle_a \,, \qquad
|j_\smp3^{-I}\rangle \equiv j_\smp3^{-I} \prod_{a=1}^3 |0\rangle_a \,,
\\
\label{pm1NN1} &&  \langle \Phi_\smp3| \equiv  \langle
\psi(p_1,\alpha_1)|\langle \psi(p_2,\alpha_2)|\langle
\phi(p_3,\alpha_3)|\,,\qquad\qquad
\\
\label{delfun01} && d\Gamma_3 \equiv (2\pi)^{d-1} \delta^{d-1}(\sum_{a=1}^3
p_a) \prod_{a=1}^3 \frac{d^{d-1} p_a}{(2\pi)^{(d-1)/2}} \,. \eeq
Here and below, the indices $a,b=1,2,3$ label three interacting fields
involved in cubic vertices.

For generosity, we assume that the bra-vectors $\langle\psi(p_1,\alpha_1)|$,
$\langle\psi(p_2,\alpha_2)|$ describe two fermionic fields having different
mass values and internal charges. Note that, in general, the generators
$P_\smp3^-$ \rf{pm1} are not hermitian. Hermitian $P_\smp3^-$ can be obtained
in two obvious ways. The first hermitian representative of $P_\smp3^-$ is
given by $P_\smp3^- + \hbox{h.c.}$, while the second hermitian representative
of $P_\smp3^-$ is given by $\i P_\smp3^- + \hbox{h.c.}$. So, in general,
there are two families of hermitian generators $P_\smp3^-$. At the end of
Section \ref{seubsub-11-11}, we present some examples of those two families
of hermitian vertices.

Densities $p_\smp3^-$, $j_\smp3^{-I}$ in \rf{pm1},\rf{npoi4} depend on the
momenta $\Po^I$, $\beta_a$, $a=1,2,3$, and variables related to the spin
D.o.F, which we denote by $\alpha$:
\beq
\label{pmpm} && p_\smp3^- =p_\smp3^-(\Po,\beta_a;\, \alpha)\,,
\\
&& j_\smp3^{-I} = j_\smp3^{-I}(\Po,\beta_a;\, \alpha)\,,
\\
&& \hspace{1cm} \label{defpi} \Po^I \equiv
\frac{1}{3}\sum_{a=1}^3\check{\beta}_a p_a^I\,, \qquad \check{\beta}_a\equiv
\beta_{a+1}-\beta_{a+2}\,, \quad \beta_a\equiv \beta_{a+3}\,.
\eeq
We note also that the densities $p_\smp3^-$, $j_\smp3^{-I}$  should satisfy the following
equations
\beq
\label{kinsod} &&  \Jbf^{IJ} |p_\smp3^-\rangle =0\,,
\\
\label{honcon04} && (\Po^I\partial_{\Po^I} +
\sum_{a=1}^3\beta_a\partial_{\beta_a} - \Mbf^{+-\dagger}) |p_\smp3^-\rangle
=0\,,
\\
\label{neweqqqq01} && \Jbf^{IJ} |j_\smp3^{-K}\rangle + \delta^{IK}
|j_\smp3^{-J}\rangle - \delta^{JK} |j_\smp3^{-I}\rangle =0\,,
\\
\label{neweqqqq02} && \Bigl(\Po^I\partial_{\Po^I} + \sum_{a=1}^3
\beta_a\partial_{\beta_a} - \Mbf^{+-\dagger}\Bigr) |j_\smp3^{-K}\rangle =0\,,
\eeq
where we use the notation
\beq
\label{JIJp3} && \Jbf^{IJ} \equiv \Lbf^{IJ} + \Mbf^{IJ}\,,
\qquad
\Lbf^{IJ} \equiv \Po^I \partial_{\Po^J}  - \Po^J
\partial_{\Po^I} \,,\qquad  \ \ \ \
\\
&& \Mbf^{IJ}\equiv \sum_{a=1}^3 M^{\sma IJ}\,,\qquad \quad
\Mbf^{+-\dagger}\equiv \sum_{a=1}^3 M^{\sma +-\dagger}\,.
\eeq
From \rf{pm1},\rf{npoi4}, we see that the problem of finding the cubic
generators $P_\smp3^-$, $J_\smp3^{-I}$ is reduced to the problem of finding
the respective densities $p_\smp3^-$, $j_\smp3^{-I}$ subject to kinematical
symmetry restrictions given in \rf{kinsod}-\rf{neweqqqq02}. The restrictions
given in \rf{kinsod}-\rf{neweqqqq02} by themselves are not enough to
determine the densities $p_\smp3^-$, $j_\smp3^{-I}$ uniquely. We proceed
therefore to the discussion of the light-cone dynamical principle.

\subsection{ Restrictions imposed by light-cone dynamical principle}

To determine the densities $p_\smp3^-$, $j_\smp3^{-I}$ uniquely we use method
which we refer to as the {\it light-cone dynamical principle}. The light-cone
dynamical principle is summarized as the following three steps:

\noindent {\bf Step 1}. Using commutators of the Poincar\'e algebra for the
dynamical generators $P^-$ and $J^{-I}$, we find restrictions imposed on the
densities $p_\smp3^-$, $j_\smp3^{-I}$. At this step, we make sure that the
density $j_\smp3^{-I}$ can be expressed in terms of the density $p_\smp3^-$.

\noindent {\bf Step 2}. We require the densities $p_\smp3^-$, $j_\smp3^{-I}$
to be polynomials in the momentum $\Po^I$. We refer to this requirement as
the light-cone locality condition.

\noindent {\bf Step 3}. We find the density $p_\smp3^-$ that cannot be removed by
field redefinitions.

We now proceed with discussing the restrictions imposed by the light-cone
dynamical principle on the densities $p_\smp3^-$, $j_\smp3^{-I}$. In what follows, the
density $p_\smp3^-$ will be referred to as cubic interaction vertex.
We now discuss the three steps of our method in turn.

{\bf Step 1}. At this step, we find the restrictions
imposed by the Poincar\'e algebra commutators between the dynamical
generators. All that is required is to consider the commutators
\beq
\label{cubeq01} && {} [\,P^-\,,\,J^{-I}\,]=0\,,
\\
\label{cubeq02} && {} [\,J^{-I}\,,\,J^{-J}\,]=0\,.
\eeq
In the cubic approximation, commutators \rf{cubeq01} lead to the equation for
the densities $|p_\smp3^-\rangle$, $|j_\smp3^{-I}\rangle$ given by
\be\label{cubver3} \Pbf^- |j_\smp3^{-I}\rangle = -\Jbf^{-I\dagger}
|p_\smp3^-\rangle\,, \ee
where $\Pbf^-$, $\Jbf^{-I\dagger}$ are defined as

\beq
\label{cubver15} && \hspace{-1cm} \Pbf^- \equiv
\frac{\Po^I\Po^I}{2\hat{\beta}} -\sum_{a=1}^3 \frac{\mas_a^2}{2\beta_a}\,,
\\
\label{cubeq04}&& \hspace{-1cm} \Jbf^{-I\dagger} \equiv
-\frac{1}{3\hat\beta}\, \XX^I\,,
\\
&& \label{cubeq06}  \XX^I \equiv X^{IJ} \Po^J + X^I +
X\partial_{\Po^I}\,,
\\
\label{harver02} && X^{IJ} \equiv \sum_{a=1}^3 \check\beta_a((
\beta_a\partial_{\beta_a} - M^{\sma +-\dagger}) \delta^{IJ} - M^{\sma IJ})\,,
\\[3pt]
\label{harver03} && X^I \equiv \sum_{a=1}^3 \frac{3\hat\beta \mas_a}{\beta_a}
M^{\sma I}\,,
\\
\label{harver04} && X \equiv -\sum_{a=1}^3 \frac{\hat{\beta}
\check{\beta}_a \mas_a^2}{2\beta_a}\,,
\\
\label{cubver16}&& \hat{\beta} \equiv \beta_1\beta_2\beta_3\,. \eeq
Taking \rf{cubeq04} into account, we can rewrite Eq.\rf{cubver3} as
\be\label{cubver13} |j_\smp3^{-I}\rangle =  \frac{1}{3\hat\beta \Pbf^-} \XX^I
|p_\smp3^-\rangle\,, \ee
i.e., as we have promised, the density $j_\smp3^{-I}$ is expressed in terms
of vertex $p_\smp3^-$ \rf{pmpm}. Plugging $j_\smp3^{-I}$ \rf{cubver13} into
\rf{cubeq02}, we make sure that, in the cubic approximation, commutators
\rf{cubeq02} are satisfied automatically. Thus, we see that commutators
\rf{cubeq01},\rf{cubeq02} amount to relation \rf{cubver13}.

Before proceeding to the second step of our method, we note that plugging
$j_\smp3^{-I}$ \rf{cubver13} into Eqs.\rf{neweqqqq01}, \rf{neweqqqq02}, we
make sure that these equations are satisfied provided the density $p_\smp3^-$
is satisfied Eqs.\rf{kinsod}, \rf{honcon04}. Also, we note that, we exhausted
all commutators of the Poincar\'e algebra in the cubic approximation.
Equations for cubic vertex \rf{kinsod}, \rf{honcon04} and relation
\rf{cubver13} provide the complete list of restrictions imposed by
commutators of the Poincar\'e algebra on the densities $p_\smp3^-$,
$j_\smp3^{-I}$. We see that the restrictions imposed by commutators of the
Poincar\'e algebra by themselves are not sufficient to determine the cubic
interaction vertex $p_\smp3^-$ uniquely. To find the cubic interaction vertex $p_\smp3^-$
we proceed to the next steps of our  method.

{\bf Step 2}.  At this step, we impose the light-cone locality condition: we require
the densities $p_\smp3^-$, $j_\smp3^{-I}$ to be polynomials in the momentum $\Po^I$.
As regards the vertex $p_\smp3^-$, we require this vertex to be local (i.e.
polynomial in $\Po^I$) from the very beginning. However it is clear from
relation \rf{cubver13} that a local $p_\smp3^-$ does not lead automatically
to a local density $j_\smp3^{-I}$. From \rf{cubver13}, we see that the
light-cone locality condition for $j_\smp3^{-I}$ amounts to the equation
\be\label{loccon01} \XX{}^I |p_\smp3^-\rangle  = \Pbf^- |V^I\rangle\,, \ee
where a vertex $|V^I\rangle$ is restricted to be polynomial in
$\Po^I$.

{\bf Step 3}. The last requirement we impose on the vertex $p_\smp3^-$ is
motivated by the desire to deal with vertex that cannot be removed by field
redefinitions. Using local (i.e. polynomial in the transverse momentum
$\Po^I$) field redefinitions, we can remove in the cubic vertex
$p_\smp3^-$ those terms that are proportional to $\Pbf^-$ (see Appendix B in
Ref.\cite{Metsaev:2005ar}.). Since we are interested in the cubic vertex
$p_\smp3^-$ that cannot be removed by field redefinitions, we impose the
following restriction:
\be\label{cubver17} |p_\smp3^-\rangle \ne \Pbf^- |V\rangle\,,\ee
where $\Pbf^-$ is given in \rf{cubver15}, and a vertex $|V\rangle$ is
restricted to be polynomial $\Po^I$.

Altogether, equations \rf{cubver13}-\rf{cubver17} exhaust restrictions
imposed by the light-cone dynamical principle.

{\bf Complete system of equations for cubic interaction vertex}.  We now
collect equations imposed by the kinematical symmetries and the light-cone
dynamical principle on vertex $p_\smp3^-$ \rf{pmpm}:
\beq
\label{basequnew0001} && \hspace{-2cm} \Jbf^{IJ} |p_\smp3^-\rangle = 0\,,
\hspace{6cm} so(d-2) \hbox{ invariance},
\\
\label{basequnew0002} && \hspace{-2cm} (\Po^I\partial_{\Po^I} +
\sum_{a=1}^3\beta_a\partial_{\beta_a} - \Mbf^{+-\dagger}) | p_\smp3^- \rangle
=0\,,\hspace{1.5cm} so(1,1) \hbox{ invariance},
\\
&& \hspace{-2cm}  \XX{}^I |p_\smp3^-\rangle  = \Pbf^- |V^I \rangle\,,
\nonumber\\
&&  \hspace{-2cm} |p_\smp3^-\rangle \ne \Pbf^- |V\rangle\,,
\nonumber\\[-10pt]
\label{cubver13nn} && \hspace{6.3cm} \hbox{ light-cone dynamical principle}.
\\[-10pt]
&&  \hspace{-2cm}  |j_\smp3^{-I}\rangle =  \frac{1}{3\hat\beta}
|V^I\rangle\,,
\nonumber\\
&& \hspace{-2cm} |p_\smp3^-\rangle\,,\ \ |V\rangle\,,\ \  |V^I\rangle \hbox{
are polynomials in } \Po^I\,,
\nonumber
\eeq
Equations \rf{basequnew0001},\rf{basequnew0002} reflect the invariance of the vertex
$|p_\smp3^-\rangle$ under the respective $so(d-2)$ and $so(1,1)$ rotations.
Equations given in \rf{cubver13nn}  summarize the restrictions obtained by
using the light-cone dynamical principle.

To summarize, equations \rf{basequnew0001}-\rf{cubver13nn} constitute a
complete system of equations on vertex $p_\smp3^-$ \rf{pmpm}. These
equations allow us to determine the cubic vertex uniquely (up to coupling
constants).

{\bf General structure of cubic vertices in Lorentz covariant and light-cone
gauge approaches}. Manifestly Lorentz invariant vertices are beyond the scope
of this paper. However for the reader convenience and in order to introduce
some terminology we discuss general structure of covariant vertices which are
related to light-cone gauge vertices we study in this paper. Let
$\Psi=\Psi(x)$ be a set of fermionic arbitrary spin Dirac fields of the
Lorentz algebra, while $\Phi = \Phi(x)$ is a set of arbitrary spin bosonic
fields of the Lorentz algebra, where the argument $x$ stands for space-time
coordinates. Let $\gamma$ be Dirac $\gamma$-matrices, while $\partial$ stands
for derivatives w.r.t the space-time coordinates. In general, vertices for
massless and massive fields involve contributions with different powers of
$\partial$. Therefore, to classify cubic vertices we need two labels at
least. We can use labels $k_\min^\cov$ and $k_\max^\cov$ which are the
respective minimal and maximal numbers of $\partial$ appearing in a
manifestly Lorentz invariant vertex. Using this notation we note that, in
Lorentz covariant approach, on-shell cubic vertices involving two fermionic
fields and one bosonic field can schematically be presented as power series
expansion in $\partial$,
\be \label{manold-09122011-01} \LL^\cov  = \bar\Psi M_\max
\partial^{k_\max^\cov} \Psi\, \Phi + \ldots + \bar\Psi M_\min
\partial^{k_\min^\cov} \Psi\, \Phi\,, \ee
where matrices $M_\max$ and $M_\min$ are constructed out of flat metric
tensor and Dirac $\gamma$-matrices. Let us restrict our attention to vertices
that do not involve Levi-Civita antisymmetric tensor and matrix  $\Gamma_*$
(for definition the matrix $\Gamma_*$, see Appendix A). We shall refer to
such vertices as parity invariant vertices. For the parity invariant vertices,
the matrix $M_\max$ involves either even number of gamma-matrices or odd
number of gamma-matrices. Using this, we separate vertices into two groups:
K-vertices and F-vertices. This is to say that if matrix $M_\max$ in
\rf{manold-09122011-01} involves even number of $\gamma$-matrices then the
vertex is referred to as K-vertex, while, if matrix $M_\max$  in
\rf{manold-09122011-01} involves odd number of $\gamma$-matrices then the
vertex is referred to as F-vertex. For example, using the nomenclature of K-
and F-vertices, we note that the covariant vertex $\bar\psi\psi\phi$ is
referred to as K-vertex, while the covariant vertex describing
electro-magnetic interaction $\bar\psi\gamma^A\psi\phi^A$ is referred to as
F-vertex in this paper.

In this paper, we restrict our attention to the light-cone vertices
corresponding to the parity invariant cubic vertices
\rf{manold-09122011-01}. Such light-cone gauge cubic vertices can also be
presented in the form similar to one in \rf{manold-09122011-01},
\be \label{manold-10122011-02} \LL^\lc  = \bar\psi M_\max^\lc
\partial^{k_\max^\lc} \psi\, \phi + \ldots + \bar\psi M_\min^\lc
\partial^{k_\min^\lc} \psi\, \phi\,, \ee
where light-cone frame matrices $M_\max^\lc$ and $M_\min^\lc$ are
constructed out of the flat delta-Kronecker of the $so(d-2)$ algebra and light-cone
frame Dirac $\gamma$-matrices. To classify light-cone vertices we also use
nomenclature of K-vertices and F-vertices.  This is to say the light-cone
counterparts of the covariant K- and F-vertices will also be refereed to as
K- and F-vertices respectively. Note however that, for the case of light-cone
K-vertex in \rf{manold-10122011-02}, the matrix $M_\max^\lc$ involves odd
number of $\gamma$-matrices, while,  for the case of light-cone F-vertex in
\rf{manold-10122011-02}, the matrix $M_\max^\lc$ involves even number of
$\gamma$-matrices.

For the examples of on-shell manifestly Lorentz invariant vertices considered
in this paper and their light-cone counterparts, we find the following
relation:
\be k_\max^\cov = k_\max^\lc -1\,.\ee
We note also that for vertices involving two massless fermionic and one
massless bosonic fields one has the relations $k_\max^\cov = k_\min^\cov$,
$k_\max^\lc = k_\min^\lc$. Taking this into account, we introduce the notation
$k^\cov$ for a number of derivatives appearing in on-shell Lorentz invariant
vertices of massless fields and notation $k^\lc$ for a number of transverse
derivatives appearing in light-cone gauge vertices
of massless fields to obtain the following relation%
\footnote{ For the case of vertices involving only massless bosonic fields,
one has the relation $k^\cov = k^\lc$.}
\be k^\cov = k^\lc -1\,,\hspace{2cm} \hbox{ for massless fields}.\ee
Throughout this paper we drop the superscript $\lc$ and use the following
simplified notation:
\be k \equiv k^\lc\,, \qquad k_\max\equiv k_\max^\lc\,, \qquad  k_\min\equiv k_\min^\lc\,.\ee

We now turn to equations for cubic vertices in
\rf{basequnew0001}-\rf{cubver13nn}. Up to this point our treatment has been
applied to vertices for massive as well as massless fields. From now on, we
separately consider vertices involving only massless fields, vertices
involving both massless and massive fields, and vertices involving only
massive fields. Depending on the values of mass parameters of fields entering
cubic vertices, we can separately consider cubic vertices with the following
mass values:
\beq
\label{25012010-man01} &&  \mas_1|_\smF  = 0\,, \qquad \mas_2|_\smF  =
0,\qquad \mas_3|_\smB = 0\,;
\\[9pt]
\label{25012010-man02}&&  \mas_1|_\smF  = \mas_2|_\smF  = 0,\qquad
\mas_3|_\smB  \ne  0\,;
\\
\label{25012010-man03} &&  \mas_1|_\smB  = \mas_2|_\smF  = 0,\qquad
\mas_3|_\smF  \ne  0\,;
\\[9pt]
\label{25012010-man04} && \mas_1|_\smF = \mas_2|_\smF \equiv \mas   \ne 0
,\qquad \mas_3|_\smB = 0\,;
\\
\label{25012010-man05} && \mas_1|_\smB = \mas_2|_\smF \equiv \mas   \ne 0
,\qquad \mas_3|_\smF = 0\,;
\\[9pt]
\label{25012010-man06} && \mas_1|_\smF \ne 0 ,\qquad \mas_2|_\smF\ne 0,\qquad
\mas_1|_\smF \ne \mas_2|_\smF, \qquad \mas_3|_\smB= 0\,;
\\
\label{25012010-man07} && \mas_1|_\smB \ne 0 ,\qquad \mas_2|_\smF\ne 0,\qquad
\mas_1|_\smB \ne \mas_2|_\smF, \qquad \mas_3|_\smF= 0\,;
\\[9pt]
\label{25012010-man08} &&  \mas_1|_\smF  \ne  0\,, \qquad \mas_2|_\smF  \ne
0,\qquad \mas_3|_\smB \ne  0\,.
\eeq
The notation $\mas_a|_\smB$ implies that $\mas_a$ is a mass parameter of
bosonic field carrying the external line index $a$, while the notation
$\mas_b|_\smF$ implies that $\mas_b$ is a mass parameter of fermionic field
carrying the external line index $b$. Vertices with mass parameters in
\rf{25012010-man01} describe interaction of massless fields, while the ones
in \rf{25012010-man08} describe interaction of massive fields. Vertices with
mass parameters given in \rf{25012010-man02},\rf{25012010-man03} describe
interaction of two massless fields with one massive field, while the vertices
in \rf{25012010-man04}- \rf{25012010-man07} describe interaction of two
massive fields with one massless field.

We now present solutions to defining equations
\rf{basequnew0001}-\rf{cubver13nn} for cubic vertices with mass parameters
given in \rf{25012010-man01}-\rf{25012010-man08} in turn. The solutions can
be found by using the same methods we used to analyze cubic vertices for
bosonic fields in Ref.\cite{Metsaev:2005ar} (see Section 4.2 and Appendix D
in Ref.\cite{Metsaev:2005ar}). Therefore to avoid the repetitions we just
present the results for cubic vertices and discuss their properties.

\newsection{ Cubic vertices for massless fermionic and bosonic
fields} \label{Solcubintversec}

In this section, we discuss the cubic interaction vertex for the massless
mixed-symmetry fields \rf{25012010-man01}. We consider vertex involving two
massless mixed-symmetry fermionic fields having the mass parameters
$\mas_1=0$ and $\mas_2=0$ and one massless mixed-symmetry bosonic field
having the mass parameter $\mas_3=0$:
\be\label{0001}
\begin{array}{lll}
\mas_1|_\smF = 0\,, \qquad & \mas_2|_\smF = 0\,, \qquad & \mas_3|_\smB=0\,.
\end{array}
\ee
Equations for the vertex involving three massless fields are obtainable from
\rf{cubver13nn} by letting $\mas_a \rightarrow 0 $, $a=1,2,3$. We find the
following two types of solutions for cubic vertices:
\beq
&& \label{0002k} p_\smp3^- = K^\smonetwo V^\Krm(B_n^\sma;\, Z_{mnq})\,,
\hspace{2cm} \hbox{ K-vertex};
\\[3pt]
&& \label{0002f} p_\smp3^- = F_n V^\Frm(B_n^\sma; \, Z_{mnq})\,,
\hspace{2.5cm} \hbox{ F-vertex};
\eeq
where $V^\Krm$, $V^\Frm$ are arbitrary polynomials of $B_n^\sma$ and $Z_{mnq}$
and we use the notation
\beq
\label{manold-12122011-02} &&  K^{\smonetwo} = \frac{1}{\beta_1\beta_2}
p_{\theta_1} \Po^I\gamma^I \gamma_* \eta_2\,,
\\[3pt]
\label{maslesfdef01}&&  F_n = \frac{1}{\beta_1\beta_2} p_{\theta_1}\Bigl(
\frac{\check{\beta}_3}{\beta_3} \Po^I  - \gamma^{IJ}\Po^J\Bigr) \eta_2
\alpha_n^{\smthree I}\,,
\eeq
\beq
\label{0003}
&& B_n^\sma \equiv \frac{\alpha_n^{\sma I}\Po^I}{\beta_a}\,,\qquad a=1,2,3\,,
\\[3pt]
\label{massZdef01} && Z_{mnq} \equiv  B_m^\smone \alpha_{nq}^\smtwothree +
B_n^\smtwo \alpha_{qm}^\smthreeone + B_q^\smthree \alpha_{mn}^\smonetwo \,,
\\[10pt]
\label{amnabdef}
&&\hspace{1cm} \alpha_{mn}^\smab \equiv \alpha_m^{\sma I}\alpha_n^{\smb I}\,.
\eeq
We recall that, for the mixed-symmetry fields, the subindices $m,n,q$ take
the following values:
\be\label{20012010-man01} m,n,q = 1,\ldots, \nu\,, \ee
where $\nu$ is arbitrary integer $\nu > 1$. Note that the quantities
$B_n^\sma$, $\alpha_{mn}^\smab$ and $Z_{mnq}$ are the respective degree-1, 2,
and 3 homogeneous polynomials in the oscillators. In what follows, we refer
to degree-1, 2, and 3 homogeneous polynomials in the oscillators as linear,
quadratic, and cubic forms respectively.

We now make comments on the solution obtained.

\noindent {\bf i)}  We see that solution for vertices in \rf{0002k},
\rf{0002f} is governed by the prefactors $K^\smonetwo$, $F_n$, and the
generating functions $V^\Krm$, $V^\Frm$. The generating functions are
arbitrary polynomials of the linear forms $B_n^\sma$ and the cubic forms
$Z_{mnq}$.

\noindent {\bf ii)} Equations for vertices allow solutions for  the
generating functions $V^\Krm$, $V^\Frm$ that depend on the quadratic forms
$\alpha_{mn}^\smoneone$, $\alpha_{mn}^\smtwotwo$,
$\alpha_{mn}^\smthreethree$. However, in view of tracelessness constraint
(see \rf{manold-26122011-01},\rf{fer12mix}), such quadratic forms do not
contribute to the Hamiltonian. Therefore we drop dependence on the quadratic
forms $\alpha_{mn}^\smaa$ in the generating functions  $V^\Krm$, $V^\Frm$.

\noindent {\bf iii)} The prefactors, the linear forms $B_n^\sma$, and cubic
forms $Z_{mnq}$ are homogeneous polynomials in momentum $\Po^I$. Appearance
of the prefactors, linear forms and cubic forms that are homogeneous
polynomials in $\Po^I$ is  characteristic feature of vertices for massless
fields.

\noindent {\bf iv)} We note that the prefactor $K^\smonetwo$ involves only
odd number of transverse $\gamma$-matric, while the prefactors $F_n$ involve
only even number of the transverse $\gamma$-matrices.

\noindent {\bf v)} Solution for the generating functions  $V^\Krm$, $V^\Frm$
given in \rf{0002k}, \rf{0002f} is complete solution, while the solution for
the prefactors $K^\smonetwo$, $F_n$ given in
\rf{manold-12122011-02},\rf{maslesfdef01} is not complete solution. This is
to say that for the vertices of mixed-symmetry fields there are extra
solutions for the prefactors that involve contributions of higher than second
order in $\gamma$-matrices. We note also for the vertices of totally
symmetric fields those extra solutions are trivial.

To understand the remaining interesting properties of the vertices we consider
cubic vertices for totally symmetric massless fields.

\subsection{ Cubic vertices for massless totally symmetric fields }

We now consider the parity invariant cubic interaction vertices for massless
totally symmetric fields. To this end it is sufficient to use one sort of
oscillators, i.e. to set $\nnu = 1$ in \rf{20012010-man01}. To simplify our
presentation of cubic vertices we drop oscillator's subscript $n=1$ and use
the simplified notation $\alpha^I\equiv \alpha_1^I$. Then the cubic vertex is
obtained from solution given in \rf{0002k},\rf{0002f} by using the
identifications
\be\label{simnot01} \alpha^{\sma I} \equiv \alpha_1^{\sma I}\,,\qquad
a=1,2,3\,,\ee
and ignoring contribution of oscillators carrying a subscript $n>1$. Adopting
notation \rf{simnot01} for prefactor $F\equiv F_1$, linear forms $B^\sma
\equiv B_1^\sma$ \rf{0003}, quadratic forms $\alpha^\smab \equiv
\alpha_{11}^\smab$ \rf{amnabdef},  and cubic form $Z\equiv Z_{111}$
\rf{massZdef01}, we get
\beq
&& \label{0002ksym} p_\smp3^- = K^\smonetwo V^\Krm(B^\sma;\, Z)\,,
\hspace{2cm} \hbox{ K-vertex};
\\[3pt]
&& \label{0002fsym} p_\smp3^- = F V^\Frm(B^\sma; \, Z)\,,  \hspace{2.5cm}
\hbox{ F-vertex};
\eeq
where $V^\Krm$, $V^\Frm$ are arbitrary polynomials of $B^\sma$ and $Z$ and we
use the notation
\beq
\label{manold-12122011-01} &&  K^\smonetwo = \frac{1}{\beta_1\beta_2}
p_{\theta_1} \Po^I\gamma^I \gamma_* \eta_2\,,
\\[3pt]
\label{maslesfdef01sym} &&  F = \frac{1}{\beta_1\beta_2} p_{\theta_1}\Bigl(
\frac{\check{\beta}_3}{\beta_3} \Po^I  - \gamma^{IJ}\Po^J\Bigr) \eta_2
\alpha^{\smthree I}\,,
\\[3pt]
\label{0003sym}
&& B^\sma \equiv \frac{\alpha^{\sma I}\Po^I}{\beta_a}\,,\qquad a=1,2,3\,,
\\[3pt]
&\label{massZdef01sym} & Z\equiv  B^\smone \alpha^\smtwothree +  B^\smtwo
\alpha^\smthreeone + B^\smthree \alpha^\smonetwo \,,
\\[5pt]
\label{amnabdefsym}
&& \hspace{1cm} \alpha^\smab \equiv \alpha^{\sma I}\alpha^{\smb I}\,.
\eeq
We now make comment on the solution obtained.

\noindent {\bf i)}  Solution for vertices in \rf{0002ksym}, \rf{0002fsym} is
governed by prefactors $K^\smonetwo$, $F$ and by generating functions
$V^\Krm$, $V^\Frm$. The generating functions are arbitrary polynomials of the
three linear forms $B^\sma$ and the one cubic form $Z$. As compared to the
case for mixed-symmetry fields, the number of linear forms and cubic forms is
decreased and this simplifies considerably the study of the solution.

\noindent {\bf ii)} Equations for vertices allow solutions for $V^\Krm$,
$V^\Frm$ that depend on the quadratic forms  $\alpha^\smoneone$,
$\alpha^\smtwotwo$, $\alpha^\smthreethree$ \rf{amnabdefsym}. However, due to
traceless constraint (see \rf{manold-07122011-07},\rf{fer12}), such quadratic
forms do not contribute to the Hamiltonian. Therefore we drop dependence on
the quadratic forms $\alpha^\smaa$ in the generating functions  $V^\Krm$,
$V^\Frm$.

\noindent {\bf iii)} Solution for the generating functions  $V^\Krm$,
$V^\Frm$ given in \rf{0002ksym},\rf{0002fsym} is complete solution. We note
that, in contrast to the vertices for mixed-symmetry fields,  solution for
the prefactors given in \rf{manold-12122011-01}, \rf{maslesfdef01sym} is also
complete solution.

\subsubsection{ Cubic vertices for totally symmetric
fields with fixed but arbitrary spin values}\label{SolcubintversecN1}

Vertices \rf{0002ksym}, \rf{0002fsym} describe interaction of towers of
massless totally symmetric bosonic and fermionic fields
\rf{manold-07122011-09}, \rf{fer07}. We now consider vertices involving two
massless totally symmetric spin $s^\smone+\half$ and $s^\smtwo+\half$
fermionic fields and one massless totally symmetric spin-$s^\smthree$ bosonic
field:

\be
\begin{array}{lll}
\mas_1 = 0\,, \qquad & \mas_2 = 0\,, \qquad &  \mas_3=0\,,
\\[5pt]
s^\smone + \half \,, & s^\smtwo+\half \,, & s^\smthree\,.
\end{array}
\ee

The massless spin $s^\smone+\half$ and $s^\smtwo + \half$ fermionic fields
are described by the respective ket-vectors $|\psi_{s^\smone }\rangle$ and
$|\psi_{s^\smtwo}\rangle$, while the massless spin-$s^\smthree$ bosonic field
is described by the ket-vector $|\phi_{s^\smthree }\rangle$. The ket-vectors
of massless fermionic fields are obtainable from \rf{fer01} by replacement
$s\rightarrow s^\sma $, $\alpha^I\rightarrow \alpha^{\sma I}$, $a=1,2$, in
\rf{fer01}, while the ket-vector of massless bosonic field is obtainable from
\rf{manold-07122011-03} by replacement $s\rightarrow s^\smthree $,
$\alpha^I\rightarrow \alpha^{\smthree I}$ in \rf{manold-07122011-03}. Taking
into account that the ket-vectors for massless fields
$|\psi_{s^\smone}\rangle$, $|\psi_{s^\smtwo}\rangle$,
$|\phi_{s^\smthree}\rangle$ are the respective degree-$s^\sma $, $a=1,2,3$,
homogeneous polynomials in the oscillators $\alpha^{\sma I}$, it is easy to
see that the vertices we are interested in must satisfy the equations

\be \label{sec05nn2} (\alpha^{\sma I}\bar\alpha^{\sma I}  - s^\sma
)|p_\smp3^-\rangle  = 0\,,\qquad a=1,2,3, \ee
which tell us that the vertices should be degree-$s^\sma $ homogeneous
polynomials in the oscillators $\alpha^{\sma I}$. Taking into account that
the prefactor $F$ and linear forms $B^\sma$ are degree-1 homogeneous
polynomials in the oscillators, while the cubic form $Z$ is the degree-3
homogeneous polynomial in the oscillators (see
\rf{maslesfdef01sym}-\rf{massZdef01sym}), we find the general solution of
Eqs.\rf{sec05nn2},
\beq\label{0006k}
&&\hspace{-1.7cm} p_\smp3^-(s^\smone+\half,s^\smtwo+\half,s^\smthree;k) =
K^\smonetwo Z^{\frac{1}{2}(\sbf - k+1)} \prod_{a=1}^3 (B^\sma)^{s^\sma +
\frac{1}{2}(k - \sbf -1) }\,, \hspace{0.6cm} \hbox{ for K-vertex};
\\
&&\hspace{-1.7cm}\label{0006f}
p_\smp3^-(s^\smone+\half,s^\smtwo+\half,s^\smthree;k) = F Z^{\frac{1}{2}(\sbf
- k)} (B^\smthree)^{-1} \prod_{a=1}^3 (B^\sma)^{s^\sma + \frac{1}{2}(k -
\sbf) }\,,  \hspace{0.6cm} \hbox{ for F-vertex};
\\
\label{0007} && \hspace{4cm}  \sbf \equiv \sum_{a=1}^3 s^\sma \,, \eeq
where integer $k$ is a freedom in our solution. The integer $k$ labels all
possible cubic vertices that can be built for massless spin $s^\smone+\half$,
$s^\smtwo+\half$, $s^\smthree$ fields and has a clear physical
interpretation. Taking into account that the prefactors  $K^\smonetwo$, $F$,
and the forms $B^\sma$, $Z$ are degree-1 homogeneous polynomials in the
momentum $\Po^I$, it is easy to see that vertices \rf{0006k},\rf{0006f} are
degree-$k$ homogeneous polynomials in $\Po^I$. To summarize, the vertex
$p_\smp3^-(s^\smone+\half,s^\smtwo+\half,s^\smthree;k)$ describes interaction
of two massless fermionic spin $s^\smone+\half$, $s^\smtwo+\half$ fields and
one bosonic massless spin-$s^\smthree$ field and the vertex is degree-$k$
homogeneous polynomial in the momentum $\Po^I$. In Lorentz covariant
approach, gauge invariant vertices corresponding to our light-cone vertices
\rf{0006k},\rf{0006f} contain $k-1$ number of the derivatives with respect to
space-time coordinates.

\noindent {\bf Restrictions on spin values and number of derivatives in cubic
vertices}. We now discuss the restrictions to be imposed on spin values
$s^\smone$, $s^\smtwo$, $s^\smthree$ and the integer $k$. The powers of the
forms $B^\sma$ and $Z$ in \rf{0006k},\rf{0006f} must be non--negative
integers. For this to be the case, it is necessary to impose the following
restrictions on the allowed spin values $s^\smone$, $s^\smtwo$, $s^\smthree$
and the number of powers of the momentum $\Po^I$ (the number of the
transverse derivatives):
\beq
&& \label{00012k} \sbf - 2s_\min + 1\leq k \leq \sbf + 1 \,,\qquad
s_\min\equiv \min (s^\smone,s^\smtwo,s^\smthree)\,,
\nonumber
\\
&& \sbf- k \hbox{ odd integer}, \hspace{6cm} \hbox{ for K-vertex}
\\[5pt]
&& \label{00012f} \sbf - 2 \min(s^\smone,s^\smtwo, s^\smthree -1) \leq k \leq
\sbf\,,
\nonumber
\\
&& \sbf- k \hbox{ even integer},
\nonumber\\
&& s^\smthree \geq 1 \quad \hbox{ when } \quad s^\smthree = s_\min\,,
\hspace{3.9cm} \hbox{ for F-vertex}\,.
\eeq
We now consider K- and F- vertices for the space-time dimensions $d>4$ and
$d=4$ in turn.

\noindent {\bf Case $d>4$}. Restrictions \rf{00012k}, \rf{00012f} lead to a
surprisingly simple result for values of allowed $k$ for cubic vertices
$p_\smp3^-(s^\smone+\half,s^\smtwo+\half,s^\smthree;k)$. Indeed, we see from
\rf{00012k} and \rf{00012f} that for spin values $s^\smone$, $s^\smtwo$,
$s^\smthree$, the integer $k$ takes the values
\beq
&& \label{kval01k} \hspace{-2cm} k = \sbf+1,\, \sbf - 1,\, \ldots\, , \sbf -
2s_\min+1\,, \hspace{3.5cm} \hbox{ for K-vertices in } d> 4\,;
\\
&& \label{kval01f} \hspace{-2cm} k = \sbf,\, \sbf - 2,\, \ldots\, ,
\sbf - 2\min(s^\smone,s^\smtwo,s^\smthree -1)\,, \hspace{1.9cm} \hbox{ for
F-vertices in } d> 4\,.
\eeq
Note that, for F-vertices, we should keep in mind the restriction $
s^\smthree \geq 1$ when $s^\smthree = s_\min$. Vertices
\rf{0006k},\rf{0006f}, with $k$ in \rf{kval01k},\rf{kval01f}, constitute the
complete list of parity invariant cubic vertices for $d>4$. Relations
\rf{kval01k},\rf{kval01f} imply that given spin values $s^\smone$,
$s^\smtwo$, $s^\smthree$, the numbers of parity invariant cubic interaction
vertices that can be constructed are given by
\beq
&& \hspace{-1cm} \Nsf = s_\min  + 1\,, \hspace{5cm} \hbox{ for K-vertices in
} d> 4\,;
\\
&& \hspace{-1cm} \Nsf = \min(s^\smone,s^\smtwo,s^\smthree -1) + 1\,,
\hspace{2cm} \hbox{ for F-vertices in } d> 4\,.\qquad
\eeq

\noindent {\bf Case $d=4$}. For $d=4$, the number of allowed vertices is
decreased. This is, if $d=4$, then for spin values
$s^\smone$, $s^\smtwo$, $s^\smthree$, the integer $k$ takes the values%
\footnote{ There is the simple rule which allows to choose nontrivial
vertices in $4d$. This, an appearance of $K^\smonetwo Z B^\smthree$-factor in
K-vertex leads to trivial K-vertex, while appearance of $F B^\smone
B^\smtwo$-factor in F-vertex leads to trivial F-vertex. This rule can be
approved by using helicity formalism in Ref.\cite{Bengtsson:1983pd}.}
\beq
\label{kval02k}
&& \hspace{-1cm} k = \sbf + 1,\,\, \sbf - 2s_\min+1\,, \hspace{1cm} \hbox{
when } s^\smthree = s_\min \hspace{1.4cm} \hbox{ for K-vertices in } d=4.
\\
&& \hspace{-1cm} k = \sbf+1,\,\, \hspace{3.6cm} \hbox{ when } \ s^\smthree >
s_\min\,,\hspace{1cm}  \hbox{ for K-vertices in } d=4;
\\
&& \hspace{-1cm} k = \sbf - 2s_\min\,, \hspace{3.1cm} \hbox{ when }
s^\smthree
> s_\min\,, \hspace{1cm} \hbox{ for F-vertices in } d=4\,.
\eeq
This implies that, in $4d$, there are two parity invariant K-vertices when $s^\smthree =
s_\min > 0$ and one parity invariant K-vertex when $s^\smthree = s_\min=0$ or $s^\smthree
> s_\min$. Also we note that, in $4d$, there is only one parity invariant
F-vertex when $s^\smthree > s_\min$ and there are no parity invariant F-vertices when
$s^\smthree = s_\min$.

Formulas \rf{0006k},\rf{0006f} provide a surprisingly simple form for the
vertices of massless higher-spin fields and also give a simple form for the
vertices of the well-known massless low-spin fields having spin values
$0,\half,1,\frac{3}{2},2$. By way of example, we consider light-cone gauge
cubic vertices and their Lorentz covariant counterparts that describe
interaction of massless spin-$\half$, -$\frac{3}{2}$ fermionic fields with
massless low-spin and higher-spin bosonic fields.

\noindent {\bf i}) Cubic vertex for massless spin-$\half$ fermionic field
$\psi$ and massless scalar field $\phi$ takes the form
\be p_\smp3^-(\half,\half,0;1) = K^\smonetwo \sim \bar\psi\psi \phi\,.  \ee
\noindent {\bf ii}) The minimal interaction of massless spin-$\half$
fermionic field $\psi$ with a massless spin-1 bosonic field $\phi^A$ is given
by
\be p_\smp3^-(\half,\half,1;1) = F \sim \bar\psi \gamma^A \psi \phi^A\,, \ee
\noindent {\bf iii}) The non-minimal interaction of massless spin-$\half$
fermionic field with massless spin-1 bosonic field takes the form
\be \label{manold-29122011-01}
p_\smp3^-(\half,\half,1;2) = K^\smonetwo B^\smthree \sim \bar\psi \gamma^{AB}
\psi F^{AB}\,, \qquad F^{AB} \equiv \partial^A\phi^B -
\partial^B\phi^A\,. \ee
\noindent {\bf iv}) The gravitational interaction of massless spin-$\half$
fermionic field $\psi$ is given by
\be \label{manold-25122011-05} p_\smp3^-(\half,\half,2;2) = F B^\smthree \sim
\bar\psi \gamma^A D^A \psi|_\smp3 \,, \ee
where $D^A$ stands for covariant derivative w.r.t. Lorentz connection and the
subscript $[3]$ of covariant Lagrangian is used to indicate the cubic vertex.

\noindent {\bf v}) The non-minimal interaction of massless spin-$\half$
fermionic field $\psi$ with massless spin-2 bosonic field $\phi^{AB}$ takes
the form
\be p_\smp3^-(\half,\half,2;3) = K^\smonetwo (B^\smthree)^2 \sim  \bar\psi
\partial^A
\partial^B \psi \phi^{AB}\,. \ee

\noindent {\bf vi}) Our approach gives simple expressions for cubic vertices
of massless spin-$\half$ fermionic field and massless spin-$s$ bosonic field.
Namely, there are the following two cubic vertices for massless spin-$\half$
fermionic field $\psi$ and  massless totally symmetric arbitrary spin-$s$
bosonic field $\phi^{A_1\ldots A_s}$:
\beq
&& p_\smp3^-(\half,\half,s;s+1) = K^\smonetwo (B^\smthree)^s \sim  \bar\psi
\partial^{A_1} \ldots \partial^{A_s} \psi \phi^{A_1\ldots A_s} \,,
\\[3pt]
&& p_\smp3^-(\half,\half,s;s) = F (B^\smthree)^{s-1} \sim  \bar\psi
\gamma^{A_1}
\partial^{A_2} \ldots \partial^{A_s} \psi \phi^{A_1\ldots A_s}\,.
\eeq

\noindent {\bf vii}) Higher-derivative interaction of massless
spin-$\frac{3}{2}$ fermionic field $\psi^A$ (gravitino field) with massless
scalar field $\phi$ takes the form
\be \label{PsiABdef3-2}  p_\smp3^-(\frac{3}{2},\frac{3}{2},0;3) = K^\smonetwo B^\smone B^\smtwo
\sim \bar\Psi^{AB} \Psi^{AB} \phi  \,, \qquad \Psi^{AB} \equiv \partial^A\psi^B-\partial^B\psi^A\,,
\ee
where $\Psi^{AB}$ is field strength of the spin-$\frac{3}{2}$ fermionic field
$\psi^A$.

\noindent {\bf viii}) The non-minimal interaction vertices of massless
spin-$\frac{3}{2}$ fermionic field $\psi^A$ with massless spin-1 bosonic
field $\phi^A$ having the field strength $F^{AB}$ as in
\rf{manold-29122011-01} take the form
\beq
&& p_\smp3^-(\frac{3}{2},\frac{3}{2},1;2) = K^\smonetwo Z \sim
(\bar\psi^A\psi^B -\frac{1}{4}\psi^C\gamma^{AB}\psi^C)F^{AB}\,,
\\
&& p_\smp3^-(\frac{3}{2},\frac{3}{2},1;3) = F B^\smone B^\smtwo \sim
\bar\psi^A \partial^A \gamma^B \psi^C F^{BC}\,,
\\
&& p_\smp3^-(\frac{3}{2},\frac{3}{2},1;4) = K^\smonetwo B^\smone B^\smtwo
B^\smthree \sim  \bar\psi^A \partial^A \partial^B \psi^C F^{BC}\,.
\eeq
\noindent {\bf ix}) The minimal gravitational interaction of massless spin-$\frac{3}{2}$
fermionic field $\psi^A$ is given by
\be \label{3-2gravver01} p_\smp3^-(\frac{3}{2},\frac{3}{2},2;2) = F Z \sim
\bar\psi^A\gamma^{ABC} D_{Lor}^B \psi^C|_\smp3\,. \ee

\noindent {\bf x}) Light-cone gauge cubic vertices for two massless
spin-$\frac{3}{2}$ fermionic fields and one massless arbitrary spin-$s$,
$s\geq 2$, bosonic field take the form
\beq
\label{3-23-21ver01} &&  p_\smp3^-(\frac{3}{2},\frac{3}{2},s;s) = F Z
(B^\smthree)^{s-2}\,,
\\
&&  p_\smp3^-(\frac{3}{2},\frac{3}{2},s;s+1) = K^\smonetwo Z
(B^\smthree)^{s-1}\,,
\\[3pt]
&&  p_\smp3^-(\frac{3}{2},\frac{3}{2},s;s+2) = F B^\smone B^\smtwo
(B^\smthree)^{s-1}\,,
\\[3pt]
\label{3-23-21ver04} &&  p_\smp3^-(\frac{3}{2},\frac{3}{2},s;s+3) =
K^\smonetwo B^\smone B^\smtwo (B^\smthree)^s\,.
\eeq
Using the notation $\LL(\frac{3}{2},\frac{3}{2},s;k)$ for Lorentz covariant
vertices corresponding to the respective light-cone gauge vertices
$p_\smp3^-(\frac{3}{2},\frac{3}{2},s;k)$ in
\rf{3-23-21ver01}-\rf{3-23-21ver04}, we obtain
\beq
\label{3-23-2sver01} &&  {\rm i}^{s-1} \LL(\frac{3}{2},\frac{3}{2},s;s) =
\bar\psi^{A_1} \gamma^{A_2}
\partial^{A_3}\ldots \partial^{A_s} \psi^B \partial^B \phi^{A_1\ldots A_s}
\nonumber\\
&& \hspace{4cm}+ \bar\psi^{C} \gamma^{A_1} \partial^{A_2}\ldots
\partial^{A_{s-1}} \Psi^{CA_s} \phi^{A_1\ldots A_s} \,,
\\
&&  {\rm i}^{s-1}  \LL(\frac{3}{2},\frac{3}{2},s;s+1) = \bar\psi^{A_1}
\partial^{A_2}\ldots \partial^{A_s} \psi^B \partial^B \phi^{A_1\ldots A_s}
\nonumber\\
&& \hspace{4cm}+ \bar\psi^C  \partial^{A_1}\ldots
\partial^{A_{s-1}} \Psi^{CA_s} \phi^{A_1\ldots A_s} \,,
\\
&&  {\rm i}^{s-1}  \LL(\frac{3}{2},\frac{3}{2},s;s+2) = \bar\psi^B
\gamma^{A_1} \partial^{A_2 }\ldots \partial^{A_s} \psi^C \partial^B
\partial^C \phi^{A_1\ldots A_s}\,,
\\
\label{3-23-2sver04} &&  {\rm i}^{s-1}  \LL(\frac{3}{2},\frac{3}{2},s;s+3) =
\bar\psi^B
\partial^{A_1 }\ldots \partial^{A_s} \psi^C \partial^B \partial^C
\phi^{A_1\ldots A_s}\,,
\eeq
where $\phi^{A_1\ldots A_s}$ stands for massless totally symmetric spin-$s$
bosonic field. We note that covariant vertices in
\rf{3-23-2sver01}-\rf{3-23-2sver04} are gauge invariant only under
linearized on-shell gauge symmetries. Also, note that, for $s=2$, the vertex
in \rf{3-23-2sver01} is nothing but the vertex of the minimal gravitational
interaction of massless spin-$\frac{3}{2}$ fermionic field \rf{3-2gravver01}.

Thus, we see that the light-cone gauge approach does indeed provide a simple
representation for interaction vertices. Another attractive property of the
light-cone approach is that it allows treating the interaction vertices on an
equal footing. Formulas \rf{0006k},\rf{0006f} provide a convenient
representation for other well-known cubic interaction vertices of massless
low-spin fields. These vertices and their Lorentz covariant counterparts are
collected in Table I.

\bigskip
\noindent{\sf Table I. Cubic vertices for massless fermionic and bosonic
low-spin fields. \small In the 4th column, $\phi$, $\psi$, $\phi^A$,
$\psi^A$, $\phi^{AB}$ stand for the respective spin $0,\,1/2\,, 1\,, 3/2\,,2$
massless fields. $F^{AB}$ and $R^{ABCE}$ stand for the respective Yang-Mills
field strength and the Riemann tensor, while $\Psi^{AB}=
\partial^A\psi^B-\partial^B\psi^A$ is field strength of spin 3/2 field
$\psi^A$. $D^A$ and $D_{Lor}^A$ stand for the respective Yang-Mills covariant
derivative and Lorentz covariant derivative. $\omega_A{}^{BC}$ stands for the
linearized Lorentz connection, $\omega_A{}^{BC}= - \omega_A{}^{BC}$. Most of
the covariant vertices in the Table are invariant only under linearized
on-shell gauge transformations.}
{\small
\begin{center}
\begin{tabular}{|c|c|c|c|}
\hline        &&&
\\[-3mm]
Spin values  and & dimension &  Light-cone vertex & Manifestly Lorentz
\\
number of derivatives   &    of     &        & invariant vertex
\\
& space-time & $p_\smp3^-(s^\smone+\half,s^\smtwo+\half,s^\smthree;k)$ &
\\
$ s^\smone+\half , s^\smtwo+\half , s^\smthree ;\, k $  &   & &
\\ [2mm]\hline
&&&
\\[-3mm]
\ \ \ \ \ \ $ \half,\half,0;\, 1$   &  $d\geq 4$ & $ K^\smonetwo $  & $
\bar\psi \psi \phi  $
\\[2mm]\hline
&&&
\\[-3mm]
\ \ \ \ \ \ $ \half,\half,1;\, 1$   &  $d\geq 4$ & $ F $  & $
\bar\psi\gamma^A\psi \phi^A  $
\\[2mm]\hline
&&&
\\[-3mm]
\ \ \ \ \ \ $ \half,\half,1;\,2 $   & $d\geq 4$ & $ K^\smonetwo B^\smthree $
& $ \bar\psi \gamma^{AB}\psi F^{AB} $
\\[2mm]\hline
&&&
\\[-3mm]
\ \ \ \ \ \ $ \half,\half,2;\,2$   & $d\geq 4$ &  $F B^\smthree$ & $
(\bar\psi \gamma^A D^A \psi)_\smp3$
\\[2mm]\hline
&&&
\\[-3mm]
\ \ \ \ \ \ $\half,\half,2;\, 3$   & $d\geq 4$ &  $ K^\smonetwo
(B^\smthree)^2 $ & $\bar\psi \gamma^{AB}\partial^C\psi \omega^{CAB} $
\\[2mm]\hline
&&&
\\[-3mm]
\ \ \ \ \ \ $ \frac{3}{2},\frac{3}{2},0;\,3$   & $d\geq 4$ & $ K^\smonetwo
B^\smone B^\smtwo $ & $ \bar\Psi^{AB}\Psi^{AB} \phi $
\\[2mm]\hline
&&&
\\[-3mm]
\ \ \ \ \ \ $ \frac{3}{2},\frac{3}{2},1;\,2$   & $d\geq 4$ & $ K^\smonetwo Z
$ & $(\bar\psi^A\psi^B -\frac{1}{4}\psi^C\gamma^{AB}\psi^C)F^{AB}$
\\[2mm]\hline
&&&
\\[-3mm]
\ \ \ \ \ \ $ \frac{3}{2},\frac{3}{2},1;\,3$   & $d >  4$ & $  F B^\smone
B^\smtwo $ & $\bar\psi^A \partial^A \gamma^B \psi^C F^{BC} $
\\[2mm]\hline
&&&
\\[-3mm]
\ \ \ \ \ \ $ \frac{3}{2},\frac{3}{2},1;\,4$   & $d\geq 4$ & $ K^\smonetwo
B^\smone B^\smtwo B^\smthree $ & $ \bar\psi^A \partial^A \partial^B \psi^C
F^{BC} $
\\[2mm]\hline
&&&
\\[-3mm]
\ \ \ \ \ \ $ \frac{3}{2},\frac{3}{2},2;\,2$   & $d\geq 4$ &  $F Z  $ &
$(\bar\psi^A \gamma^{ABC} D_{Lor}^B \psi^C)_\smp3 $
\\[2mm]\hline
&&&
\\[-3mm]
\ \ \ \ \ \ $ \frac{3}{2},\frac{3}{2},2;\,3 $   & $d > 4$ & $ K^\smonetwo Z
B^\smthree $ & $\bar\psi^C \partial^A \Psi^{BC} \phi^{AB}$
\\[2mm]\hline
&&&
\\[-3mm]
\ \ \ \ \ \ $ \frac{3}{2},\frac{3}{2},2;\,4 $   & $d > 4$ & $ F B^\smone
B^\smtwo B^\smthree $ &
$\bar\psi^A\gamma^B \Psi^{CE}\partial^A \omega^{BCE}$
\\[2mm]\hline
&&&
\\[-3mm]
\ \ \ \ \ \ $ \frac{3}{2},\frac{3}{2},2;\,5 $   &  $d\geq 4$ &  $ K^\smonetwo
B^\smone B^\smtwo (B^\smthree)^2 $  & $\bar\Psi^{AB} \Psi^{CE} R^{ABCE}$
\\[2mm]\hline
\end{tabular}
\end{center}
}

\medskip

\noindent {\bf Non-minimal gravitational interaction of higher-spin fields}.
Using results in \rf{kval01k},\rf{kval01f}, we can present all interaction
vertices of massless higher-spin fermionic field with massless spin-2 field
(graviton field). Indeed, the interaction of a massless spin-$(s+\half)$
fermionic field with massless spin-2 field is described by the vertex
$p_\smp3^-(s^\smone+\half,s^\smtwo+\half,s^\smthree;k)$ with
$s^\smone=s^\smtwo=s \geq 2$, $s^\smthree=2$. For these values of $s^\sma$,
we obtain
\be \label{manold-25122011-01} \sbf = 2s+2\,, \qquad \quad \sbf
-2\min(s^\smone,s^\smtwo,s^\smthree -1)  = 2s\,,\qquad s_\min=2\,.\qquad \ee
Using \rf{manold-25122011-01} in \rf{kval01k},\rf{kval01f}, we get the
following allowed values of $k$:
\beq
\label{manold-25122011-02} &&  k = 2s-1\,, \ 2s+ 1\,,\ 2s+3\,, \hspace{3cm}
\hbox{K-vertices};
\\
\label{manold-25122011-03} && k = 2s\,, \ 2s+ 2\,, \hspace{5.2cm}
\hbox{F-vertices}.
\eeq
We recall that minimal gravitational interaction of low-spin massless
fermionic fields is described by F-vertices having $k=2$ (see
\rf{manold-25122011-05},\rf{3-2gravver01}). However we note, using
\rf{manold-25122011-02},\rf{manold-25122011-03}, that, for $s \geq 2$, the
minimal gravitational interaction, i.e. the case $k=2$, is not allowed. Thus,
we see that it is the relations
\rf{manold-25122011-02},\rf{manold-25122011-03} that leave no place for the
minimal gravitational interaction of massless higher-spin fermionic fields.
On the other hand, we see, from relations
\rf{manold-25122011-02},\rf{manold-25122011-03}, that there are three type K-
and two type F-vertices describing the non-minimal (higher-derivative)
gravitational interactions of massless higher-spin fermionic fields and the
graviton field.

\bigskip

\noindent{\sf Table II. Cubic vertices for massless fermionic
spin-$\frac{5}{2}$ field and massless bosonic spin 0,1,2 fields. \small In
the 4th column, $\phi$, $\phi^A$ stand for the respective spin $0,\, 1$
massless fields. $F^{AB}$ and $R^{ABCE}$ stand for the respective Yang-Mills
field strength and the Riemann tensor, while $\Psi^{ABCE}$ stands for field
strength of spin-$\frac{5}{2}$ field. $\omega_A{}^{BC}$ and $\Omega_A{}^{BC}$
stand for the respective linearized Lorentz connection of spin-2 and
spin-$\frac{5}{2}$ fields, $\omega_A{}^{BC}= - \omega_A{}^{BC}$,
$\Omega_A{}^{BC}= - \Omega_A{}^{BC}$. Most of the covariant vertices in the
Table are invariant only under linearized on-shell gauge transformations.}
{\small
\begin{center}
\begin{tabular}{|l|c|c|c|}
\hline        &&&
\\[-3mm]
Spin values  & dimension &  Light-cone  & Manifestly Lorentz
\\
and number   &    of     & vertex       & invariant vertex
\\
of derivatives & space-time &
$p_\smp3^-(\frac{5}{2},\frac{5}{2},s^\smthree;k)$ &
\\
$ \frac{5}{2}, \frac{5}{2}, s^\smthree ;\, k $  &   & &
\\ [2mm]\hline
&&&
\\[-3mm]
\ \ \ \ \ \ $ \frac{5}{2},\frac{5}{2},0;\, 5$   &  $d\geq 4$ & $ K^\smonetwo
(B^\smone B^\smtwo)^2 $ & $\bar\Psi^{ABCE}\Psi^{ABCE}\phi $
\\ [2mm]\hline
&&&
\\[-3mm]
\ \ \ \ \ \ $ \frac{5}{2},\frac{5}{2},1;\, 4$   &  $d\geq 4$ & $ K^\smonetwo
Z B^\smone B^\smtwo $ & $F^{AB} (\bar\Omega^{A,CE} \Omega^B{}_{C E} -
\bar\Omega^{C,E A} \Omega_{C,E}{}^B)$
\\[2mm]\hline
&&&
\\[-3mm]
\ \ \ \ \ \ $ \frac{5}{2},\frac{5}{2},1;\,5 $   & $d > 4$ & $ F
(B^\smone)^2(B^\smtwo)^2 $ & $\bar{\Psi}^{ABCD}\gamma^E \Psi^{ABCD} \phi^E$
\\[2mm]\hline
&&&
\\[-3mm]
\ \ \ \ \ \ $ \frac{5}{2},\frac{5}{2},1;\,6$   & $d\geq 4$ &  $ K^\smonetwo
(B^\smone)^2(B^\smtwo)^2 B^\smthree$ & $\bar{\Psi}^{ACDE} \Psi^{BCDE} F^{AB}$
\\[2mm]\hline
&&&
\\[-3mm]
\ \ \ \ \ \ $\frac{5}{2},\frac{5}{2},2;\, 3$   & $d\geq 4$ &  $ K^\smonetwo
Z^2 $ &
\\[2mm]\hline
&&&
\\[-3mm]
\ \ \ \ \ \ $ \frac{5}{2},\frac{5}{2},2;\,4$   & $d > 4$ & $ F Z B^\smone
B^\smtwo  $ & See Ref.\cite{Metsaev:2006ui}
\\[2mm]\hline
&&&
\\[-3mm]
\ \ \ \ \ \ $ \frac{5}{2},\frac{5}{2},2;\,5$   & $d >  4$ & $  K^\smonetwo Z
B^\smone B^\smtwo B^\smthree $ &
\\[2mm]\hline
&&&
\\[-3mm]
\ \ \ \ \ \ $ \frac{5}{2},\frac{5}{2},2;\,6$   & $d > 4$ & $ F (B^\smone)^2
(B^\smtwo)^2 B^\smthree $ & $\bar\Psi^{ACDE}\gamma^F \Psi^{BCDE}
\omega^{FAB}$
\\[2mm]\hline
&&&
\\[-3mm]
\ \ \ \ \ \ $ \frac{5}{2},\frac{5}{2},2;\,7$   & $d\geq 4$ &  $ K^\smonetwo
(B^\smone B^\smtwo B^\smthree)^2 $ &
$\bar{\Psi}^{AB}_{CD}\Psi^{CD}_{EF}R^{EF}_{AB}$
\\[2mm]\hline
\end{tabular}
\end{center}
}

\medskip

We note that K-vertices with $k=\sbf+1$ correspond to gauge theory cubic
interaction vertices built entirely in terms of gauge field
strengths.%
\footnote{Our result for the vertex $p_\smp3^-(\frac{5}{2},\frac{5}{2},2;7)$
(see last row in Table II) implies that there is only one Lorentz covariant
vertex $\bar{\Psi}^{AB}_{CD}\Psi^{CD}_{EF}R^{EF}_{AB}$ that gives a
non-trivial contribution to the 3-point scattering amplitude. Similar
statement for the case of vertex $R^{AB}_{CD} R^{CD}_{EF}R^{EF}_{AB}$ was
proved in Ref.\cite{Metsaev:1986yb}.}
The vertices with $k<\sbf+1$ cannot be built entirely in terms of gauge field
strengths. It is the vertices with $k<\sbf+1$ that are difficult to construct
in Lorentz covariant approaches. The light-cone approach treats all vertices
on an equal footing.

\newsection{ Cubic vertices for two massless fermionic fields and one\\
massive bosonic field}\label{secMMO}

We now consider cubic interaction vertex for three mixed-symmetry fields
with the mass values
\be\label{00m1mix} \mas_1|_\smF = \mas_2|_\smF = 0,\qquad \mas_3|_\smB \ne
0\,, \ee
i.e. the {\it massless} fermionic fields carry external line indices $a=1,2$,
while the {\it massive} bosonic field corresponds to $a=3$. Equations for the
vertex involving two massless fields can be obtained from Eqs.\rf{cubver13nn}
in the limit as $\mas_1 \rightarrow 0 $, $\mas_2 \rightarrow 0 $. The
solution for K- and F-vertices is found to be
\beq
&& \label{00m2kmix} p_\smp3^- = K^\smonetwo V^\Krm ( B_m^\smthree;\,
Q_{mn}^\smaaplusone)\,, \hspace{2cm} \hbox{ K-vertex} \,,
\\
&& \label{00m2fmix} p_\smp3^- = F_n V^\Frm ( B_m^\smthree;\,
Q_{mn}^\smaaplusone)\,, \hspace{2.5cm}  \hbox{ F-vertex}\,,
\eeq
where generating functions $V^\Krm$, $V^\Frm$ are arbitrary polynomials of
forms $B_m^\smthree$, $Q_{mn}^\smaaplusone$ and  we use the notation%
\footnote{ We recall that the short notation like $p_\smp3^-(Q^\smaaplusone)$
is used to indicate a dependence of $p_\smp3^-$ on $Q^\smonetwo$,
$Q^\smtwothree$, $Q^\smthreeone$.}
\beq
\label{00m6fer01mix} &&  K^{\smonetwo} \equiv \frac{1}{\beta_1\beta_2}
p_{\theta_1} \Po^I\gamma^I \gamma_* \eta_2\,,
\\[3pt]
\label{00m6fer02mix} &&  F_n \equiv \frac{1}{\beta_1\beta_2}
p_{\theta_1}\Bigl(( \frac{\check{\beta}_3}{\beta_3} \Po^I  -
\gamma^{IJ}\Po^J) \alpha_n^{\smthree I} + \frac{2\beta_1\beta_2}{\beta_3}
\mas_3 \zeta_n^\smthree \Bigr) \eta_2 \,,
\eeq
\beq
&& \label{00m6mix} B_n^\sma \equiv \frac{\alpha_n^{\sma I}
\Po^I}{\beta_a}\,,\quad a=1,2;\qquad \ \ \
\\
&& B_n^\smthree\equiv \frac{\alpha_n^{\smthree I}\Po^I}{\beta_3} -
\frac{\check{\beta}_3}{2\beta_3} \mas_3 \zeta_n^\smthree\,,
\\
\label{00m3mix} && Q_{mn}^\smonetwo \equiv \alpha_{mn}^\smonetwo -
\frac{2}{\mas_3^2} B_m^\smone  B_n^\smtwo \,,
\\
\label{00m4mix}&& Q_{mn}^\smtwothree \equiv \alpha_{mn}^\smtwothree -
\frac{\zeta_n^\smthree}{\mas_3} B_m^\smtwo + \frac{2}{\mas_3^2} B^\smtwo_m
B_n^\smthree\,,
\\
\label{00m5mix}&& Q_{mn}^\smthreeone \equiv \alpha_{mn}^\smthreeone +
\frac{\zeta_m^\smthree}{\mas_3} B_n^\smone + \frac{2}{\mas_3^2} B_m^\smthree
B_n^\smone\,. \eeq
The quadratic forms $\alpha_{mn}^\smab$ are defined in \rf{amnabdef}.

From \rf{00m2kmix},\rf{00m2fmix}, we see that the prefactor $K^\smonetwo$ is
homogeneous polynomial in $\Po^I$, while the prefactors $F_n$, the linear
forms $B_n^\smthree$ and the quadratic forms $Q_{mn}^\smaaplusone$ are
non-homogeneous polynomials in $\Po^I$. This implies that, in general, the
cubic vertices we obtained are non-homogeneous polynomials in $\Po^I$. The
appearance of massive field interaction vertices involving different powers
of derivatives is a well-known fact (see e.g. Refs.\cite{GR}). We see that
the light-cone approach gives a simple explanation to this phenomenon by
means of the prefactors $F_n$, the linear forms $B_n^\smthree$, and the
quadratic forms $Q_{mn}^\smaaplusone$. We note that solution for the
generating functions $V^\Krm$, $V^\Frm$ given in \rf{00m2kmix}, \rf{00m2fmix}
is complete solution, while the solution for the prefactors $K^\smonetwo$,
$F_n$ given in \rf{00m6fer01mix},\rf{00m6fer02mix} is not complete solution.
Namely, for the vertices of mixed-symmetry fields there are extra solutions
for the prefactors that involve contributions of higher than second order in
$\gamma$-matrices. We note also for the vertices of totally symmetric fields
those extra solutions are trivial. To understand the remaining characteristic
properties of solution \rf{00m2kmix},\rf{00m2fmix}, we restrict our attention
to cubic vertices for totally symmetric fields.

\subsection{ Cubic vertices for totally symmetric two massless fermionic fields
and one massive bosonic field}

We now restrict ourselves to cubic vertices for two massless totally
symmetric fermionic fields and one massive totally symmetric bosonic field
with mass parameters as in \rf{00m1mix}. To consider the totally symmetric
fields it is sufficient to use one sort of oscillators. Therefore we set
$\nnu = 1$ in \rf{00m2kmix},\rf{00m2fmix}. To simplify the formulas we drop
the oscillator's subscript $n=1$ and use the simplified notation for
oscillators: $\alpha^I \equiv \alpha_1^I$, $\zeta \equiv \zeta_1$. The cubic
vertex for totally symmetric fields under consideration can then be obtained
from solution in \rf{00m2kmix},\rf{00m2fmix} by making the identifications
\beq
\label{00m7} && \alpha^{\sma I} \equiv \alpha_1^{\sma I}, \quad  \ a =1,2\,;
\qquad \alpha^{\smthree I} \equiv  \alpha_1^{\smthree I} \qquad
\zeta^\smthree \equiv \zeta_1^\smthree\,, \eeq
and ignoring the contribution of oscillators carrying a subscript $n>1$.
Adopting simplified notation \rf{00m7} for the prefactors and forms in
\rf{00m6mix}-\rf{00m5mix}:
\be \label{00m9}  F\equiv F_1\,,\qquad B^\sma \equiv B_1^\sma\,, \qquad
Q^\smab \equiv Q_{11}^\smab\,, \qquad \alpha^\smab \equiv
\alpha_{11}^\smab\,, \ee
we see that vertices \rf{00m2kmix},\rf{00m2fmix} take the form
\beq
&& \label{00m2ksymgen} p_\smp3^- = K^\smonetwo V^\Krm ( B^\smthree;\,
Q^\smaaplusone)\,, \hspace{2cm}  \hbox{ K-vertex} \,,
\\
&& \label{00m2fsymgen} p_\smp3^- = F V^\Frm ( B^\smthree;\,
Q^\smaaplusone)\,, \hspace{2.5cm} \hbox{ F-vertex}\,,
\eeq
where we use the notation
\beq
\label{00m6fer02} &&  F \equiv \frac{1}{\beta_1\beta_2} p_{\theta_1}\Bigl((
\frac{\check{\beta}_3}{\beta_3} \Po^I  - \gamma^{IJ}\Po^J) \alpha^{\smthree
I} + \frac{2\beta_1\beta_2}{\beta_3} \mas_3 \zeta^\smthree \Bigr) \eta_2 \,,
\eeq
\beq
&& \label{00m6} B^\sma \equiv \frac{\alpha^{\sma I} \Po^I}{\beta_a}\,,\qquad
a=1,2;\qquad \ \ \
\\
&& B^\smthree\equiv \frac{\alpha^{\smthree I}\Po^I}{\beta_3} -
\frac{\check{\beta}_3}{2\beta_3} \mas_3 \zeta^\smthree\,,\eeq
\beq
\label{00m3} && Q^\smonetwo \equiv \alpha^\smonetwo - \frac{2}{\mas_3^2}
B^\smone  B^\smtwo \,,
\\
\label{00m4}&& Q^\smtwothree \equiv \alpha^\smtwothree -
\frac{\zeta^\smthree}{\mas_3} B^\smtwo + \frac{2}{\mas_3^2} B^\smtwo
B^\smthree\,,
\\
\label{00m5}&& Q^\smthreeone \equiv \alpha^\smthreeone +
\frac{\zeta^\smthree}{\mas_3} B^\smone + \frac{2}{\mas_3^2} B^\smthree
B^\smone\,, \eeq
while the quadratic forms $\alpha^\smab$ and prefactor $K^{\smonetwo}$ are
defined in \rf{amnabdefsym} and \rf{00m6fer01mix} respectively.

Vertices \rf{00m2ksymgen},\rf{00m2fsymgen} describe an interaction of the
towers of totally symmetric massive bosonic fields in \rf{manold-07122011-09}
and massless fermionic fields in \rf{fer07}.  As we have already noted
solution for vertices given in \rf{00m2ksymgen}, \rf{00m2fsymgen} is the
complete solution. To understand better the characteristic properties of
vertices for totally symmetric fields \rf{00m2ksymgen},\rf{00m2fsymgen}, we
consider vertices for fields with fixed but arbitrary spin values.

\subsubsection{ Cubic vertices for totally symmetric fields with fixed but arbitrary
spin values}

We now turn to cubic vertices for totally symmetric fields with fixed spin
values. This is to say that we consider vertices involving two massless spin
$s^\smone + \half$ and $s^\smtwo + \half$ fermionic fields and one massive
spin-$s^\smthree$ bosonic field having mass parameter $\mas_3$:
\be
\begin{array}{lll}
\mas_1 = 0\,, \qquad & \mas_2 = 0\,, \qquad &  \mas_3 \ne 0\,,
\\[5pt]
s^\smone + \half \,, & s^\smtwo+\half \,, & s^\smthree\,.
\end{array}
\ee
The massless spin-$(s^\smone+\half)$ and -$(s^\smtwo+\half)$ fermionic fields
are described by the respective ket-vectors $|\psi_{s^\smone }\rangle$ and
$|\psi_{s^\smtwo }\rangle$, while the massive spin-$s^\smthree$ bosonic field
is described by the ket-vector $|\phi_{s^\smthree }\rangle$. The ket-vectors
of massless fields $|\psi_{s^\sma }\rangle$, $a=1,2$, can be obtained from
\rf{fer01} by the replacement $s\rightarrow s^\sma $, $\alpha^I\rightarrow
\alpha^{\sma I}$, $a=1,2$, in \rf{fer01}, while the ket-vector of the massive
field $|\phi_{s^\smthree }\rangle$ can be obtained from
\rf{manold-07122011-04} by the replacement $s\rightarrow s^\smthree $,
$\alpha^I\rightarrow \alpha^{\smthree I}$, $\zeta\rightarrow \zeta^\smthree$
in \rf{manold-07122011-04}. Taking into account that the ket-vectors
$|\psi_{s^\sma }\rangle$, $a=1,2$, are the respective degree-$s^\sma $
homogeneous polynomials in the oscillators $\alpha^{\sma I}$, while the
ket-vector $|\phi_{s^\smthree }\rangle$ is a degree-$s^\smthree$ homogeneous
polynomial in the oscillators $\alpha^{\smthree I}$, $\zeta^\smthree$, it is
easy to understand that the vertices we are interested in must satisfy the
equations
\beq
&& \label{00m11}  (\alpha^{\sma I}\bar\alpha^{\sma I}  - s^\sma )
|p_\smp3^-\rangle = 0\,,\qquad a=1,2\,,
\\
&& \label{00m12}  (\alpha^{\smthree I} \bar\alpha^{\smthree I}  +
\zeta^\smthree\bar\zeta^\smthree - s^\smthree ) |p_\smp3^-\rangle = 0\,.
\eeq
These equations tell us that the vertices must be degree-$s^\sma $
homogeneous polynomials in the respective oscillators. Taking into account
that the prefactor $F$ and the linear form $B^\smthree$ are degree-1
homogeneous polynomials in the oscillators, while the quadratic forms
$Q^\smaaplusone$ are degree-2 homogeneous polynomials in the oscillators, we
find the general solution of Eqs.\rf{00m11}, \rf{00m12} as:
\beq
&& \label{intver30k} p_\smp3^-(s^\smone+\half,s^\smtwo+\half,s^\smthree;w)  =
K^\smonetwo (B^\smthree)^w \prod_{a=1}^3 (Q^\smaaplusone)^{y^\smaplustwo }
\,, \qquad \hbox{ K-vertex}\,,
\\
&& \label{intver30f} p_\smp3^-(s^\smone+\half,s^\smtwo+\half,s^\smthree;w)  =
F (B^\smthree)^w \prod_{a=1}^3 (Q^\smaaplusone)^{y^\smaplustwo } \,,
\hspace{1.3cm} \hbox{ F-vertex}\,,
\eeq
where integers $y^\sma $ are expressible in terms of the $s^\sma $ and
integer $w$ by the relations
\beq
&& \label{yexp01k} y^\sma  = \frac{\sbf - w}{2} -s^\sma \,,\quad a=1,2\,,
\nonumber\\
&& y^\smthree  = \frac{\sbf + w}{2} -s^\smthree \,, \hspace{3.5cm} \hbox{ for
K-vertex};
\\[7pt]
&& \label{yexp01f} y^\sma  = \frac{\sbf - w -1}{2} -s^\sma \,,\quad a=1,2\,,
\nonumber\\
&& y^\smthree  = \frac{\sbf + w + 1}{2} -s^\smthree \,, \hspace{3cm} \hbox{
for F-vertex};
\eeq
The integer $w$ expresses the freedom of the solution and labels all possible
cubic interaction vertices that can be constructed for the fields under
consideration. For vertices \rf{intver30k},\rf{intver30f} to be sensible, we
impose the restrictions
\beq
\label{restr01k}&&  w\geq 0\,,\qquad y^\sma  \geq 0\,, \quad a=1,2,3;
\nonumber\\
&& \sbf - w \qquad \hbox{ even integer}\,, \hspace{3.5cm} \hbox{ for
K-vertex};
\\
\label{restr01f}&&  w \geq 0\,,\qquad y^\sma  \geq 0\,, \quad a=1,2,3;
\nonumber\\
&& \sbf - w  \qquad \hbox{ odd integer}\,, \hspace{3.5cm} \hbox{ for
F-vertex};
\eeq
which amount to the requirement that the powers of all forms in
\rf{intver30k},\rf{intver30f} be non--negative integers. We note that using
relations \rf{yexp01k}, \rf{yexp01f} allows rewriting the restrictions
\rf{restr01k},\rf{restr01f} as%
\footnote{ If $w=0$, then restrictions \rf{restr03NNN1k} become the
restrictions well known in the angular momentum theory: $ |s^\smone -
s^\smtwo| \leq s^\smthree \leq s^\smone + s^\smtwo$, while restriction
\rf{restr03NNN1f} takes the form $|s^\smone - s^\smtwo| \leq s^\smthree -1
\leq s^\smone + s^\smtwo$ .}
\beq
&& \hspace{-2cm} \label{restr03NNN1k} \max(0, s^\smthree - s^\smone -
s^\smtwo) \leq w \leq s^\smthree - |s^\smone - s^\smtwo|\,,  \hspace{3.5cm}
\hbox{ for K-vertex};
\\
&&  \hspace{-2cm}\label{restr03NNN1f} \max(0, s^\smthree - s^\smone -
s^\smtwo-1) \leq w \leq s^\smthree-1 - |s^\smone - s^\smtwo|\,,  \hspace{2cm}
\hbox{ for F-vertex}.
\eeq
As compared to vertices for three massless fields \rf{0006k},\rf{0006f}, the
vertices given in \rf{intver30k},\rf{intver30f} are non-homogeneous
polynomials in $\Po^I$. An interesting property of vertices
\rf{intver30k},\rf{intver30f} is that the maximal and minimal numbers of
powers of the momentum $\Po^I$, denoted by $k_\max$ and $k_\min$
respectively, are independent of $w$ and given by%
\footnote{ Expressions for $K^\smonetwo$, $F$, $B^\smthree$, and
$Q^\smaaplusone$ \rf{00m9} imply that $k_\max = 1+ w + 2\sum_{a=1}^3 y^\sma$.
Taking expressions for $y^\sma $ \rf{yexp01k}, \rf{yexp01f} into account, we
find $k_\max$ in \rf{kmaxN1k},\rf{kmaxN1f}.}
\beq
&& \label{kmaxN1k} k_\max= \sbf+1\,, \qquad \ k_\min =1\,,  \hspace{2.8cm}
\hbox{ for K-vertex};
\\
&& \label{kmaxN1f} k_\max= \sbf\,,  \qquad\qquad k_\min = 0\,, \hspace{2.8cm}
\hbox{ for F-vertex}.
\eeq

\newsection{ Cubic vertices for one massless fermionic field, one massless\\
bosonic field and one massive fermionic field}

We proceed with the cubic interaction vertex for three mixed-symmetry fields
with the mass values
\be\label{00m1n} \mas_1|_\smB = \mas_2|_\smF = 0,\qquad \mas_3|_\smF \ne 0\,,
\ee
i.e. the {\it massless} bosonic field carries external line index $a=1$, the
{\it massless} fermionic field carries external line index $a=2$, and the
{\it massive} fermionic field corresponds to $a=3$. Equations for the vertex
involving two massless fields can be obtained from Eqs.\rf{cubver13nn} in the
limit as $\mas_1 \rightarrow 0 $, $\mas_2 \rightarrow 0 $. The solution for
K- and F-vertices we found is given by
\beq
&& \label{00m2knmix} p_\smp3^- = K^\smtwothree V^\Krm ( B_n^\smthree;\,
Q_{mn}^\smaaplusone)\,, \hspace{2cm} \Krm \hbox{ vertex} \,,
\\
&& \label{00m2fnmix} p_\smp3^- = F_n V^\Frm ( B_n^\smthree;\,
Q_{mn}^\smaaplusone)\,, \hspace{2.5cm} \Frm \hbox{ vertex}\,,
\eeq
where generating functions $V^\Krm$, $V^\Frm$ are arbitrary polynomials of
forms $B_n^\smthree$, $Q_{mn}^\smaaplusone$ and  we use the notation%
\beq
\label{00m6fer01nmix} &&  K^{\smtwothree} \equiv \frac{1}{\beta_2\beta_3}
p_{\theta_2} ( \Po^I\gamma^I \gamma_*  + \beta_2 \mas_3\Bigr) \eta_3\,,
\\[3pt]
\label{00m6fer02nf0mix} &&  F_n \equiv \frac{1}{\beta_2\beta_3}
p_{\theta_2}\Bigl( \frac{\check{\beta}_1}{\beta_1} \Po^I  - \gamma^{IJ}\Po^J
+ \beta_2 \mas_3 \gamma^I\gamma_* \Bigr) \eta_3 \alpha_n^{\smone I} -
\frac{2}{\mas_3} K^\smtwothree B_n^\smone\,,
\eeq
\beq
&& \label{00m6nmix} B_n^\sma \equiv \frac{\alpha_n^{\sma I}
\Po^I}{\beta_a}\,,\quad a=1,2;\qquad \ \ \
\\
&& B_n^\smthree\equiv \frac{\alpha_n^{\smthree I}\Po^I}{\beta_3} -
\frac{\check{\beta}_3}{2\beta_3} \mas_3 \zeta_n^\smthree\,,
\\
\label{00m3nmix} && Q_{mn}^\smonetwo \equiv \alpha_{mn}^\smonetwo -
\frac{2}{\mas_3^2} B_m^\smone  B_n^\smtwo \,,
\\
\label{00m4nmix}&& Q_{mn}^\smtwothree \equiv \alpha^\smtwothree -
\frac{\zeta_n^\smthree}{\mas_3} B_m^\smtwo + \frac{2}{\mas_3^2} B_m^\smtwo
B_n^\smthree\,,
\\
\label{00m5nmix}&& Q_{mn}^\smthreeone \equiv \alpha_{mn}^\smthreeone +
\frac{\zeta_m^\smthree}{\mas_3} B_n^\smone + \frac{2}{\mas_3^2} B_m^\smthree
B_n^\smone\,.
\eeq
The quadratic forms $\alpha_{mn}^\smab$ are defined in \rf{amnabdef}.

We note that the prefactors $K^\smtwothree$ and $F_n$, the linear forms
$B_n^\smthree$, and quadratic forms $Q_{mn}^\smaaplusone$ are non-homogeneous
polynomials in the momentum $\Po^I$. This implies that the cubic interaction
vertices given in \rf{00m2knmix}, \rf{00m2fnmix} are also non-homogeneous
polynomials in the momentum $\Po^I$. We note that solution for $V^\Krm$,
$V^\Frm$ given in \rf{00m2knmix}, \rf{00m2fnmix} is complete solution, while
the solution for the prefactors $K^\smtwothree$, $F_n$ given in
\rf{00m6fer01nmix},\rf{00m6fer02nf0mix} is not complete solution. This is to
say that for the vertices of mixed-symmetry fields there are extra solutions
for the prefactors that involve contributions of higher than second order in
$\gamma$-matrices.

\subsection{ Cubic vertices for totally symmetric one massless fermionic field,
one massless bosonic field and one massive fermionic field}

We proceed with the study of vertices for three totally symmetric fields with
the mass values given in \rf{00m1n}, i.e. the {\it massless} bosonic field
carries external line index $a=1$, the {\it massless} fermionic field carries
external line index $a=2$, and the {\it massive} fermionic field corresponds
to $a=3$. To consider the totally symmetric fields it is sufficient to use
one sort of oscillators. Therefore we set $\nnu = 1$ in
\rf{00m2knmix},\rf{00m2fnmix}. To simplify the formulas we drop the
oscillator's subscript $n=1$ and use the simplified notation for oscillators:
$\alpha^{\sma I} \equiv \alpha_1^{\sma I}$, $\zeta^\sma \equiv \zeta_1^\sma$.
The cubic interaction vertices for totally symmetric fields under
consideration can then be obtained from solution
\rf{00m2knmix},\rf{00m2fnmix}. The vertices take the form
\beq
&& \label{00m2kn} p_\smp3^- = K^\smtwothree V^\Krm ( B^\smthree;\,
Q^\smaaplusone)\,, \hspace{2cm} \Krm \hbox{ vertex} \,,
\\
&& \label{00m2fn} p_\smp3^- = F V^\Frm ( B^\smthree;\, Q^\smaaplusone)\,,
\hspace{2.5cm} \Frm \hbox{ vertex}\,,
\eeq
where we use the notation
\be
\label{00m6fer02n} F \equiv \frac{1}{\beta_2\beta_3} p_{\theta_2}\Bigl(
\frac{\check{\beta}_1}{\beta_1} \Po^I  - \gamma^{IJ}\Po^J + \beta_2 \mas_3
\gamma^I\gamma_* \Bigr) \eta_3 \alpha^{\smone I} - \frac{2}{\mas_3}
K^\smtwothree B^\smone\,,
\ee

\beq
&& \label{00m6n} B^\sma \equiv \frac{\alpha^{\sma I} \Po^I}{\beta_a}\,,\qquad
a=1,2;\qquad \ \ \
\\
&& B^\smthree\equiv \frac{\alpha^{\smthree I}\Po^I}{\beta_3} -
\frac{\check{\beta}_3}{2\beta_3} \mas_3 \zeta^\smthree\,,\eeq
\beq
\label{00m3n} && Q^\smonetwo \equiv \alpha^\smonetwo - \frac{2}{\mas_3^2}
B^\smone  B^\smtwo \,,
\\
\label{00m4n}&& Q^\smtwothree \equiv \alpha^\smtwothree -
\frac{\zeta^\smthree}{\mas_3} B^\smtwo + \frac{2}{\mas_3^2} B^\smtwo
B^\smthree\,,
\\
\label{00m5n}&& Q^\smthreeone \equiv \alpha^\smthreeone +
\frac{\zeta^\smthree}{\mas_3} B^\smone + \frac{2}{\mas_3^2} B^\smthree
B^\smone\,, \eeq
while $\alpha^\smab$ and $K^{\smtwothree}$ are defined in \rf{amnabdefsym}
and \rf{00m6fer01nmix} respectively.

Vertices \rf{00m2kn},\rf{00m2fn} describe an interaction of the towers of
bosonic and fermionic fields.  We note that solution for vertices given in
\rf{00m2kn},\rf{00m2fn} provide the complete list of parity invariant cubic
vertices for totally symmetric fields with the mass parameters shown in
\rf{00m1n}. To understand the remaining characteristic properties of solution
\rf{00m2kn},\rf{00m2fn}, we consider vertices for totally symmetric fields
with fixed spin values.

\subsubsection{ Vertices for totally symmetric fields with fixed but arbitrary
spin values}

In this section, we restrict ourselves to cubic vertices for totally
symmetric fields with fixed spin values. This is to say that we consider
vertices for one massless spin-$s^\smone$ bosonic field, one massless
spin-$(s^\smtwo+\half)$ fermionic field and one massive
spin-$(s^\smthree+\half)$ fermionic field. This, we consider vertices
involving fields with the following spin and mass values:
\be
\begin{array}{lll}
\mas_1 = 0\,, \qquad & \mas_2 = 0\,, \qquad &  \mas_3 \ne 0\,,
\\[5pt]
s^\smone\,, & s^\smtwo+\half \,, & s^\smthree+\half\,.
\end{array}
\ee
The massless spin-$s^\smone$ bosonic field and massless
spin-$(s^\smtwo+\half)$ fermionic field are described by the respective
ket-vectors $|\phi_{s^\smone }\rangle$ and $|\psi_{s^\smtwo }\rangle$, while
the massive spin-$(s^\smthree+\half)$ fermionic field is described by the
ket-vector $|\psi_{s^\smthree }\rangle$. The ket-vectors of massless fields
$|\phi_{s^\smone }\rangle$, $|\psi_{s^\smtwo }\rangle$ can be obtained from
the respective expressions in \rf{manold-07122011-03},\rf{fer01} by the
replacement $s\rightarrow s^\sma $, $\alpha^I\rightarrow \alpha^{\sma I}$,
$a=1,2$, while the ket-vector of the massive field $|\psi_{s^\smthree
}\rangle$ can be obtained from \rf{fer01} by the replacement $s\rightarrow
s^\smthree $, $\alpha^I\rightarrow \alpha^{\smthree I}$, $\zeta\rightarrow
\zeta^\smthree$. Taking into account that the ket-vectors
$|\phi_{s^\smone}\rangle$ and $|\psi_{s^\smtwo }\rangle$ are the respective
degree-$s^\smone $ and degree-$s^\smtwo$ homogeneous polynomials in the
oscillators $\alpha^{\sma I}$, $a=1,2$, while the ket-vector
$|\psi_{s^\smthree }\rangle$ is a degree-$s^\smthree$ homogeneous polynomial
in the oscillators $\alpha^{\smthree I}$, $\zeta^\smthree$, it is easy to
understand that the vertices we are interested in must satisfy the equations
\beq
&& \label{00m11n}  (\alpha^{\sma I}\bar\alpha^{\sma I}  - s^\sma )
|p_\smp3^-\rangle = 0\,,\qquad a=1,2,
\\[3pt]
&& \label{00m12n}  (\alpha^{\smthree I} \bar\alpha^{\smthree I}  +
\zeta^\smthree\bar\zeta^\smthree - s^\smthree ) |p_\smp3^-\rangle = 0\,.
\eeq
These equations tell us that the vertices must be a degree-$s^\sma $
homogeneous polynomials in the respective oscillators. Taking into account
that the prefactor $F$ and the form $B^\smthree$ are degree-1 homogeneous
polynomials in the oscillators, while the forms $Q^\smaaplusone$ are degree-2
homogeneous polynomials in the oscillators, we find the general solution of
Eqs.\rf{00m11n}, \rf{00m12n} as:
\beq
&& \label{intver30kn} p_\smp3^-(s^\smone,s^\smtwo+\half,s^\smthree+\half;w)
= K^\smtwothree (B^\smthree)^w \prod_{a=1}^3 (Q^\smaaplusone)^{y^\smaplustwo
} \,, \qquad \Krm \hbox{ vertex} \,,
\\[5pt]
&& \label{intver30fn} p_\smp3^-(s^\smone,s^\smtwo+\half,s^\smthree+\half;w)
= F (B^\smthree)^w \prod_{a=1}^3 (Q^\smaaplusone)^{y^\smaplustwo } \,,
\hspace{1.3cm} \Frm \hbox{ vertex}\,,
\eeq
where integers $y^\sma $ are expressible in terms of the $s^\sma $ and
integer $w$ by the relations
\beq
&& \label{yexp01kn} y^\sma  = \frac{\sbf - w}{2} -s^\sma \,,\quad a=1,2\,,
\nonumber\\
&& y^\smthree  = \frac{\sbf + w}{2} -s^\smthree \,, \hspace{3.5cm} \hbox{ for
K-vertex};
\\[7pt]
&& \label{yexp01fn}
y^\smone  = \frac{\sbf - w + 1}{2} -s^\smone \,,
\nonumber\\
&& y^\smtwo  = \frac{\sbf - w - 1}{2} -s^\smtwo \,,
\nonumber\\
&& y^\smthree  = \frac{\sbf + w - 1}{2} - s^\smthree \,, \hspace{3cm} \hbox{
for F-vertex};
\eeq
The integer $w$ expresses the freedom of the solution and labels all possible
cubic interaction vertices that can be constructed for the fields under
consideration. For vertices \rf{yexp01kn}, \rf{yexp01fn} to be sensible, we
impose the restrictions
\beq
\label{restr01kn}&&  w\geq 0\,,\qquad y^\sma  \geq 0\,, \quad a=1,2,3;
\nonumber\\
&& \sbf - w \qquad \hbox{ even integer}\,, \hspace{3.5cm} \hbox{ for
K-vertex};
\\
\label{restr01fn}&&  w \geq 0\,,\qquad y^\sma  \geq 0\,, \quad a=1,2,3;
\nonumber\\
&& \sbf - w  \qquad \hbox{ odd integer}\,, \hspace{3.5cm} \hbox{ for
F-vertex};
\eeq
which amount to the requirement that the powers of all forms in
\rf{intver30kn},\rf{intver30fn} be non--negative integers. We note that using
relations \rf{yexp01kn}, \rf{yexp01fn} allows rewriting the restrictions
\rf{restr01kn},\rf{restr01fn} as%
\footnote{ If $w=0$, then restrictions \rf{restr03NNN1kn} becomes the
restrictions well known in the angular momentum theory: $ |s^\smone -
s^\smtwo| \leq s^\smthree \leq s^\smone + s^\smtwo$, while restriction
\rf{restr03NNN1fn} takes the form $ |s^\smone - s^\smtwo-1| \leq s^\smthree
\leq s^\smone + s^\smtwo-1$}
\beq
&& \hspace{-2cm} \label{restr03NNN1kn} \max(0, s^\smthree - s^\smone -
s^\smtwo) \leq w \leq s^\smthree - |s^\smone - s^\smtwo|\,,  \hspace{3.5cm}
\hbox{ for K-vertex};
\\
&&  \hspace{-2cm}\label{restr03NNN1fn} \max(0, s^\smthree - s^\smone -
s^\smtwo + 1) \leq w \leq s^\smthree  - |s^\smone - s^\smtwo - 1 |\,,
\hspace{2cm} \hbox{ for F-vertex}.
\eeq
The maximal and minimal numbers of powers of the momentum $\Po^I$, denoted by
$k_\max$ and $k_\min$ respectively, are independent of $w$ and given by%
\beq
&& \label{kmaxN1kn} k_\max= \sbf+1\,,  \qquad \ k_\min=0\,, \hspace{2.8cm}
\hbox{ for K-vertex};
\\
&& \label{kmaxN1fn} k_\max= \sbf\,,  \qquad \qquad k_\min=0\,, \hspace{2.8cm}
\hbox{ for F-vertex}.
\eeq

\newsection{ Cubic vertices for one massless bosonic field and two massive\\
fermionic fields with the same mass values}\label{equalmasses}

The case under consideration is most interesting because it involves the
Yang-Mills and gravitational interactions of massive arbitrary spin fermionic
fields as particular cases. We now consider the cubic interaction vertex for
mixed-symmetry fields having the following mass values,
\be\label{eqmas00007NN1} \mas_1|_\smF = \mas_2|_\smF \equiv \mas  \ne 0
,\qquad \mas_3|_\smB = 0\,, \ee
i.e. the {\it massive} fermionic fields having the same values of mass
parameter carry external line indices $a=1,2$, while the {\it massless}
bosonic field corresponds to $a=3$. The solution for  K- and F-cubic vertices
we found is given by
\beq
\label{intvereqmas01kmix} && p_\smp3^- = K^\smonetwo V^\Krm (L_n^\smone,
L_n^\smtwo, B_n^\smthree;\, Q_{mn}^\smonetwo\,;\, Z_{mnq})\,,  \hspace{2.6cm}
\hbox{ K-vertex};
\\
\label{intvereqmas01fmix} && p_\smp3^- = F_{0\, n}^\smthree V^\Frm
(L_n^\smone, L_n^\smtwo, B_n^\smthree;\, Q_{mn}^\smonetwo\,;\, Z_{mnq})\,,
\hspace{2.8cm} \hbox{ F-vertex};
\eeq
where generating functions $V^\Krm$, $V^\Frm$ are arbitrary polynomials of
linear, quadratic, and cubic forms. We use the notation%
\beq
\label{Kzerdefmix} && K^\smonetwo = \frac{1}{\beta_1\beta_2}p_{\theta_1}
\Bigl( \Po^I\gamma^I \gamma_* + \mas \check{\beta}_3\Bigr) \eta_2
\\[3pt]
\label{fzerdefmix} && F_{0\,n}^\smthree =
\frac{1}{\beta_1\beta_2}p_{\theta_1} \Bigl(\frac{\check{\beta}_3}{\beta_3}
\Po^I -\gamma^{IJ} \Po^J - \mas \beta_3 \gamma^I \gamma_*\Bigr) \eta_2
\alpha_n^{\smthree I}\,,
\eeq
\beq
\label{eqmas00007mix} &&  L_n^\smone \equiv B_n^\smone - \frac{1}{2}\mas
\zeta_n^\smone\,,\qquad\quad
%
%
L_n^\smtwo \equiv B_n^\smtwo + \frac{1}{2}\mas \zeta_n^\smtwo\,,
\\
\label{eqmas00009mix} && B_n^\sma \equiv \frac{\alpha_n^{\sma
I}\Po^I}{\beta_a} - \frac{\check{\beta}_a}{2\beta_a} \mas
\zeta_n^\sma\,,\qquad a=1,2;
\\
\label{eqmas00010mix} && B_n^\smthree\equiv \frac{\alpha_n^{\smthree
I}\Po^I}{\beta_3}\,,
\eeq
\beq
\label{eqmas00006mix} &&  Q_{mn}^\smonetwo \equiv \alpha_n^\smonetwo -
\frac{\zeta_n^\smtwo}{\mas} B_m^\smone + \frac{\zeta_m^\smone}{\mas}
B_n^\smtwo\,,
\\[5pt]
\label{eqmas00005mix} && Z_{mnq} \equiv L_m^\smone \alpha_{nq}^\smtwothree +
L_n^\smtwo \alpha_{qm}^\smthreeone + B_q^\smthree ( \alpha_{mn}^\smonetwo -
\zeta_m^\smone\zeta_n^\smtwo )\,,
\eeq
where the quadratic forms $\alpha_{mn}^\smab$ are defined in \rf{amnabdef}.

From \rf{Kzerdefmix}-\rf{eqmas00005mix}, we see that only the linear forms
$B_n^\smthree$ \rf{eqmas00010mix} are homogeneous polynomials in the momentum
$\Po^I$, while the prefactors $K^\smonetwo$, $F_{0\,n}^\smthree$, the linear
forms $L_n^\smone$, $L_n^\smtwo$, the quadratic forms $Q_{mn}^\smonetwo$, and
the cubic forms $Z_{mnq}$ are non-homogeneous polynomials in $\Po^I$. Because
the prefactors $K^\smonetwo$, $F_{0\,n}^\smthree$ are non-homogeneous
polynomials in $\Po^I$ all cubic interaction vertices for fields with mass
parameters as in \rf{eqmas00007NN1} are also non-homogeneous polynomials in
$\Po^I$. Note also that, because the linear forms $B_n^\smthree$ are
homogeneous polynomials in $\Po^I$, the minimal number of powers of $\Po^I$
in cubic vertices \rf{intvereqmas01kmix},\rf{intvereqmas01fmix} is not equal
to zero in general. As before, solution for the generating functions
$V^\Krm$, $V^\Frm$ given in \rf{intvereqmas01kmix}, \rf{intvereqmas01fmix} is
complete solution, while the solution for the prefactors $K^\smonetwo$,
$F_{0\,n}^\smthree$ given in \rf{Kzerdefmix},\rf{fzerdefmix} is not complete
solution. Namely, for the vertices of mixed-symmetry fields there are extra
solutions for the prefactors that involve contributions of higher than second
order in $\gamma$-matrices.

To discuss the remaining important properties of the solutions given
\rf{intvereqmas01kmix},\rf{intvereqmas01fmix} we restrict our attention to
cubic vertices for the totally symmetric fields.

\subsection{ Cubic vertices for totally symmetric one massless bosonic field and two
massive fermionic fields with the same mass values}

We proceed with the study of vertices for three totally symmetric fields with
the mass values given in \rf{eqmas00007NN1}, i.e. the {\it massive} fermionic
fields having the same values of mass parameter carry external line indices
$a=1,2$, while the {\it massless} bosonic field carries external line index
$a=3$. As before, to consider the totally symmetric fields, it is sufficient
to use one sort of oscillators. Therefore we set $\nnu = 1$ in
\rf{intvereqmas01kmix},\rf{intvereqmas01fmix} and ignore the contribution of
the oscillators with $n>1$. Also, to simplify the formulas we drop the
oscillator's subscript $n=1$ and use the simplified notation for oscillators:
$\alpha^{\sma I} \equiv \alpha_1^{\sma I}$, $\zeta^\sma \equiv \zeta_1^\sma$.
In doing so, we obtain, using the solutions given in
\rf{intvereqmas01kmix},\rf{intvereqmas01fmix}, the cubic interaction vertices
for the totally symmetric fields,
\beq
\label{intvereqmas01k} && p_\smp3^- = K^\smonetwo V^\Krm (L^\smone, L^\smtwo,
B^\smthree;\, Q^\smonetwo\,;\, Z)\,,  \hspace{2.6cm} \hbox{ K-vertex};
\\
\label{intvereqmas01f} && p_\smp3^- = F_0^\smthree V^\Frm (L^\smone,
L^\smtwo, B^\smthree;\, Q^\smonetwo\,;\, Z)\,,  \hspace{2.8cm} \hbox{
F-vertex};
\eeq
where we use the notation
\beq
\label{fzerdef} && F_0^\smthree = \frac{1}{\beta_1\beta_2}p_{\theta_1}
\Bigl(\frac{\check{\beta}_3}{\beta_3} \Po^I -\gamma^{IJ} \Po^J - \mas \beta_3
\gamma^I \gamma_*\Bigr) \eta_2 \alpha^{\smthree I}\,,
\\[3pt]
\label{eqmas00007} &&  L^\smone \equiv B^\smone - \frac{1}{2}\mas
\zeta^\smone\,,\qquad\quad
%
%
L^\smtwo \equiv B^\smtwo + \frac{1}{2}\mas \zeta^\smtwo\,,
\\
\label{eqmas00009} && B^\sma \equiv \frac{\alpha^{\sma I}\Po^I}{\beta_a} -
\frac{\check{\beta}_a}{2\beta_a} \mas \zeta^\sma\,,\qquad a=1,2;
\\
\label{eqmas00010} && B^\smthree\equiv \frac{\alpha^{\smthree
I}\Po^I}{\beta_3}\,,
\\
\label{eqmas00006} &&  Q^\smonetwo \equiv \alpha^\smonetwo -
\frac{\zeta^\smtwo}{\mas} B^\smone + \frac{\zeta^\smone}{\mas} B^\smtwo\,,
\\[5pt]
\label{eqmas00005} && Z \equiv L^\smone \alpha^\smtwothree + L^\smtwo
\alpha^\smthreeone + B^\smthree ( \alpha^\smonetwo -
\zeta^\smone\zeta^\smtwo )\,,
\eeq
while the quadratic forms $\alpha^\smab$ and the prefactor $K^\smonetwo$ are
defined in \rf{amnabdefsym} and \rf{Kzerdefmix} respectively.  Solutions for
vertices given in \rf{intvereqmas01k},\rf{intvereqmas01f} are the complete
solutions, i.e., vertices given in \rf{intvereqmas01k},\rf{intvereqmas01f}
provide the complete list of parity invariant cubic vertices for the totally
symmetric fields with the mass parameters given in \rf{eqmas00007NN1}.

To discuss the remaining important properties of the solutions given in
\rf{intvereqmas01k},\rf{intvereqmas01f} we restrict our attention to cubic
vertices for totally symmetric fields with fixed spin values.

\subsubsection{ Cubic vertices for  totally symmetric fields with fixed but
arbitrary spin values}\label{nnn-sce-01}

In this section, we restrict ourselves to cubic interaction vertices for the
totally symmetric fields with fixed but arbitrary spin values and with mass
values given in \rf{eqmas00007NN1}. This is to say that we consider vertices
involving two massive spin $s^\smone + \half$ and $s^\smtwo + \half$
fermionic fields having the same mass parameter $\mas$ and one massless
spin-$s^\smthree$ bosonic field,
\be \label{manold-28122011-01}
\begin{array}{lll}
\mas_1 = \mas\,, \qquad & \mas_2 = \mas\,, \quad \mas\ne 0\,, \qquad &
\mas_3 = 0\,,
\\[5pt]
s^\smone + \half \,, & s^\smtwo+\half \,, & s^\smthree\,.
\end{array}
\ee
Two massive spin $s^\smone+\half$ and $s^\smtwo+\half$ fermionic fields are
described by the respective ket-vectors $|\psi_{s^\smone }\rangle$ and
$|\psi_{s^\smtwo }\rangle$, while one massless spin-$s^\smthree$ bosonic
field is described by the ket-vector $|\phi_{s^\smthree }\rangle$. The
ket-vectors of massive fields $|\psi_{s^\sma }\rangle$, $a=1,2$, can be
obtained from \rf{fer01} by the replacement $s\rightarrow s^\sma $,
$\alpha^I\rightarrow \alpha^{\sma I}$, $\zeta\rightarrow \zeta^\sma$,
$a=1,2$, while the ket-vector of massless field $|\phi_{s^\smthree }\rangle$
can be obtained from \rf{manold-07122011-03} by the replacement $s\rightarrow
s^\smthree $, $\alpha^I\rightarrow \alpha^{\smthree I}$. Taking into account
that the ket-vectors $|\psi_{s^\sma }\rangle$, $a=1,2$, are the respective
degree-$s^\sma $ homogeneous polynomials in the oscillators $\alpha^{\sma
I}$, $\zeta^\sma $, while the ket-vector $|\phi_{s^\smthree }\rangle$ is a
degree-$s^\smthree$ homogeneous polynomial in the oscillator
$\alpha^{\smthree I}$, it is easy to understand that the vertices we are
interested in must satisfy the equations
\beq
&& \label{mm018}  (\alpha^{\sma I}\bar\alpha^{\sma I} +\zeta^\sma
\bar\zeta^\sma  - s^\sma ) |p_\smp3^-\rangle = 0\,,\qquad a=1,2\,,
\\[3pt]
&& \label{mm019}  (\alpha^{\smthree I} \bar\alpha^{\smthree I}  -
s^\smthree ) |p_\smp3^-\rangle = 0\,. \eeq
These equations tell us that the vertices must be a degree-$s^\sma $
homogeneous polynomials in the respective oscillators. Taking into account
that the prefactor $F_0^\smthree$ and the forms $L^\smone$, $L^\smtwo$,
$B^\smthree$ given in \rf{fzerdef}-\rf{eqmas00010} are degree-1 homogeneous
polynomials in the oscillators, while quadratic form $Q^\smonetwo$
\rf{eqmas00006} and cubic form $Z$ \rf{eqmas00005} are the respective
degree-2 and degree-3 homogeneous polynomials in the oscillators, we find the
general solution of Eqs.\rf{mm018}, \rf{mm019} as
\beq
&& \label{intvereqmass01k} p_\smp3^-(s^\smone
+\half,s^\smtwo+\half,s^\smthree\,;\,k_\min,k_\max)
\nonumber\\
&&\qquad\qquad   = K^\smonetwo (L^\smone)^{w^\smone } (L^\smtwo)^{w^\smtwo }
(B^\smthree)^{w^\smthree } (Q^\smonetwo)^{y^\smthree } Z^y\,, \hspace{1.5cm}
\hbox{ K-vertex};
\\[3pt]
&& \label{intvereqmass01f} \qquad \qquad  = F_0^\smthree (L^\smone)^{w^\smone
} (L^\smtwo)^{w^\smtwo } (B^\smthree)^{w^\smthree } (Q^\smonetwo)^{y^\smthree
} Z^y\,,  \hspace{1.7cm} \hbox{ F-vertex}\,,
\eeq
where the parameters $w^\smone  $, $w^\smtwo $, $w^\smthree $, $y^\smthree $,
$y$ are given by
\beq
\label{xadefmm0k}
&& w^\smone   = k_\max - k_\min - s^\smtwo  - 1 \,,
\nonumber\\
&& w^\smtwo  = k_\max - k_\min - s^\smone - 1 \,,
\nonumber\\
&& w^\smthree  = k_\min\,,
\nonumber\\
&& y^\smthree =  \sbf - 2s^\smthree  - k_\max + 2 k_\min + 1\,,
\nonumber\\
&& y= s^\smthree  -k_\min\,, \hspace{5cm} \hbox{ for K-vertices};
\eeq
\beq
\label{xadefmm0f}
&& w^\smone   = k_\max - k_\min - s^\smtwo -1 \,,
\nonumber\\
&& w^\smtwo  = k_\max - k_\min - s^\smone -1 \,,
\nonumber\\
&& w^\smthree  = k_\min\,,
\nonumber\\
&& y^\smthree =  \sbf - 2s^\smthree  - k_\max + 2 k_\min +2 \,,
\nonumber\\
&& y= s^\smthree  -k_\min -1\,, \hspace{4.5cm} \hbox{ for F-vertices};
\eeq
and $\sbf$ is defined in \rf{0007}. Integers $k_\min$ and $k_\max$ in
\rf{intvereqmass01k}-\rf{xadefmm0f} are the freedom in our solution. In
general, vertices \rf{intvereqmass01k},\rf{intvereqmass01f} are
non-homogeneous polynomials in the momentum $\Po^I$ and the integers $k_\min$
and $k_\max$ are the respective minimal and maximal numbers of powers of the
momentum $\Po^I$ in \rf{intvereqmass01k},\rf{intvereqmass01f}. For vertices
\rf{intvereqmass01k},\rf{intvereqmass01f} to be sensible, we should impose
the restrictions
\be
\label{restrict01} w^\sma  \geq 0\,, \qquad a=1,2,3; \qquad  y^\smthree \geq
0\,,\qquad y \geq 0\,,
\ee
which amount to requiring the powers of all forms in
\rf{intvereqmass01k},\rf{intvereqmass01f} to be non--negative integers. Using
\rf{xadefmm0k},\rf{xadefmm0f}, restrictions \rf{restrict01} can be rewritten
in a more convenient form as
\beq
&& \label{restrict04k} k_\min  + \max_{a=1,2} s^\sma +1 \leq k_\max \leq
\sbf- 2s^\smthree  + 2 k_\min + 1\,,
\nonumber\\
&& 0 \leq k_\min \leq s^\smthree \,, \hspace{5.1cm} \hbox{ for K-vertices};
\\[5pt]
&& \label{restrict04f} k_\min  + \max_{a=1,2} s^\sma  + 1 \leq k_\max \leq
\sbf- 2s^\smthree  + 2 k_\min + 2\,,
\nonumber\\
&&  0\leq k_\min \leq s^\smthree -1\,, \hspace{4.5cm} \hbox{ for F-vertices}.
\eeq

Cubic vertices in \rf{intvereqmass01k}, \rf{intvereqmass01f} and restrictions
on allowed values of $k_\max$ and $k_\min$ in \rf{restrict04k},
\rf{restrict04f} provide the complete list of parity invariant cubic vertices
for the totally symmetric fields with spin and mass values given in
\rf{manold-28122011-01}.

\subsubsection{ Interaction of massive arbitrary spin fermionic field with
massless scalar field}

We now use results in Section \ref{nnn-sce-01} to discuss cubic vertices for
two fermionic fields and one scalar field. In doing so, we present the list
of all parity invariant cubic vertices for the totally symmetric massive
arbitrary spin-$(s+\half)$ fermionic field interacting with the massless
scalar field. This is to say that we consider vertices \rf{intvereqmass01k},
\rf{intvereqmass01f} with the mass and spin values given by
\be \label{intvereqmass02N1sca}
\begin{array}{lll}
&& \mas_1 = \mas\,, \qquad \quad\mas_2 = \mas\,, \qquad \qquad \ \ \mas_3 =
0\,,
\\[5pt]
&& s^\smone + \half \,, \qquad\quad  s^\smtwo+\half \,, \qquad  \qquad\quad
s^\smthree=0\,,
\end{array}
\ee
where $\mas \ne 0$, $s^\smone  = s^\smtwo = s$.

Allowed spin values and powers of derivatives $k_\min$, $k_\max$ should
satisfy restrictions in \rf{restrict04k}, \rf{restrict04f}. Plugging
$s^\smthree =0$ in the last restriction in \rf{restrict04f}, we learn that
F-vertices are not allowed. Plugging $s^\smthree =0$ in the last restriction
in \rf{restrict04k}, we learn that all allowed K-vertices take $k_\min=0$.
Plugging $k_\min=0$ into the 1st restrictions in \rf{restrict04k}, we
summarize the allowed values of $k_\min$ and $k_\max$ for K-vertices:
\be
\label{xxxN1N1fsca} k_\min=0\,, \qquad s + 1 \leq \ k_\max \leq \ 2s+1\,,
\qquad \hbox{ for K-vertices}.\qquad
\ee
Using $k_\min=0$ in \rf{intvereqmass01k},\rf{xadefmm0k}, we get the following
cubic interactions K-vertices:

\be \label{minintss1fsca} p_\smp3^-(s+\half,s+\half,0; 0, k_\max ) =
K^\smonetwo (L^\smone L^\smtwo)^{k_\max - s - 1}
(Q^\smonetwo)^{2s+1-k_\max} \,.\ee
Vertices \rf{minintss1fsca} supplemented with inequalities for allowed values
of $k_\max$ in \rf{xxxN1N1fsca} constitute the complete list of the parity
invariant cubic vertices that can be built for two totally symmetric massive
spin-$(s+\half)$ fermionic fields and one massless scalar field.

For the case of $s=0$, we obtain only one allowed value of $k_\max = 1$ in
\rf{xxxN1N1fsca} and the corresponding vertex given by
\be \label{minintss1fsca02} p_\smp3^-(\half,\half,0; 0, 1) = K^\smonetwo \ee
describes interaction of spin-$\half$ fermionic field with scalar field.
Appropriate covariant vertex is well known, $\LL_\smp3 =\bar\psi\psi\phi$.

For the case of $s>0$, we obtain $s+1$ interaction vertices labelled by
allowed values of $k_\max$ in \rf{xxxN1N1fsca}. Note that quadratic form $Q^\smonetwo$
\rf{eqmas00006} does not have a smooth massless limit ($\mas \rightarrow 0$).
This implies that the vertex for the massive spin $s+\half \geq
\frac{3}{2}$ fermionic field \rf{minintss1fsca} does not admit a sensible massless limit
when $k_\max < 2s+1$. In other words, there are $s$ vertices with $s + 1\leq
k_\max < 2s+1$ which do not have a smooth massless limit and one vertex with
$k_\max = 2s+1$ which has smooth massless limit. For example, covariant vertex
for one scalar field and two spin-$\frac{3}{2}$ fermionic fields which has smooth
massless limit takes the form $\LL = \bar\Psi^{AB}\Psi^{AB}\phi$, where
$\Psi^{AB}$ is defined in \rf{PsiABdef3-2}.

\subsubsection{ Yang-Mills interaction of massive arbitrary spin fermionic
field}

We now apply our results in Section \ref{nnn-sce-01} to the discussion of the
Yang-Mills interaction of the massive arbitrary spin fermionic fields.%
\footnote{ Discussion of gauge invariant formulation of massive arbitrary
spin fermionic fields in constant electromagnetic field may be found in
\cite{Klishevich:1998yt}.}
Since the Yang-Mills interaction of massive fermionic fields are described by
F-vertices we restrict our attention to the F-vertices in what follows. We
first present the list of {\it all} cubic F-vertices for the massive
arbitrary spin-$(s+\half)$ fermionic field interacting with the massless
spin-1 field (Yang-Mills field). This is to say that we consider F-vertices
\rf{intvereqmass01f} with
\be \label{intvereqmass02N1}
\begin{array}{lll}
&& \mas_1 = \mas\,, \qquad \quad\mas_2 = \mas\,, \qquad \qquad \ \ \mas_3 =
0\,,
\\[5pt]
&& s^\smone + \half \,, \qquad\quad  s^\smtwo+\half \,, \qquad  \qquad\quad
s^\smthree=1\,,
\end{array}
\ee
where $\mas \ne 0$, $s^\smone  = s^\smtwo = s$. The 2nd restrictions in
\rf{restrict04f} lead to one allowed value of $k_\min=0$. Plugging such
$k_\min$ into the 1st inequalities in \rf{restrict04f}, we summarize the
allowed values of $k_\min$ and $k_\max$ for F-vertices:
\beq
\label{xxxN1N1f} &&  k_\min=0\,,\qquad s + 1 \leq \ k_\max \leq \ 2s+1 \,,\quad
\qquad s\geq 0\,,\qquad \quad \hbox{ for F-vertices}. \qquad
\eeq
We now discuss those vertices from the list in \rf{xxxN1N1f} that correspond
to the Yang-Mills interaction of the massive arbitrary spin fermionic field. We
consider various spin fields in turn.

\noindent {\bf a}) Spin-$\half$ field ($s=0$). Plugging $s=0$ into
\rf{xxxN1N1f} we obtain $k_\max=1$. Thus we see that the cubic vertex of the
Yang-Mills interaction of the massive spin-$\half$ fermionic  field has
$k_\min=0$, $k_\max=1$. Using such $k_\min$, $k_\max$ in
\rf{intvereqmass01f},\rf{xadefmm0f}, we get the Yang-Mills interaction vertex
for the massive spin-$\half$ fermionic field
\be\label{minint001f} p_\smp3^-(\half,\half,1; 0, 1 ) =  F_0^\smthree \,.\ee

\noindent {\bf b}) Spin $s+\half \geq \frac{3}{2}$ field. All vertices with
$k_\min$, $k_\max$ as in \rf{xxxN1N1f} are candidates for the Yang-Mills
interaction of the fermionic spin $s +\half \geq \frac{3}{2}$ field. We
impose an additional requirement, which allows us to choose one suitable
vertex: given value of $s$, we look for the vertex with the minimal value of
$k_\max$. It is easy to see that such a vertex is given by $k_\max = s+1$.
Plugging $k_\min=0$,  $k_\max = s+1$ into
\rf{intvereqmass01f},\rf{xadefmm0f}, we get the Yang-Mills
interaction of the massive spin $s+\half \geq \frac{3}{2}$ field,%
\footnote{ A gauge invariant description of the electromagnetic interaction
of the massive spin-2 field was obtained in Ref.\cite{Klishevich:1997pd}. The
derivation of the electromagnetic interaction of massive spin $s=2,3$ fields
from string theory is given in Refs.\cite{Argyres:1989cu}. In these
references, the electromagnetic field is treated as an external
(non-dynamical) field.}
\be \label{minintss1f} p_\smp3^-(s+\half,s+\half,1; 0, s ) = F_0^\smthree
(Q^\smonetwo)^s \,,\qquad s\geq 1\,.\ee

A few remarks are in order.

i) The prefactor $F_0^\smthree$ \rf{fzerdef} has smooth massless limit ($\mas
\rightarrow 0$). Therefore, the Yang-Mills interaction of the massive
spin-$\half$ fermionic field given in \rf{minint001f} has a smooth massless
limit, as this should be. This interaction in the massless limit coincides
with the respective interaction of the massless spin-$\half$ fermionic field
in Table I.

ii) The form $Q^\smonetwo$  \rf{eqmas00006} does not have a smooth massless
limit ($\mas \rightarrow 0$). This implies that the Yang-Mills interaction of
the massive spin $s+\half > \frac{3}{2}$ field \rf{minintss1f} does not admit
the massless limit. In the light-cone approach, it is the contribution of
$Q^\smonetwo$ that explains why the Yang-Mills interaction of the massive
spin $s+\half>\frac{3}{2}$ field does not admit the massless limit. As was
expected, the Yang-Mills interaction of the massive spin
$s+\half>\frac{3}{2}$ field \rf{minintss1f} involves higher derivatives. The
appearance of the higher derivatives in \rf{minintss1f} can be seen from the
expression for $Q^\smonetwo$ \rf{eqmas00006}.

\subsubsection{ Gravitational interaction of massive arbitrary spin fermionic
field}

We now apply our results in Section \ref{nnn-sce-01} to discuss the
gravitational interaction of massive arbitrary spin fermionic field. We
first present the list of {\it all} cubic F-vertices for the massive
spin-$(s+\half)$ fermionic field interacting with massless spin-2 field.
This is, we consider F-vertices \rf{intvereqmass01f} with
\be \label{intvereqmass02}
\begin{array}{lll}
&& \mas_1 = \mas\,, \qquad \quad\mas_2 = \mas\,, \qquad \qquad \ \ \mas_3 =
0\,,
\\[5pt]
&& s^\smone + \half \,, \qquad\quad  s^\smtwo+\half \,, \qquad  \qquad\quad
s^\smthree=2\,,
\end{array}
\ee
where $\mas \ne 0$, $s^\smone  = s^\smtwo = s$. The 2nd restrictions in
\rf{restrict04f} lead to two allowed values of $k_\min$: $k_\min =0,1$.
Plugging these values of $k_\min$ into the 1st restrictions in
\rf{restrict04f}, we obtain two families of F-vertices
\beq
\label{intvereqmass03f} && k_\min = 1 \,, \qquad  s + 2 \leq \ k_\max \leq
\ 2s+2 \,,\qquad \ \quad s\geq 0\,;
\\[3pt]
\label{intvereqmass04f} && k_\min = 0 \,, \qquad  s+1 \leq \ k_\max \leq \
\ \ 2s \,,\qquad\qquad \ \ s\geq 1\,;
\eeq
We now discuss those vertices in \rf{intvereqmass03f},\rf{intvereqmass04f}
that correspond to the gravitational interaction of the massive arbitrary
spin fermionic fields. We consider various spin fields in turn.

\noindent {\bf a}) Spin-$\half$ field ($s=0$). The gravitational interaction
of the massive spin-$\half$ fermionic field is given by \rf{intvereqmass03f}.
Plugging $s=0$ in \rf{intvereqmass03f}, we obtain the well-known relation
$k_\max=2$, which tells us that the cubic vertex of the gravitational
interaction of the massive spin-$\half$ fermionic field is a degree-2
non-homogeneous polynomial in the derivatives. Formulas
\rf{intvereqmass01f},\rf{xadefmm0f} lead to the gravitational interaction of
the massive spin-$\half$ fermionic field,
\be \label{intvereqmass06f} p_\smp3^-(\half,\half,2; 1, 2 ) = F_0^\smthree
B^\smthree \,.\ee

\noindent {\bf b}) Spin $\frac{3}{2}$ field ($s=1$). The gravitational
interaction vertex of the massive spin-$\frac{3}{2}$ fermionic field is given in
\rf{intvereqmass04f}. Plugging $s=1$ in \rf{intvereqmass04f}, we obtain
$k_\max=2$. Formulas \rf{intvereqmass01f},\rf{xadefmm0f} then lead to the
gravitational interaction of the massive spin-$\frac{3}{2}$ field
\be \label{intvereqmass07f} p_\smp3^-(\frac{3}{2},\frac{3}{2},2; 0, 2 ) =
F_0^\smthree Z\,.\ee

\noindent {\bf c}) Higher-spin fermionic fields, $(s+\half)\geq \frac{5}{2}$.
All vertices given in \rf{intvereqmass03f},\rf{intvereqmass04f} are
candidates for the gravitational interaction of the higher-spin fermionic
fields. We should impose some additional requirement that would allow us to
choose one suitable vertex. Our additional requirement is that we look for
vertex that has  the minimal value of $k_\max$. It can be seen that such a
vertex is given by \rf{intvereqmass04f} with $k_\min=0$, $k_\max = s+1$.
Plugging such $k_\min$, $k_\max$ into \rf{intvereqmass01f},\rf{xadefmm0f} we
get cubic vertex for the gravitational interaction of the massive higher-spin
fermionic field,
\be \label{intvereqmass08f} p_\smp3^-(s+\half,s+\half,2; 0, s+1 ) =
F_0^\smthree Z (Q^\smonetwo)^{s-1} \,,\qquad s\geq 2\,.\ee

A few remarks are in order.

i) Since prefactor $F_0^\smthree$ \rf{fzerdef}, and the forms $B^\smthree$,
$Z$ \rf{eqmas00010},\rf{eqmas00005} have a smooth massless limit, $\mas
\rightarrow 0$, the gravitational interactions of the massive spin-$\half$,
-$\frac{3}{2}$ fermionic fields \rf{intvereqmass06f},\rf{intvereqmass07f}
have smooth massless limit, as they should. These gravitational interactions
in the massless limit reduce to the corresponding interactions of the
massless spin-$\half$,-$\frac{3}{2}$ fermionic fields given in Table I.

ii) Since the form $Q^\smonetwo$ \rf{eqmas00006} does not have a smooth
massless limit ($\mas \rightarrow 0$), the gravitational interaction of the
massive higher-spin fermionic field \rf{intvereqmass08f} does not admit a
sensible massless limit; it is the form $Q^\smonetwo$ that explains why the
gravitational interaction of the massive higher-spin field does not admit the
massless limit. Higher derivatives in the gravitational interaction of the
massive higher-spin fermionic field are related to the contribution of
$Q^\smonetwo$ \rf{eqmas00006}.%
\footnote{ Gauge invariant formulations of the gravitational interaction of
massive fields are studied e.g. in Ref.\cite{Cucchieri:1994tx}. Interesting
discussion of various aspects of the massive spin 2 field in gravitational
background may be found in Refs.\cite{Buchbinder:1999ar}.}

\subsubsection{ Interaction of arbitrary spin massless bosonic field with two
massive spin-$\half$ fermionic fields}

Using results in Section \ref{nnn-sce-01}, we now discuss the interaction
vertices for massless  arbitrary spin-$s$ bosonic field and massive
spin-$\half$ fermionic fields.%
\footnote{ Discussion of vertices for arbitrary spin massless bosonic field
and two massive scalar fields in framework of BRST invariant approach may be
found in Ref.\cite{Fotopoulos:2007yq}.}
This is, we consider K- and F-vertices
\rf{intvereqmass01k},\rf{intvereqmass01f} with the spin and mass values given by
\be \label{intvereqmass02nn}
\begin{array}{lll}
&& \mas_1 = \mas\,, \qquad \quad\mas_2 = \mas\,, \qquad \qquad \ \ \mas_3 =
0\,,
\\[5pt]
&& s^\smone + \half \,, \qquad\quad  s^\smtwo+\half \,, \qquad  \qquad\quad
s^\smthree=s\,,
\end{array}
\ee
where $\mas \ne 0$, $s^\smone  = s^\smtwo = 0$. Plugging these spin values
into \rf{restrict04k}, \rf{restrict04f}, we find that $k_\min$ and $k_\max$
are fixed uniquely and therefore there are only one allowed K-vertex and only
one allowed F-vertex,
\beq
\label{intvereqmass03fnn} && k_\min = s \,, \ \ \quad\qquad  k_\max = s+1\,,
\hspace{2cm} \hbox{ for K-vertices},
\\[3pt]
\label{intvereqmass04fnn}
&& k_\min = s-1 \,, \qquad  k_\max = s\,, \hspace{2cm} \quad \ \ \hbox{ for
F-vertices}.
\eeq
Plugging \rf{intvereqmass03fnn} and \rf{intvereqmass04fnn} into
\rf{intvereqmass01k}-\rf{xadefmm0f} we obtain the respective K- and F- cubic
interaction vertices
\beq
\label{man0ld-310122011-15} && p_\smp3^-(\half,\half,s; s, s+1 ) =
K^\smonetwo (B^\smthree)^s \,,
\\
\label{man0ld-310122011-16} && p_\smp3^-(\half,\half,s; s-1, s ) =
F_0^\smthree (B^\smthree)^{s-1}\,.
\eeq
Vertices \rf{man0ld-310122011-15},\rf{man0ld-310122011-16} have smooth
massless limit.

\newsection{ Cubic vertices for one massless fermionic field, one massive
fermionic field and one bosonic field with the same mass
value}\label{tsymMM0}

As before, we start with vertices for mixed-symmetry fields and consider the
cubic interaction vertices for one massless fermionic field and two massive
fields (bosonic and fermionic) having the same mass values,
\be\label{eqmas00007NN1n} \mas_1|_\smB = \mas_2|_\smF \equiv \mas  \ne 0
,\qquad \mas_3|_\smF = 0\,, \ee
i.e. the {\it massive} bosonic field carries external line index $a=1$, one
{\it massive} fermionic field carries external line index $a=2$, while the
{\it massless} fermionic field corresponds to $a=3$. We find the following
solution for K- and F-vertices:
\beq
\label{intvereqmas01knmix} && p_\smp3^- = K^\smtwothree V^\Krm (L_n^\smone,
L_n^\smtwo, B_n^\smthree;\, Q_{mn}^\smonetwo\,;\, Z_{mnq})\,,  \hspace{2.4cm}
\hbox{ K-vertex};
\\
\label{intvereqmas01fnmix} && p_\smp3^- = F_n V^\Frm (L_n^\smone, L_n^\smtwo,
B_n^\smthree;\, Q_{mn}^\smonetwo\,;\, Z_{mnq})\,,  \hspace{2.8cm} \hbox{
F-vertex};
\eeq
where generating functions $V^\Krm$, $V^\Frm$ are arbitrary polynomials of
linear, quadratic, and cubic forms. We use the notation%
\beq
\label{Kzerdefnmix}  && K^\smtwothree = \frac{1}{\beta_2\beta_3}p_{\theta_2}
\Bigl( \Po^I\gamma^I \gamma_* -  \mas \beta_3\Bigr) \eta_3\,,
\\[3pt]
\label{fzerdefnmix} && F_n = \frac{1}{\beta_2\beta_3}p_{\theta_2}
\Bigl(\frac{\check{\beta}_1}{\beta_1} \Po^I -\gamma^{IJ} \Po^J + \mas \beta_3
\gamma^I \gamma_*\Bigr) \eta_3 \alpha_n^{\smone I}
\nonumber\\[3pt]
&& \qquad + \frac{1}{\beta_2\beta_3}p_{\theta_2} \Bigl( -
\Po^I\gamma^I\gamma_* +  \frac{\check{\beta}_1}{\beta_1} \beta_3 \mas \Bigr)
\eta_3 \zeta_n^\smone\,,
\eeq
\beq
\label{eqmas00007nmix} &&  L_n^\smone \equiv B_n^\smone - \frac{1}{2}\mas
\zeta_n^\smone\,,\qquad\quad
%
%
L_n^\smtwo \equiv B_n^\smtwo + \frac{1}{2}\mas \zeta_n^\smtwo\,,
\\
\label{eqmas00009nmix} && B_n^\sma \equiv \frac{\alpha_n^{\sma
I}\Po^I}{\beta_a} - \frac{\check{\beta}_a}{2\beta_a} \mas
\zeta_n^\sma\,,\qquad a=1,2;
\\
\label{eqmas00010nmix} && B_n^\smthree\equiv \frac{\alpha_n^{\smthree
I}\Po^I}{\beta_3}\,,
\eeq
\beq
\label{eqmas00006nmix} &&  Q_{mn}^\smonetwo \equiv \alpha_{mn}^\smonetwo -
\frac{\zeta_n^\smtwo}{\mas} B_m^\smone + \frac{\zeta_m^\smone}{\mas}
B_n^\smtwo\,,
\\[5pt]
\label{eqmas00005nmix} && Z_{mnq} \equiv L_m^\smone \alpha_{nq}^\smtwothree +
L^\smtwo \alpha^\smthreeone + B_q^\smthree ( \alpha_{mn}^\smonetwo -
\zeta_m^\smone\zeta_n^\smtwo )\,,
\eeq
where the quadratic forms $\alpha_{mn}^\smab$ are defined in \rf{amnabdef}.

Comparing vertices \rf{intvereqmas01knmix},\rf{intvereqmas01fnmix} with the
ones in \rf{intvereqmas01kmix},\rf{intvereqmas01fmix}, we see that generating
functions $V^\Krm$,  $V^\Frm$ given in
\rf{intvereqmas01knmix},\rf{intvereqmas01fnmix} coincide with ones given in
\rf{intvereqmas01kmix},\rf{intvereqmas01fmix} in Section \ref{equalmasses}.
Vertices in \rf{intvereqmas01knmix},\rf{intvereqmas01fnmix} and the ones in
\rf{intvereqmas01kmix},\rf{intvereqmas01fmix} are distinguished only by
prefactors given in \rf{Kzerdefnmix},\rf{fzerdefnmix} and
\rf{Kzerdefmix},\rf{fzerdefmix} respectively. Note however that, as in
Section \ref{equalmasses}, the prefactors in
\rf{Kzerdefnmix},\rf{fzerdefnmix} are also non-homogeneous polynomials in
$\Po^I$. This implies that, in general, cubic interaction vertices for fields
with mass parameters as in \rf{eqmas00007NN1n} are also non-homogeneous
polynomials in $\Po^I$. As before, solution for $V^\Krm$, $V^\Frm$ given in
\rf{intvereqmas01knmix}, \rf{intvereqmas01fnmix} is the complete solution,
while the solution for the prefactors $K^\smtwothree$, $F_n$  given in
\rf{Kzerdefnmix},\rf{fzerdefnmix} is not complete solution. Namely, for the vertices
of mixed-symmetry fields there are extra solutions for the prefactors that involve
contributions of higher than second order in $\gamma$-matrices.

To discuss the remaining important properties of solution
\rf{intvereqmas01knmix},\rf{intvereqmas01fnmix} we restrict our attention to
cubic vertices for totally symmetric fields.

\subsection{ Cubic vertices for totally symmetric fields }

In this section, we study vertices for three totally symmetric fields with
the mass values given in \rf{eqmas00007NN1n}, i.e. the {\it massive} bosonic
field carries external line index $a=1$, one {\it massive} fermionic field
carries external line index $a=2$, while the {\it massless} fermionic field
corresponds to $a=3$. Because for the studying the totally symmetric fields
it is sufficient to use one sort of oscillators, we set $\nnu = 1$ in
\rf{intvereqmas01knmix},\rf{intvereqmas01fnmix} and ignore the contribution
of the oscillators with $n>1$. To simplify the formulas we drop the
oscillator's subscript $n=1$ and use the simplified notation for oscillators:
$\alpha^{\sma I} \equiv \alpha_1^{\sma I}$, $\zeta^\sma \equiv \zeta_1^\sma$.
In doing so, we obtain, using the solution given in
\rf{intvereqmas01knmix},\rf{intvereqmas01fnmix}, the cubic interaction
vertices for totally symmetric fields,
\beq
\label{intvereqmas01kn} && p_\smp3^- = K^\smtwothree V^\Krm (L^\smone,
L^\smtwo, B^\smthree;\, Q^\smonetwo\,;\, Z)\,,  \hspace{2.6cm} \hbox{
K-vertex};
\\
\label{intvereqmas01fn} && p_\smp3^- = F V^\Frm (L^\smone, L^\smtwo,
B^\smthree;\, Q^\smonetwo\,;\, Z)\,,  \hspace{3.2cm} \hbox{ F-vertex};
\eeq
where we use the notation
\beq
\label{Kzerdefn}  && K^\smtwothree \equiv
\frac{1}{\beta_2\beta_3}p_{\theta_2} \Bigl( \Po^I\gamma^I \gamma_* -  \mas
\beta_3\Bigr) \eta_3\,,
\\[3pt]
\label{fzerdefn} && F \equiv \frac{1}{\beta_2\beta_3}p_{\theta_2}
\Bigl(\frac{\check{\beta}_1}{\beta_1} \Po^I -\gamma^{IJ} \Po^J + \mas \beta_3
\gamma^I \gamma_*\Bigr) \eta_3 \alpha^{\smone I}
\nonumber\\[3pt]
&& \qquad + \frac{1}{\beta_2\beta_3}p_{\theta_2} \Bigl( -
\Po^I\gamma^I\gamma_* +  \frac{\check{\beta}_1}{\beta_1} \beta_3 \mas \Bigr)
\eta_3 \zeta^\smone\,,
\eeq
\beq
\label{eqmas00007n} &&  L^\smone \equiv B^\smone - \frac{1}{2}\mas
\zeta^\smone\,,\qquad\quad
%
%
L^\smtwo \equiv B^\smtwo + \frac{1}{2}\mas \zeta^\smtwo\,,
\\
\label{eqmas00009n} && B^\sma \equiv \frac{\alpha^{\sma I}\Po^I}{\beta_a} -
\frac{\check{\beta}_a}{2\beta_a} \mas \zeta^\sma\,,\qquad a=1,2;
\\
\label{eqmas00010n} && B^\smthree\equiv \frac{\alpha^{\smthree
I}\Po^I}{\beta_3}\,,
\\[5pt]
\label{eqmas00006n} &&  Q^\smonetwo \equiv \alpha^\smonetwo -
\frac{\zeta^\smtwo}{\mas} B^\smone + \frac{\zeta^\smone}{\mas} B^\smtwo\,,
\\[5pt]
\label{eqmas00005n} && Z \equiv L^\smone \alpha^\smtwothree + L^\smtwo
\alpha^\smthreeone + B^\smthree ( \alpha^\smonetwo -
\zeta^\smone\zeta^\smtwo )\,.
\eeq
Solution in \rf{intvereqmas01kn},\rf{intvereqmas01fn} is the complete
solution, i.e., vertices given in \rf{intvereqmas01kn},\rf{intvereqmas01fn}
provide the complete list of parity invariant cubic vertices for the totally
symmetric fields with the mass parameters given in \rf{eqmas00007NN1n}. To
discuss the remaining important properties of solution in
\rf{intvereqmas01kn},\rf{intvereqmas01fn} we restrict our attention to cubic
vertices for totally symmetric fields with fixed spin values.

\subsubsection{ Cubic vertices for totally symmetric fields with fixed but arbitrary
spin values}\label{secfefereqmasbos}

We now restrict ourselves to cubic vertices for the totally symmetric fields
with fixed but arbitrary spin values and with mass values given in
\rf{eqmas00007NN1n}. Vertices \rf{intvereqmas01kn},\rf{intvereqmas01fn}
describe interaction of the towers of fermionic and bosonic fields. We next
consider the vertices involving one massive spin-$s^\smone$ bosonic field,
one massive spin-$(s^\smtwo + \half)$ fermionic field and one massless
spin-$(s^\smthree + \half)$ fermionic field, where bosonic and fermionic
fields have the same mass parameter $\mas\ne 0$:
\be \label{manold-30122011-01}
\begin{array}{lll}
\mas_1 = \mas\,, \qquad & \mas_2 = \mas\,, \quad \qquad \qquad & \mas_3 =
0\,,
\\[5pt]
s^\smone \,, & s^\smtwo+\half \,, & s^\smthree+\half\,.
\end{array}
\ee
Massive spin-$s^\smone$ bosonic field and massive spin-$(s^\smtwo+\half)$
fermionic field are described by the respective ket-vectors $|\phi_{s^\smone
}\rangle$ and $|\psi_{s^\smtwo }\rangle$, while the massless
spin-$(s^\smthree+\half)$ fermionic field is described by ket-vector
$|\psi_{s^\smthree }\rangle$. The ket-vectors $|\phi_{s^\smone}\rangle$,
$|\psi_{s^\smtwo }\rangle$ can be obtained from the respective expressions in
\rf{manold-07122011-04},\rf{fer01} by the replacement $s\rightarrow s^\sma $,
$\alpha^I\rightarrow \alpha^{\sma I}$, $\zeta\rightarrow \zeta^\sma$,
$a=1,2$, while the ket-vector $|\psi_{s^\smthree }\rangle$ can be obtained
from \rf{fer01} by the replacement $s\rightarrow s^\smthree $,
$\alpha^I\rightarrow \alpha^{\smthree I}$. Taking into account that the
ket-vectors $|\phi_{s^\smone }\rangle$, $|\psi_{s^\smtwo }\rangle$ are the
respective degree-$s^\sma $ homogeneous polynomials in the oscillators
$\alpha^{\sma I}$, $\zeta^\sma $, $a=1,2$, while the ket-vector
$|\psi_{s^\smthree }\rangle$ is a degree-$s^\smthree$ homogeneous polynomial
in the oscillator $\alpha^{\smthree I}$, we see that the vertices we are
interested in must satisfy the equations
\beq
&& \label{mm018n}  (\alpha^{\sma I}\bar\alpha^{\sma I} +\zeta^\sma
\bar\zeta^\sma  - s^\sma ) |p_\smp3^-\rangle = 0\,,\qquad a=1,2\,,
\\[3pt]
&& \label{mm019n}  (\alpha^{\smthree I} \bar\alpha^{\smthree I}  - s^\smthree
) |p_\smp3^-\rangle = 0\,. \eeq
These equations tell us that the vertices must be a degree-$s^\sma $
homogeneous polynomial in the respective oscillators. Taking into account
that the prefactor $F$ and the forms $L^\smone$, $L^\smtwo$, $B^\smthree$
\rf{fzerdefn}-\rf{eqmas00010n} are degree-1 homogeneous polynomials in the
oscillators, while forms $Q^\smonetwo$,$Z$, \rf{eqmas00006n},\rf{eqmas00005n}
are the respective degree-2 and degree-3 homogeneous polynomials in the
oscillators we find the general solution of Eqs.\rf{mm018n}, \rf{mm019n} as
\beq
&& \label{intvereqmass01kn}
p_\smp3^-(s^\smone,s^\smtwo+\half,s^\smthree+\half\,;\,k_\min,k_\max)
\nonumber\\
&&\qquad\qquad   = K^\smtwothree (L^\smone)^{w^\smone } (L^\smtwo)^{w^\smtwo
} (B^\smthree)^{w^\smthree } (Q^\smonetwo)^{y^\smthree } Z^y\,,
\hspace{1.2cm} \hbox{ K-vertex};
\\[3pt]
&& \label{intvereqmass01fn} \qquad \qquad  = F (L^\smone)^{x^\smone  }
(L^\smtwo)^{x^\smtwo } (B^\smthree)^{x^\smthree } (Q^\smonetwo)^{y^\smthree }
Z^y\,,  \hspace{2cm} \hbox{ F-vertex}\,,\qquad
\eeq
where the parameters $w^\smone  $, $w^\smtwo $, $w^\smthree $, $y^\smthree $,
$y$ are given by
\beq
\label{xadefmm0kn}
&& w^\smone   = k_\max - k_\min - s^\smtwo  -1 \,,
\nonumber\\
&& w^\smtwo  = k_\max - k_\min - s^\smone -1 \,,
\nonumber\\
&& w^\smthree  = k_\min\,,
\nonumber\\
&& y^\smthree =  \sbf - 2s^\smthree  - k_\max + 2 k_\min + 1\,,
\nonumber\\
&& y= s^\smthree  -k_\min\,, \hspace{5cm} \hbox{ for K-vertices};
\eeq
\beq
\label{xadefmm0fn}
&& w^\smone   = k_\max - k_\min - s^\smtwo -1 \,,
\nonumber\\
&& w^\smtwo  = k_\max - k_\min - s^\smone \,,
\nonumber\\
&& w^\smthree  = k_\min\,,
\nonumber\\
&& y^\smthree =  \sbf - 2s^\smthree  - k_\max + 2 k_\min \,,
\nonumber\\
&& y= s^\smthree  -k_\min\,, \hspace{4.5cm} \hbox{ for F-vertices};
\eeq
and $\sbf$ is defined in \rf{0007}. Integers $k_\min$ and $k_\max$ in
\rf{intvereqmass01kn}-\rf{xadefmm0fn} are the freedom in our solution. In
general, vertices \rf{intvereqmass01kn},\rf{intvereqmass01fn} are
non-homogeneous polynomials in the momentum $\Po^I$ and the integers $k_\min$
and $k_\max$ are the respective minimal and maximal numbers of powers of the
momentum $\Po^I$ in \rf{intvereqmass01kn},\rf{intvereqmass01fn}. For vertices
\rf{intvereqmass01kn},\rf{intvereqmass01fn} to be sensible, we should impose
the restrictions
\be
\label{restrict01n} x^\sma  \geq 0\,, \quad a=1,2,3; \qquad  y^\smthree \geq
0\,,\qquad y \geq 0\,,
\ee
which amount to requiring the powers of all forms in
\rf{intvereqmass01kn},\rf{intvereqmass01fn} to be non--negative integers.
Using \rf{xadefmm0kn},\rf{xadefmm0fn}, restrictions \rf{restrict01n} can be
rewritten in a more convenient form as
\beq
&& \label{restrict04kn} k_\min  + \max_{a=1,2} s^\sma +1 \leq k_\max \leq
\sbf- 2s^\smthree  + 2 k_\min + 1\,,
\nonumber\\
&& 0 \leq k_\min \leq s^\smthree \,, \hspace{5.5cm} \hbox{ for K-vertices};
\\[5pt]
&& \label{restrict04fn} k_\min  + \max (s^\smone-1,s^\smtwo)  + 1 \leq
k_\max \leq \sbf- 2s^\smthree  + 2 k_\min\,,
\nonumber\\
&&  0\leq k_\min \leq s^\smthree \,, \hspace{5.5cm} \hbox{ for F-vertices}\,.
\eeq

Cubic vertices in \rf{intvereqmass01kn}, \rf{intvereqmass01fn} and
restrictions on the  allowed values of $k_\max$ and $k_\min$ in
\rf{restrict04kn},\rf{restrict04fn} provide the complete list of parity
invariant cubic vertices for the totally symmetric fields with spin and mass
values given in \rf{manold-30122011-01}.

\newsection{ Cubic vertices for one massless bosonic field and two massive fermionic
fields with different mass values}\label{sec-09-001}

We now consider the cubic vertices for mixed-symmetry fields with
the following mass values:
\be \label{mm08} \mas_1|_\smF \ne 0 ,\qquad \mas_2|_\smF\ne  0,\qquad \mas_1
\ne \mas_2, \qquad \mas_3|_\smB= 0, \ee
i.e. the {\it massive} fermionic fields carry external line indices $a=1,2$,
while the {\it massless} bosonic field corresponds to $a=3$. Equations for
the vertex involving one massless field can be obtained from
Eqs.\rf{cubver13nn} in the limit as $\mas_3 \rightarrow 0 $. The K- and
F-vertices we found take the form
\beq
&& \label{mm09ex01kmix} p_\smp3^- = K^\smonetwo V^\Krm(L_n^\smone,
L_n^\smtwo;\, Q_{mn}^\smaaplusone)\,, \hspace{2cm} \hbox{ K-vertex};
\\[3pt]
&& \label{mm09ex01fmix} p_\smp3^- = F_n V^\Frm(L_n^\smone, L_n^\smtwo;\,
Q_{mn}^\smaaplusone)\,, \hspace{2.5cm} \hbox{ F-vertex};
\eeq
where generating functions $V^\Krm$, $V^\Frm$ are arbitrary polynomials of
linear and quadratic forms. We use the notation
\beq
\label{mm012fmix} && \hspace{-1.5cm} K^\smonetwo \equiv
\frac{1}{\beta_1\beta_2}p_{\theta_1}\Bigl( \Po^I\gamma^I\gamma_* -
\mas_1\beta_2 + \mas_2\beta_1 \Bigr)\eta_2\,,
\\
\label{mm012fmixa1} &&  \hspace{-1.5cm} F_n \equiv
\frac{1}{\beta_1\beta_2}p_{\theta_1}\Bigl( \frac{\check{\beta_3}}{\beta_3}
\Po^I - \gamma^{IJ}\Po^J +( \mas_1\beta_2 + \mas_2\beta_1) \gamma^I \gamma_*
\Bigr) \eta_2 \alpha_n^{\smthree I} - \frac{2}{\mas_1 + \mas_2} K^\smonetwo
B_n^\smthree\,,
\eeq
\beq
\label{mm012mix} && L_n^\smone \equiv B_n^\smone -
\frac{\mas_2^2}{2\mas_1}\zeta^\smone\,,\qquad  L_n^\smtwo \equiv B_n^\smtwo +
\frac{\mas_1^2}{2\mas_2}\zeta_n^\smtwo\,,
\\
&& B_n^\sma \equiv \frac{\alpha_n^{\sma I}\Po^I}{\beta_a}-
\frac{\check{\beta}_a}{2\beta_a} \mas_a \zeta_n^\sma\,,\qquad a=1,2;
\\
\label{mm013ex01mix}&& B_n^\smthree\equiv \frac{\alpha_n^{\smthree
I}\Po^I}{\beta_3}\,,
\\
\label{mm09mix}&& Q_{mn}^\smonetwo \equiv \alpha_{mn}^\smonetwo -
\frac{\zeta_n^\smtwo}{\mas_2} B_m^\smone
 + \frac{\zeta_m^\smone}{\mas_1} B_n^\smtwo\,,
\\
\label{mm010mix}&& Q_{mn}^\smtwothree \equiv \alpha_{mn}^\smtwothree -
\frac{\mas_2\zeta_m^\smtwo}{\mas_1^2 - \mas_2^2} B_n^\smthree -
\frac{2}{\mas_1^2 - \mas_2^2} B_m^\smtwo B_n^\smthree\,,
\\
\label{mm011mix}&& Q_{mn}^\smthreeone \equiv \alpha_{mn}^\smthreeone -
\frac{\mas_1\zeta_n^\smone}{\mas_1^2 - \mas_2^2} B_m^\smthree +
\frac{2}{\mas_1^2 - \mas_2^2} B_m^\smthree B_n^\smone\,, \eeq
and the quadratic forms and $\alpha_{mn}^\smab$ are defined in \rf{amnabdef}.

The following remarks are in order.
\\
\noindent {\bf i}) We note appearance of expressions like $\mas_1^2 -
\mas_2^2$ in the denominators of quadratic forms $Q_{mn}^\smtwothree$,
$Q_{mn}^\smthreeone$, \rf{mm010mix}, \rf{mm011mix}, i.e., we see that the quadratic
forms $Q_{mn}^\smtwothree$, $Q_{mn}^\smthreeone$ are singular as
$\mas_1\rightarrow \mas_2$. For this reason, we considered the case of
$\mas_1 = \mas_2$ separately in Section \ref{equalmasses}.
\\
\noindent {\bf ii}) From \rf{mm012fmix}-\rf{mm011mix}, we see that the prefactors
$K^\smonetwo$, $F_n$, the linear forms $L_n^\smone$, $L_n^\smtwo$, and the
quadratic forms $Q_{mn}^\smaaplusone$ are non-homogeneous polynomials in
$\Po^I$. This implies that, in general, the cubic vertices given in
\rf{mm09ex01kmix},\rf{mm09ex01fmix} are a non-homogeneous polynomials in
$\Po^I$.
\\
\noindent {\bf iii}) Solution for generating functions $V^\Krm$, $V^\Frm$
given in \rf{mm09ex01kmix}, \rf{mm09ex01fmix} is complete solution, while the
solution for the prefactors $K^\smonetwo$, $F_n$ given in
\rf{mm012fmix},\rf{mm012fmixa1} is not complete solution. Namely, for the vertices of
mixed-symmetry fields there are extra solutions for the prefactors that involve
contribution of higher than second order in $\gamma$-matrices.

To understand the remaining characteristic properties of the solution
obtained, we consider the vertices for totally symmetric fields.

\subsection{ Cubic vertices for totally symmetric one massless bosonic field
and two massive fermionic fields with different mass values}

We proceed with the study of vertices for three totally symmetric fields with
the mass values given in \rf{mm08}, i.e. the {\it massive} fermionic fields
carry external line indices $a=1,2$, while the {\it massless} bosonic field
corresponds to $a=3$. Since for the studying the totally symmetric fields it is
sufficient to use one sort of oscillators, we set $\nnu = 1$ in
\rf{mm09ex01kmix},\rf{mm09ex01fmix} and ignore the contribution of the
oscillators with $n>1$. To simplify the formulas we drop the oscillator's
subscript $n=1$ and use the simplified notation for oscillators:
$\alpha^{\sma I} \equiv \alpha_1^{\sma I}$, $\zeta^\sma \equiv \zeta_1^\sma$.
In doing so, we obtain, using the solution given in
\rf{mm09ex01kmix},\rf{mm09ex01fmix}, the cubic interaction vertices for
totally symmetric fields,
\beq
&& \label{mm09ex01k} p_\smp3^- = K^\smonetwo V^\Krm(L^\smone, L^\smtwo;\,
Q^\smaaplusone)\,, \hspace{2cm} \hbox{ K-vertex};
\\[3pt]
&& \label{mm09ex01f} p_\smp3^- = F V^\Frm(L^\smone, L^\smtwo;\,
Q^\smaaplusone)\,, \hspace{2.6cm} \hbox{ F-vertex};
\eeq
where we use the notation
\beq
\label{23012010-man02} && \hspace{-1.5cm} F \equiv
\frac{1}{\beta_1\beta_2}p_{\theta_1}\Bigl( \frac{\check{\beta_3}}{\beta_3}
\Po^I - \gamma^{IJ}\Po^J +( \mas_1\beta_2 + \mas_2\beta_1) \gamma^I \gamma_*
\Bigr) \eta_2 \alpha^{\smthree I} - \frac{2}{\mas_1 + \mas_2} K^\smonetwo
B^\smthree\,,
\eeq
\be
\label{mm012} L^\smone \equiv B^\smone -
\frac{\mas_2^2}{2\mas_1}\zeta^\smone\,,\qquad
L^\smtwo \equiv B^\smtwo + \frac{\mas_1^2}{2\mas_2}\zeta^\smtwo\,, \ee
\beq
&& B^\sma \equiv \frac{\alpha^{\sma I}\Po^I}{\beta_a}-
\frac{\check{\beta}_a}{2\beta_a} \mas_a \zeta^\sma\,,\qquad a=1,2;
\\
\label{mm013ex01}&& B^\smthree\equiv \frac{\alpha^{\smthree
I}\Po^I}{\beta_3}\,,\eeq
\vspace{-0.5cm}
\beq
\label{mm09}&& Q^\smonetwo \equiv \alpha^\smonetwo -
\frac{\zeta^\smtwo}{\mas_2} B^\smone
 + \frac{\zeta^\smone}{\mas_1} B^\smtwo\,,
\\
\label{mm010}&& Q^\smtwothree \equiv \alpha^\smtwothree -
\frac{\mas_2\zeta^\smtwo}{\mas_1^2 - \mas_2^2} B^\smthree -
\frac{2}{\mas_1^2 - \mas_2^2} B^\smtwo B^\smthree\,,
\\
\label{mm011}&& Q^\smthreeone \equiv \alpha^\smthreeone -
\frac{\mas_1\zeta^\smone}{\mas_1^2 - \mas_2^2} B^\smthree + \frac{2}{\mas_1^2
- \mas_2^2} B^\smthree B^\smone\,. \eeq
The quadratic forms $\alpha^\smab$ and prefactor $K^\smonetwo$ are defined in
\rf{amnabdefsym} and \rf{mm012fmix} respectively. Solution given in
\rf{mm09ex01k},\rf{mm09ex01f} is the complete solution. In other words,
solution given in \rf{mm09ex01k},\rf{mm09ex01f} provides the complete list of
parity invariant cubic interaction for totally symmetric field with mass
parameters given in \rf{mm08}. To understand the remaining characteristic
properties of the solution obtained, we consider the vertices for the totally
symmetric fields with fixed spin values.

\subsubsection{ Cubic vertices for totally symmetric fields with fixed but
arbitrary mass values} \label{tsymMM0noneqmas}

Discussion of cubic interaction vertices for two massive fermionic fields
with different mass values and one massless bosonic field is similar the one
in Section \ref{equalmasses}. This, we consider vertices involving two
massive spin-$(s^\smone + \half)$ and spin-$(s^\smtwo + \half)$ fermionic
fields and one massless spin-$s^\smthree$ bosonic field, where the fermionic
fields have different values of the mass parameter:
\be
\begin{array}{lll}
\mas_1 \ne 0\,, \qquad & \mas_2 \ne 0\,, \quad \mas_1\ne \mas_2\,, \qquad &
\mas_3 = 0\,,
\\[5pt]
s^\smone+\half \,, & s^\smtwo+\half \,, & s^\smthree\,.
\end{array}
\ee
The vertices for totally symmetric spin-$(s^\smone+\half)$ and spin-$(s^\smtwo+\half)$
fermionic fields $|\psi_{s^\smone }\rangle$, $|\psi_{s^\smtwo }\rangle$ with
different mass values and one massless spin-$s^\smthree$  bosonic field
$|\phi_{s^\smthree }\rangle$ can be obtained by solving
Eqs.\rf{mm018}, \rf{mm019} with $p_\smp3^-$ given in
\rf{mm09ex01k},\rf{mm09ex01f}. We then obtain the cubic vertices
\beq
\label{pint40k} &&
p_\smp3^-(s^\smone+\half,s^\smtwo+\half,s^\smthree\,;\,w^\smone ,w^\smtwo )
\nonumber\\
&& \qquad = K^\smonetwo (L^\smone)^{w^\smone  } (L^\smtwo)^{w^\smtwo }
\prod_{a=1}^3 (Q^\smaaplusone)^{y^\smaplustwo} \,, \hspace{1cm} \hbox{
K-vertex};
\\
\label{pint40f} && \qquad = F (L^\smone)^{w^\smone} (L^\smtwo)^{w^\smtwo}
\prod_{a=1}^3 (Q^\smaaplusone)^{y^\smaplustwo}\,, \hspace{1.5cm} \hbox{
F-vertex};
\eeq
where the parameters $y^\sma $ are given by
\beq
\label{mm014k}&& y^\smone = \frac{1}{2}(s^\smtwo  + s^\smthree - s^\smone  +
w^\smone  - w^\smtwo )\,,
\nonumber\\
&& y^\smtwo = \frac{1}{2}(s^\smone  + s^\smthree - s^\smtwo  - w^\smone +
w^\smtwo )\,,
\nonumber\\
&& y^\smthree = \frac{1}{2}(s^\smone  + s^\smtwo -s^\smthree - w^\smone -
w^\smtwo )\,,\qquad \hspace{1cm} \hbox{ for K-vertex};
\\
\label{mm014f}&& y^\smone = \frac{1}{2}(s^\smtwo  + s^\smthree -s^\smone  +
w^\smone   - w^\smtwo -1)\,,
\nonumber\\
&& y^\smtwo = \frac{1}{2}(s^\smone + s^\smthree - s^\smtwo  - w^\smone   +
w^\smtwo -1)\,,
\nonumber\\
&& y^\smthree = \frac{1}{2}(s^\smone  + s^\smtwo - s^\smthree  - w^\smone -
w^\smtwo +1  )\,, \hspace{1cm} \hbox{ for F-vertex}.
\eeq
Two integers $w^\smone$, $w^\smtwo $ are the freedom of our solution. For
fixed spin values $s^\smone$, $s^\smtwo$, $s^\smthree$, these integers label
all possible cubic interaction vertices that can be built for the fields
under consideration. For vertices \rf{pint40k},\rf{pint40f} to be sensible we
impose the restrictions
\beq
&& \label{restr0101k} w^\smone  \geq 0 \,,\qquad w^\smtwo \geq 0\,, \qquad
 y^\sma  \geq 0\,,\qquad a=1,2,3;
\nonumber \\[3pt]
&& \sbf - w^\smone   - w^\smtwo\,\,  \hbox{ even integer}\,, \hspace{6cm}
\hbox{ for K-vertex};\qquad
\\[3pt]
&& \label{restr0101f} w^\smone  \geq 0 \,,\qquad w^\smtwo \geq 0\,, \qquad
y^\sma  \geq 0\,,\qquad a=1,2,3;
\nonumber \\[3pt]
&& \sbf - w^\smone   - w^\smtwo\,\,  \hbox{ odd integer}\,, \hspace{6cm}
\hbox{ for F-vertex},
\eeq
which amount to the requirement that the powers of all forms in
\rf{pint40k},\rf{pint40f} be non-negative integers. The maximal and minimal
number of powers of $\Po^I$ in \rf{pint40k},\rf{pint40f}, which we
denote by $k_\max$ and $k_\min$ respectively, are given by%
\footnote{ Expressions for $K^\smonetwo$ \rf{mm012fmix} and $F$, $L^\sma$,
$Q^\smaaplusone$ \rf{23012010-man02}-\rf{mm011} imply that $k_\max = w^\smone
+ w^\smtwo +2y^\smone + 2 y^\smtwo + y^\smthree + 1$ for K-vertex and $k_\max
= w^\smone + w^\smtwo +2y^\smone + 2 y^\smtwo + y^\smthree + 2$ for F-vertex.
Relations for $y^\sma$ \rf{mm014k},\rf{mm014f} then lead to $k_\max$ given in
\rf{kmaxmm0N1k},\rf{kmaxmm0N1f}.}
\beq
\label{kmaxmm0N1k} &&\hspace{-1cm} k_\max = \frac{1}{2}(s^\smone  + s^\smtwo   +
3s^\smthree + w^\smone  + w^\smtwo ) + 1 \,,  \qquad k_\min=0\,, \hspace{1cm} \hbox{ for
K-vertex};\qquad
\\[2pt]
\label{kmaxmm0N1f} &&\hspace{-1cm}  k_\max = \frac{1}{2}(s^\smone  + s^\smtwo
+ 3s^\smthree + w^\smone  + w^\smtwo +1)\,,  \qquad k_\min=0\,, \hspace{1cm}
\hbox{ for F-vertex}.
\eeq
We note that using \rf{mm014k},\rf{mm014f} allows rewriting restrictions
\rf{restr0101k},\rf{restr0101f} in the equivalent form%
\footnote{ If $w^\smone = w^\smtwo=0$, then restrictions \rf{mm017k} becomes
the  restrictions well known in the angular momentum theory: $ |s^\smone -
s^\smtwo| \leq s^\smthree \leq s^\smone + s^\smtwo$, while restriction
\rf{mm017f} takes the form $ |s^\smone - s^\smtwo| \leq s^\smthree -1 \leq
s^\smone + s^\smtwo$.}
\beq
&&\hspace{-2cm}  \label{mm017k} |s^\smone  -s^\smtwo  - w^\smone   + w^\smtwo
| \leq s^\smthree \leq s^\smone + s^\smtwo - w^\smone   - w^\smtwo\,,
\hspace{2.4cm} \hbox{ for K-vertex};
\\[3pt]
&& \hspace{-2cm} \label{mm017f} |s^\smone  -s^\smtwo  - w^\smone   + w^\smtwo
| +1 \leq s^\smthree \leq s^\smone + s^\smtwo - w^\smone  - w^\smtwo +1\,,
\hspace{1cm} \hbox{ for F-vertex};
\eeq

\newsection{ Cubic vertices for one massless fermionic field, one massive fermionic
field and one massive bosonic  field with different mass values}
\label{Sec1010}

We now consider the cubic interaction vertices for mixed-symmetry fields with
the following mass values:
\be \label{mm08n} \mas_1|_\smB \ne 0 ,\qquad \mas_2|_\smF\ne  0,\qquad \mas_1
\ne \mas_2, \qquad \mas_3|_\smF= 0, \ee
i.e. the {\it massive} bosonic field carries external line index $a=1$, the
{\it massive} fermionic field carries external line index $a=2$, while the
{\it massless} fermionic field corresponds to $a=3$. Equations for the vertex
involving one massless field can be obtained from Eqs.\rf{cubver13nn} in the
limit as $\mas_3 \rightarrow 0 $. The solution for K- and F-vertices we found
is given by
\beq
&& \label{mm09ex01knmix} p_\smp3^- = K^\smtwothree V^\Krm(L_n^\smone,
L_n^\smtwo;\, Q_{mn}^\smaaplusone)\,, \hspace{2cm} \hbox{ K-vertex};
\\[3pt]
&& \label{mm09ex01fnmix} p_\smp3^- = F_n V^\Frm(L_n^\smone, L_n^\smtwo;\,
Q_{mn}^\smaaplusone)\,, \hspace{2.5cm} \hbox{ F-vertex};
\eeq
where generating functions $V^\Krm$, $V^\Frm$ are arbitrary polynomials of
linear and quadratic forms. We use the notation
\beq
\label{mm012fnmix} && K^\smtwothree \equiv
\frac{1}{\beta_2\beta_3}p_{\theta_2}\Bigl( \Po^I\gamma^I\gamma_* -
\mas_2\beta_3 \Bigr)\eta_3\,,
\\
\label{mm012fnmixa1} &&  F_n \equiv
\frac{1}{\beta_2\beta_3}p_{\theta_2}\Bigl( \frac{\check{\beta_1}}{\beta_1}
\Po^I - \gamma^{IJ}\Po^J + \beta_3 \mas_2 \gamma^I \gamma_* \Bigr) \eta_3
\alpha_n^{\smone I}
\nonumber\\[3pt]
&& \qquad + \frac{1}{\beta_2\beta_3}p_{\theta_2}\Bigl( - \mas_2 \Po^I
\gamma^I \gamma_* +  \beta_3 \mas_2^2 + \frac{2\beta_2\beta_3}{\beta_1}
\mas_1^2 \Bigr) \eta_3 \frac{\zeta_n^\smone}{\mas_1}\,,
\eeq

\be
\label{mm012nmix} L_n^\smone \equiv B_n^\smone -
\frac{\mas_2^2}{2\mas_1}\alpha_n^\smone\,,\qquad
L_n^\smtwo \equiv B_n^\smtwo + \frac{\mas_1^2}{2\mas_2}\zeta_n^\smtwo\,, \ee
\beq
&& B_n^\sma \equiv \frac{\alpha_n^{\sma I}\Po^I}{\beta_a}-
\frac{\check{\beta}_a}{2\beta_a} \mas_a \zeta_n^\sma\,,\qquad a=1,2;
\\
\label{mm013ex01nmix}&& B_n^\smthree\equiv \frac{\alpha_n^{\smthree
I}\Po^I}{\beta_3}\,,
\\
\label{mm09nmix}&& Q_{mn}^\smonetwo \equiv \alpha_{mn}^\smonetwo -
\frac{\zeta_n^\smtwo}{\mas_2} B_m^\smone
 + \frac{\zeta_m^\smone}{\mas_1} B_n^\smtwo\,,
\\
\label{mm010nmix}&& Q_{mn}^\smtwothree \equiv \alpha_{mn}^\smtwothree -
\frac{\mas_2\zeta_m^\smtwo}{\mas_1^2 - \mas_2^2} B_n^\smthree -
\frac{2}{\mas_1^2 - \mas_2^2} B_m^\smtwo B_n^\smthree\,,
\\
\label{mm011nmix}&& Q_{mn}^\smthreeone \equiv \alpha_{mn}^\smthreeone -
\frac{\mas_1\zeta_n^\smone}{\mas_1^2 - \mas_2^2} B_m^\smthree +
\frac{2}{\mas_1^2 - \mas_2^2} B_m^\smthree B_n^\smone\,, \eeq
and the quadratic forms $\alpha_{mn}^\smab$ are defined in \rf{amnabdef}.
Comparing vertices \rf{mm09ex01knmix},\rf{mm09ex01fnmix} with the ones in
\rf{mm09ex01kmix},\rf{mm09ex01fmix}, we see that generating functions
$V^\Krm$, $V^\Frm$ given in \rf{mm09ex01knmix},\rf{mm09ex01fnmix} coincide
with the ones discussed in \rf{mm09ex01kmix},\rf{mm09ex01fmix} in Section
\ref{sec-09-001}. Vertices in \rf{mm09ex01knmix},\rf{mm09ex01fnmix} and the
ones in \rf{mm09ex01kmix},\rf{mm09ex01fmix} are distinguished only by the
prefactors given in \rf{mm012fnmix},\rf{mm012fnmixa1} and
\rf{mm012fmix},\rf{mm012fmixa1} respectively. Note however that, as in
Section \ref{sec-09-001}, the prefactors in \rf{mm012fnmix},\rf{mm012fnmixa1}
are also non-homogeneous polynomials in $\Po^I$. This implies that, in
general, cubic vertices for fields with mass parameters given in \rf{mm08n}
are also non-homogeneous polynomials in $\Po^I$. As before, solution for
$V^\Krm$, $V^\Frm$ given in \rf{mm09ex01knmix},\rf{mm09ex01fnmix} is complete
solution, while the solution for the prefactors $K^\smtwothree$, $F_n$  given
in \rf{mm012fnmix},\rf{mm012fnmixa1} is not complete solution. Namely, for
the vertices of mixed-symmetry fields there are extra solutions for the
prefactors that involve contributions of higher than second order in
$\gamma$-matrices.

\subsection{ Cubic vertices for totally symmetric arbitrary spin fields }

We now consider cubic vertices for totally symmetric fields
with the mass values given in \rf{mm08n} i.e., the {\it massive}
bosonic field carries external line index $a=1$, the {\it massive} fermionic
field carries external line index $a=2$, while the {\it massless} fermionic
field corresponds to $a=3$. To study the totally symmetric fields we set $\nnu = 1$ in
\rf{mm09ex01knmix},\rf{mm09ex01fnmix} and ignore the contribution of the
oscillators with $n>1$. To simplify the formulas we drop the oscillator's
subscript $n=1$ and use the simplified notation for oscillators:
$\alpha^{\sma I} \equiv \alpha_1^{\sma I}$, $\zeta^\sma \equiv \zeta_1^\sma$.
In doing so, we obtain, using the solution given
\rf{mm09ex01knmix},\rf{mm09ex01fnmix}, the cubic interaction vertices for the
totally symmetric fields,
\beq
&& \label{mm09ex01kn} p_\smp3^- = K^\smtwothree V^\Krm(L^\smone, L^\smtwo;\,
Q^\smaaplusone)\,, \hspace{2cm} \hbox{ K-vertex};
\\[3pt]
&& \label{mm09ex01fn} p_\smp3^- = F V^\Frm(L^\smone, L^\smtwo;\,
Q^\smaaplusone)\,, \hspace{2.6cm} \hbox{ F-vertex};
\eeq
where we use the notation
\beq
\label{23012010-man03} &&  F \equiv
\frac{1}{\beta_2\beta_3}p_{\theta_2}\Bigl( \frac{\check{\beta_1}}{\beta_1}
\Po^I - \gamma^{IJ}\Po^J + \beta_3 \mas_2 \gamma^I \gamma_* \Bigr) \eta_3
\alpha^{\smone I}
\nonumber\\
&& \qquad + \frac{1}{\beta_2\beta_3}p_{\theta_2}\Bigl( - \mas_2 \Po^I
\gamma^I \gamma_* +  \beta_3 \mas_2^2 + \frac{2\beta_2\beta_3}{\beta_1}
\mas_1^2 \Bigr) \eta_3 \frac{\zeta^\smone}{\mas_1}\,,
\eeq

\beq
\label{mm012n} && L^\smone \equiv B^\smone -
\frac{\mas_2^2}{2\mas_1}\zeta^\smone\,,\qquad
L^\smtwo \equiv B^\smtwo + \frac{\mas_1^2}{2\mas_2}\zeta^\smtwo\,,
\\
&& B^\sma \equiv \frac{\alpha^{\sma I}\Po^I}{\beta_a}-
\frac{\check{\beta}_a}{2\beta_a} \mas_a \zeta^\sma\,,\qquad a=1,2;
\\
\label{mm013ex01n}&& B^\smthree\equiv \frac{\alpha^{\smthree
I}\Po^I}{\beta_3}\,,
\\
\label{mm09n}&& Q^\smonetwo \equiv \alpha^\smonetwo -
\frac{\zeta^\smtwo}{\mas_2} B^\smone
 + \frac{\zeta^\smone}{\mas_1} B^\smtwo\,,
\\
\label{mm010n}&& Q^\smtwothree \equiv \alpha^\smtwothree -
\frac{\mas_2\zeta^\smtwo}{\mas_1^2 - \mas_2^2} B^\smthree -
\frac{2}{\mas_1^2 - \mas_2^2} B^\smtwo B^\smthree\,,
\\
\label{mm011n}&& Q^\smthreeone \equiv \alpha^\smthreeone -
\frac{\mas_1\zeta^\smone}{\mas_1^2 - \mas_2^2} B^\smthree + \frac{2}{\mas_1^2
- \mas_2^2} B^\smthree B^\smone\,. \eeq
The quadratic forms $\alpha^\smab$ and the prefactor $K^\smtwothree$ are
defined in \rf{amnabdefsym} and \rf{mm012fnmix} respectively.  Solution given
in \rf{mm09ex01kn},\rf{mm09ex01fn} provides the complete list of parity
invariant cubic interaction for totally symmetric field with mass parameters
given in \rf{mm08n}. We now turn  to vertices for the totally symmetric
fields with fixed spin values.

\subsubsection{ Cubic vertices for totally symmetric fields with fixed but
arbitrary spin values}

Discussion of cubic vertices for one massless fermionic field, one massive
fermionic field and one massive bosonic field with different mass values is
similar to the one in Section \ref{secfefereqmasbos}. This, we consider
vertices involving one massive spin-$s^\smone$ bosonic field, one massive
spin-$(s^\smtwo + \half)$ fermionic field and one massless
spin-$(s^\smthree+\half)$ fermionic field, where the massive fermionic and
massive bosonic fields have different values of the mass parameter:
\be \label{manold-30122011-02}
\begin{array}{lll}
\mas_1 \ne 0\,, \qquad & \mas_2 \ne 0\,, \quad \mas_1\ne \mas_2\,, \qquad &
\mas_3 = 0\,,
\\[5pt]
s^\smone \,, & s^\smtwo+\half \,, & s^\smthree+\half\,.
\end{array}
\ee
The cubic vertices for fields with the spin and mass values as in
\rf{manold-30122011-02}  can be obtained by solving Eqs.\rf{mm018n},
\rf{mm019n} with $p_\smp3^-$ given in \rf{mm09ex01kn},\rf{mm09ex01fn}. We
then obtain the cubic vertices
\beq
\label{pint40kn} && p_\smp3^-(s^\smone,s^\smtwo+\half,s^\smthree +\half\,;\,
w^\smone, w^\smtwo )
\nonumber\\
&& \hspace{3cm}= K^\smtwothree (L^\smone)^{w^\smone  } (L^\smtwo)^{w^\smtwo
} \prod_{a=1}^3 (Q^\smaaplusone)^{y^\smaplustwo}\,, \hspace{1cm} \hbox{
K-vertex};\qquad
\\
\label{pint40fn} && \hspace{3cm} = F (L^\smone)^{w^\smone}
(L^\smtwo)^{w^\smtwo} \prod_{a=1}^3 (Q^\smaaplusone)^{y^\smaplustwo} \,,
\hspace{1.5cm} \hbox{ F-vertex};
\eeq
where the parameters $y^\sma $ are given by
\beq
\label{mm014kn}&& y^\smone = \frac{1}{2}(s^\smtwo  + s^\smthree -s^\smone  +
w^\smone  - w^\smtwo )\,,
\nonumber\\
&& y^\smtwo = \frac{1}{2}(s^\smone  + s^\smthree - s^\smtwo  - w^\smone +
w^\smtwo )\,,
\nonumber\\
&& y^\smthree = \frac{1}{2}(s^\smone  + s^\smtwo -s^\smthree - w^\smone -
w^\smtwo )\,,\qquad \hspace{1cm} \hbox{ for K-vertex};
\\
\label{mm014fn}&& y^\smone = \frac{1}{2}(s^\smtwo  + s^\smthree -s^\smone  +
w^\smone  - w^\smtwo + 1)\,,
\nonumber\\
&& y^\smtwo = \frac{1}{2}(s^\smone + s^\smthree - s^\smtwo  - w^\smone   +
w^\smtwo -1)\,,
\nonumber\\
&& y^\smthree = \frac{1}{2}(s^\smone  + s^\smtwo - s^\smthree  - w^\smone -
w^\smtwo - 1  )\,, \hspace{1cm} \hbox{ for F-vertex}.
\eeq
Two integers $w^\smone$, $w^\smtwo$ are the freedom of our solution. For
fixed spin values $s^\smone$, $s^\smtwo$, $s^\smthree$, these integers label
all possible cubic interaction vertices that can be built for the fields
under consideration. For vertices \rf{pint40kn},\rf{pint40fn} to be sensible
we impose the restrictions
\beq
&& \label{restr0101kn} w^\smone  \geq 0 \,,\qquad w^\smtwo \geq 0\,, \qquad
y^\sma  \geq 0\,,\qquad a=1,2,3;
\nonumber \\[3pt]
&& \sbf - w^\smone   - w^\smtwo\,\,  \hbox{ even integer}\,, \hspace{5.3cm}
\hbox{ for K-vertex};\qquad
\\[3pt]
&& \label{restr0101fn} w^\smone  \geq 0 \,,\qquad w^\smtwo \geq 0\,, \qquad
 y^\sma  \geq 0\,,\qquad a=1,2,3;
\nonumber \\[3pt]
&& \sbf - w^\smone   - w^\smtwo\,\,  \hbox{ odd integer}\,, \hspace{5.5cm}
\hbox{ for F-vertex}.
\eeq
which amount to the requirement that the powers of all forms in
\rf{pint40kn}, \rf{pint40fn} be non--negative integers. The maximal and
minimal number of powers of $\Po^I$ in \rf{pint40kn}, \rf{pint40fn}, denoted
by $k_\max$ and $k_\min$ respectively, are given by%
\footnote{ Expressions for $K^\smonetwo$ \rf{mm012fnmix} and $F$, $L^\sma$,
$Q^\smaaplusone$ \rf{23012010-man03}-\rf{mm011n} imply that $k_\max =
w^\smone + w^\smtwo +2y^\smone + 2 y^\smtwo + y^\smthree + 1$ for K-vertex
and $k_\max = w^\smone + w^\smtwo +2y^\smone + 2 y^\smtwo + y^\smthree + 1$
for F-vertex. Relations for $y^\sma$ \rf{mm014kn},\rf{mm014fn} then lead to
$k_\max$ given in \rf{kmaxmm0N1kn},\rf{kmaxmm0N1fn}.}
\beq
\label{kmaxmm0N1kn} && \hspace{-1cm}  k_\max = \frac{1}{2}(s^\smone  +
s^\smtwo + 3s^\smthree + w^\smone  + w^\smtwo ) + 1 \,,  \qquad k_\min=0\,,
\hspace{1.3cm} \hbox{ for K-vertex};\qquad
\\[2pt]
\label{kmaxmm0N1fn} && \hspace{-1cm}  k_\max = \frac{1}{2}(s^\smone  +
s^\smtwo + 3s^\smthree + w^\smone  + w^\smtwo +1)\,,  \qquad k_\min=0\,,
\hspace{1.3cm} \hbox{ for F-vertex}.
\eeq
We note that using \rf{mm014kn},\rf{mm014fn} allows rewriting restrictions
\rf{restr0101kn},\rf{restr0101fn} in the equivalent form%
\footnote{ If $w^\smone = w^\smtwo=0$, then restriction \rf{mm017kn} becomes
the  restrictions well known in the angular momentum theory, $ |s^\smone -
s^\smtwo| \leq s^\smthree \leq s^\smone + s^\smtwo$, while restriction
\rf{mm017fn} takes the form $ |s^\smone - s^\smtwo-1| \leq s^\smthree \leq
s^\smone + s^\smtwo-1$.}
\beq
&&\hspace{-2cm}  \label{mm017kn} |s^\smone  -s^\smtwo  - w^\smone + w^\smtwo
| \leq s^\smthree \leq s^\smone + s^\smtwo -w^\smone   -w^\smtwo\,,
\hspace{2cm} \hbox{ for K-vertex};
\\[3pt]
&& \hspace{-2cm} \label{mm017fn} |s^\smone - s^\smtwo - w^\smone  + w^\smtwo
-1 |  \leq s^\smthree \leq s^\smone + s^\smtwo - w^\smone  - w^\smtwo  - 1\,,
\hspace{0.6cm} \hbox{ for F-vertex};
\eeq

\newsection{ Cubic vertices for two massive fermionic fields and one massive bosonic
field }\label{secMMM}

We finally consider the cubic interaction vertex for three mixed-symmetry
massive fields with mass parameters given by
\be \label{23012010-man01}  \mas_1|_\smF \ne 0 ,\qquad  \mas_2|_\smF \ne
0,\qquad \mas_3|_\smB \ne 0. \ee
The solution we found for K- and F-vertices is given by
\beq
&& \label{mmm1kmix} p_\smp3^- = K^\smonetwo V^\Krm (L_n^\sma ;\,
Q_{mn}^\smaaplusone)\,, \hspace{2cm} \hbox{ K-vertex};
\\[3pt]
&& \label{mmm1fmix} p_\smp3^- = F_n V^\Frm (L_n^\sma ;\,
Q_{mn}^\smaaplusone)\,, \hspace{2.5cm} \hbox{ F-vertex};
\eeq
where $V^\Krm$, $V^\Frm$ are arbitrary polynomials of
linear and quadratic forms and we use the notation
\beq
\label{kmasdefmix} && K^\smonetwo \equiv
\frac{1}{\beta_1\beta_2}p_{\theta_1}\Bigl( \Po^I\gamma^I\gamma_* - \mas_{12}
\Bigr)\eta_2\,,
\\
\label{fmasdefmix} && F_n \equiv \frac{1}{\beta_1\beta_2}p_{\theta_1}\Bigl(
\frac{\check\beta_3}{\beta_3} \Po^I - \gamma^{IJ}\Po^J +( \mas_1\beta_2 +
\mas_2\beta_1) \gamma^I \gamma_* \Bigr) \eta_2 \zeta_n^{\smthree I}
\nonumber\\
&& \quad \ \ + \ \frac{1}{\beta_1\beta_2}p_{\theta_1}\Bigl( - \check\mas_3
\Po^I\gamma^I\gamma_* + \check\mas_3 \mas_{12}  +
\frac{2\beta_1\beta_2}{\beta_3}\mas_3^2\Bigr) \eta_2
\frac{\zeta_n^\smthree}{\mas_3}\,,
\eeq
\beq
\label{mmm3mix} && L_n^\sma\equiv B_n^\sma - \frac{\mas_{a+1}^2 -
\mas_{a+2}^2}{2 \mas_a}\zeta_n^\sma\,,\qquad
\\[3pt]
&& B_n^\sma\equiv \frac{\alpha_n^{\sma I}\Po^I}{\beta_a} -
\frac{\check\beta_a}{2\beta_a}\mas_a \zeta_n^\sma\,,
\\[5pt]
\label{mmm2mix} && Q_{mn}^\smaaplusone \equiv \alpha_{mn}^\smaaplusone -
\frac{\zeta_n^\smaplusone }{\mas_{a+1}} B_m^\sma +
\frac{\zeta_m^\sma}{\mas_a} B_n^\smaplusone - \frac{\mas_{a+2}^2}{2\mas_a
\mas_{a+1}}\zeta_m^\sma \zeta_n^\smaplusone\,,
\\
&& \mas_{ab} \equiv \mas_a \beta_b -\mas_b \beta_a\,, \qquad \check{\mas}_a
\equiv \mas_{a+1} - \mas_{a+2}\,.  \eeq
The quadratic forms $\alpha_{mn}^\smab$ are defined in \rf{amnabdef}. From
\rf{fmasdefmix},\rf{mmm3mix}, \rf{mmm2mix}, we see that the prefactors $F_n$,
the linear forms $L_n^\sma$,  and the quadratic forms $Q_{mn}^\smaaplusone$
are singular in the massless limit, $\mas_a\rightarrow 0$, $a=1,2,3$. This
implies that, in general, the cubic vertices for massive fields are also
singular as $\mas_a\rightarrow 0$, $a=1,2,3$. Note that, in contrast to the
prefactors $F_n$, the prefactor $K^\smonetwo$ has smooth massless limit.
Solution for $V^\Krm$, $V^\Frm$ in \rf{mmm1kmix},\rf{mmm1fmix} is complete
solution, while the solution for the prefactors $K^\smonetwo$, $F_n$ in
\rf{kmasdefmix},\rf{fmasdefmix} is not complete solution. Namely, for the
vertices of mixed-symmetry fields there are extra solutions for the
prefactors that involve contribution of higher than second order in
$\gamma$-matrices.

\subsection{ Cubic vertices for totally symmetric
two massive fermionic fields and one massive bosonic field }

We now consider cubic vertices for totally symmetric fields with the mass
values given in \rf{23012010-man01}, i.e., the {\it massive} fermionic fields
carry external line indices $a=1,2$, while the {\it massive} bosonic field
carries external line index $a=3$. Because for the studying the totally
symmetric fields it is sufficient to use one sort of oscillators, we set
$\nnu = 1$ in \rf{mmm1kmix},\rf{mmm1fmix} and ignore the contribution of the
oscillators with $n>1$. To simplify the formulas we drop the oscillator's
subscript $n=1$ and use the simplified notation for oscillators:
$\alpha^{\sma I} \equiv \alpha_1^{\sma I}$, $\zeta^\sma \equiv \zeta_1^\sma$.
In doing so, we obtain, using the solution given in \rf{mmm1kmix}
\rf{mmm1fmix}, the cubic interaction vertices for totally symmetric fields,
\beq
&& \label{mmm1k} p_\smp3^- = K^\smonetwo V^\Krm (L^\sma ;\,
Q^\smaaplusone)\,, \hspace{2cm} \hbox{ K-vertex};
\\[3pt]
&& \label{mmm1f} p_\smp3^- = F V^\Frm (L^\sma ;\, Q^\smaaplusone)\,,
\hspace{2.5cm} \hbox{ F-vertex};
\eeq
where we use the notation
\beq
\label{fmasdef} && F \equiv  \frac{1}{\beta_1\beta_2}p_{\theta_1}\Bigl(
\frac{\check{\beta}_3}{\beta_3} \Po^I - \gamma^{IJ}\Po^J +( \mas_1\beta_2 +
\mas_2\beta_1) \gamma^I \gamma_* \Bigr) \eta_2 \alpha^{\smthree I}
\nonumber\\
&& \quad + \frac{1}{\beta_1\beta_2}p_{\theta_1}\Bigl( - \check\mas_3
\Po^I\gamma^I\gamma_* + \check\mas_3 \mas_{12}  +
\frac{2\beta_1\beta_2}{\beta_3}\mas_3^2\Bigr) \eta_2
\frac{\zeta^\smthree}{\mas_3}\,,
\\
\label{mmm3} && L^\sma\equiv B^\sma - \frac{\mas_{a+1}^2 - \mas_{a+2}^2}{2
\mas_a}\zeta^\sma\,,\qquad
\\[3pt]
&& B^\sma\equiv \frac{\alpha^{\sma I}\Po^I}{\beta_a} -
\frac{\check\beta_a}{2\beta_a}\mas_a \zeta^\sma\,,
\\[5pt]
\label{mmm2} && Q^\smaaplusone \equiv \alpha^\smaaplusone -
\frac{\zeta^\smaplusone }{\mas_{a+1}} B^\sma + \frac{\zeta^\sma}{\mas_a}
B^\smaplusone - \frac{\mas_{a+2}^2}{2\mas_a \mas_{a+1}}\zeta^\sma
\zeta^\smaplusone\,,
\\
&& \mas_{ab} \equiv \mas_a \beta_b -\mas_b \beta_a\,, \qquad \check{\mas}_a
\equiv \mas_{a+1} - \mas_{a+2}\,,  \eeq
while $\alpha^\smab$ and $K^\smonetwo$ are defined in \rf{amnabdefsym} and
\rf{kmasdefmix} respectively. We note that solution for vertices given in
\rf{mmm1k},\rf{mmm1f} is the complete solution. In other words, vertices in
\rf{mmm1k},\rf{mmm1f} constitute the complete list of parity invariant cubic
vertices for massive totally symmetric fields. We now restrict our attention to
vertices for totally symmetric fields with fixed spin values.

\subsubsection{ Cubic vertices for totally symmetric fields with fixed but
arbitrary spin values}\label{seubsub-11-11}

Vertices \rf{mmm1k},\rf{mmm1f} describe interaction of the towers of massive
totally symmetric fields \rf{manold-07122011-09}, \rf{fer07}. We next obtain
vertices for massive spin $s^\smone+\half$, $s^\smtwo+\half$ and $s^\smthree$
fields. Namely, we consider vertices involving two spin $s^\smone + \half$
and $s^\smtwo +\half$ fermionic fields having the respective mass parameters
$\mas_1$ and $\mas_2$ and one spin-$s^\smthree$ bosonic field having the mass
parameter $\mas_3$:
\be
\begin{array}{lll}
\mas_1|_\smF \ne 0\,, \qquad & \mas_2|_\smF \ne 0\,, \qquad & \mas_3|_\smB
\ne 0\,,
\\[7pt]
s^\smone + \half \,, & s^\smtwo+\half \,, & s^\smthree\,.
\end{array}
\ee
The massive spin-$(s^\smone +\half)$ and -$(s^\smtwo +\half)$ fermionic
fields are described by the respective ket-vectors $|\psi_{s^\smone
}\rangle$, $|\psi_{s^\smtwo }\rangle$, while the massive spin-$s^\smthree$
bosonic field is described by ket-vector $|\phi_{s^\smthree}\rangle$. The
ket-vectors of massive fields $|\psi_{s^\sma }\rangle$, $a=1,2$ and
$|\phi_{s^\smthree }\rangle$, can be obtained from the respective expressions
in \rf{fer01} and \rf{manold-07122011-04} by replacing $s\rightarrow s^\sma
$, $\alpha^I\rightarrow \alpha^{\sma I}$, $\zeta\rightarrow \zeta^\sma$.
Because $|\psi_{s^\smone }\rangle$, $|\psi_{s^\smtwo }\rangle$,
$|\phi_{s^\smthree }\rangle$ are respective degree-$s^\sma $ homogeneous
polynomials in $\alpha^{\sma I}$, $\zeta^\sma$, it is obvious that the
vertices we are interested in must satisfy the equations
\be \label{mmmN3} (\alpha^{\sma I}\bar\alpha^{\sma I} +
\zeta^\sma\bar\zeta^\sma  - s^\sma )|p_\smp3^-\rangle  = 0\,,\qquad
a=1,2,3, \ee
which tell us that the vertices must be degree-$s^\sma $ homogeneous
polynomials in the oscillators $\alpha^{\sma I}$, $\zeta^\sma$. Taking into
account that the prefactor $F$ and the linear forms $L^\sma$ are degree-1
homogeneous polynomials in oscillators, while the quadratic forms
$Q^\smaaplusone$ are degree-2 homogeneous polynomials in the oscillators we
obtain the general solution of Eqs.\rf{mmmN3} as
\beq
\label{intvermmm01k} && \hspace{-1.7cm} p_\smp3^-(s^\smone+\half,
s^\smtwo+\half,s^\smthree;w^\smone,w^\smtwo,w^\smthree) = K^\smonetwo
\prod_{a=1}^3 (L^\sma)^{w^\sma}(Q^\smaaplusone)^{y^\smaplustwo}\,,
\hspace{0.1cm} \hbox{ K-vertex};
\\
\label{intvermmm01f} && \hspace{-1.7cm}
p_\smp3^-(s^\smone+\half,s^\smtwo+\half,
s^\smthree;w^\smone,w^\smtwo,w^\smthree) = F \prod_{a=1}^3 (L^\sma)^{w^\sma
}(Q^\smaaplusone)^{y^\smaplustwo}\,, \hspace{0.6cm} \hbox{ F-vertex};
\eeq
where integers $y^\sma $ are expressible in terms of $s^\sma$ and three
integers $w^\sma$ labeling the freedom of our solution,
\beq
\label{mmm12k} && \hspace{-1.5cm} y^\sma  = \frac{1}{2}(\sbf + w^\sma -
w^\smaplusone - w^\smaplustwo) - s^\sma \,, \qquad\quad a=1,2,3\,,
\hspace{0.8cm} \hbox{ for K-vertex};\qquad
\\
\label{mmm12f} && \hspace{-1.5cm} y^\sma  = \frac{1}{2}(\sbf + w^\sma -
w^\smaplusone - w^\smaplustwo-1)  -s^\sma \,, \quad \ a=1,2;
\nonumber\\
&& \hspace{-1.5cm} y^\smthree  = \frac{1}{2}(\sbf + w^\smthree - w^\smone -
w^\smtwo + 1)  - s^\smthree \,, \hspace{4cm} \hbox{ for F-vertex};
\eeq
and $\sbf$ is given in \rf{0007}. The maximal and minimal numbers of powers
of $\Po^I$ in \rf{intvermmm01k},\rf{intvermmm01f}, denoted by $k_\max$ and
$k_\min$ respectively,  are given by%
\footnote{ Expressions for $K^\smonetwo$ \rf{kmasdefmix} and $F$, $L^\sma$,
$Q^\smaaplusone$ \rf{fmasdef}-\rf{mmm2} imply that $k_\max = 1+ \sum_{a=1}^3
(w^\sma + y^\sma )$. Taking $y^\sma$ \rf{mmm12k},\rf{mmm12f} into account we
then find $k_\max$ given in \rf{mmmkmaxk},\rf{mmmkmaxf}.}
\beq
&& \label{mmmkmaxk} k_\max = \frac{1}{2}\bigl(\sbf + \sum_{a=1}^3
w^\sma\bigr) + 1 \,, \qquad k_\min =0 \,, \hspace{1cm}  \hbox{ for K-vertex};\qquad
\\
&& \label{mmmkmaxf} k_\max = \frac{1}{2}\bigl(\sbf + \sum_{a=1}^3 w^\sma +
1\bigr)\,, \qquad k_\min =0 \,, \hspace{1cm} \hbox{ for F-vertex}.
\eeq
Requiring the powers of the forms $L^\sma$ and $Q^\smaaplusone$ in
\rf{intvermmm01k},\rf{intvermmm01f} to be non--negative integers gives the
restrictions
\beq
\label{mmm13k} && w^\sma  \geq 0 \,,\qquad   y^\sma  \geq 0\,, \qquad a
=1,2,3\,;
\nonumber\\
&& \sbf + \sum_{a=1}^3 w^\sma  \quad \hbox{ even integer},  \hspace{3cm}
\hbox{ for K-vertex};
\\
\label{mmm13f} && w^\sma  \geq 0 \,,\qquad   y^\sma  \geq 0\,, \qquad a
=1,2,3\,;
\nonumber\\
&& \sbf + \sum_{a=1}^3 w^\sma  \quad \hbox{ odd integer}\,,  \hspace{3cm}
\hbox{ for F-vertex}.
\eeq
Using relations \rf{mmm12k},\rf{mmm12f} allows rewriting restrictions
\rf{mmm13k},\rf{mmm13f} as%
\footnote{ If $w^\sma=0$, $a=1,2,3$, then restrictions \rf{mmm15k} becomes
the restrictions well known in the angular momentum theory: $ |s^\smone -
s^\smtwo| \leq s^\smthree \leq s^\smone + s^\smtwo$, while restriction
\rf{mmm15f} takes the form $ |s^\smone - s^\smtwo| \leq s^\smthree -1 \leq
s^\smone + s^\smtwo$.}
\beq
&&\hspace{-2cm} \label{mmm15k} s^\smthree  - s^\smone - s^\smtwo  + w^\smone
+ w^\smtwo \leq w^\smthree \leq s^\smthree - | s^\smone - s^\smtwo - w^\smone
+ w^\smtwo |\,, \hspace{0.3cm} \hbox{ for K-vertex};
\\
&& \hspace{-2cm} \label{mmm15f} s^\smthree  - s^\smone - s^\smtwo  + w^\smone
+ w^\smtwo -1 \leq w^\smthree \leq s^\smthree - | s^\smone - s^\smtwo -
w^\smone + w^\smtwo |-1\,, \,\,\,
\nonumber\\
&& \hspace{10.6cm} \hbox{ for F-vertex}.
\eeq

\noindent {\bf Examples of light-cone cubic vertices and their counterparts
in manifestly Lorentz invariant approach}. We now give examples of vertices
for particular spin values and make comment concerning hermitian properties of our
vertices.

Cubic vertices for two massive spin-$\half$ and spin-$\half$ fermionic fields
having the respective mass parameters $\mas_1$, $\mas_2$ and one massive
spin-$s$ bosonic field with the mass parameter $\mas_3$ are given by
\beq
\label{1-21-2sKver01} && p_\smp3^-(\half,\half,s; 0,0,s) = K^\smonetwo
(L^\smthree)^s \,,
\\[2pt]
\label{1-21-2sFver01} && p_\smp3^-(\half,\half,s; 0,0,s-1) = F
(L^\smthree)^{s-1}\,.
\eeq
Consider vertices \rf{1-21-2sKver01},\rf{1-21-2sFver01} for $s=1$. Manifestly
Lorentz invariant K-vertex corresponding to our light-cone K-vertex
\rf{1-21-2sKver01} with $s=1$ takes the form
\be \label{covlag1-21-2sFver04} \LL = \bar\psi_1 \gamma^{AB}\psi_2
F^{AB}\,,\qquad F^{AB} \equiv \partial^A \phi^B - \partial^B\phi^A\,,
\ee
while the manifestly Lorentz invariant F-vertex corresponding to our
light-cone F-vertex \rf{1-21-2sFver01} with $s=1$ is given by
\be \label{covlag1-21-2sFver01} \LL =
\bar\psi_1 \gamma^A \psi_2 (\phi^A + \frac{1}{\mas_3} \partial^A \phi)\,.
\ee
In \rf{covlag1-21-2sFver01}, $\phi^A$ and $\phi$ are vector and scalar fields
of the Lorentz algebra which are used in gauge invariant Stueckelberg
formulation of spin-1 massive field having mass parameter $\mas_3$. We recall
that the scalar field $\phi$ is the Stueckelberg field.

Light-cone gauge vertex \rf{1-21-2sKver01} and manifestly Lorentz invariant
vertex  \rf{covlag1-21-2sFver04} are not hermitian in general. We can obtain
hermitian light-cone gauge and covariant vertices in the standard way: by
adding or subtracting appropriate hermitian conjugated expression. In this
way, we obtain the two well-known covariant vertices
\beq
&& \label{covlag1-21-2sFver05} \LL = (\bar\psi_1 \gamma^{AB}\psi_2 +
\bar\psi_2 \gamma^{AB}\psi_1) F^{AB}\,, \
\\[3pt]
&& \label{covlag1-21-2sFver06}  {\rm i}\LL = (\bar\psi_1 \gamma^{AB}\psi_2 -
\bar\psi_2 \gamma^{AB}\psi_1) F^{AB}\,.
\eeq

In the same way we can consider vertices in \rf{1-21-2sFver01} and
\rf{covlag1-21-2sFver01}.  Namely, light-cone gauge vertex \rf{1-21-2sFver01}
and covariant vertex \rf{covlag1-21-2sFver01} are not hermitian when $\mas_1
\ne \mas_2$, i.e., when $\psi_1\ne \psi_2$. The hermitian light-cone gauge
and covariant vertices are obtained by adding or subtracting appropriate
hermitian conjugated expression,
\be \label{covlag1-21-2sFver02} \LL =
(\bar\psi_1 \gamma^A \psi_2 + \bar\psi_2 \gamma^A \psi_1)(\phi^A +
\frac{1}{\mas_3} \partial^A \phi)\,,
\ee
\be \label{covlag1-21-2sFver03}  {\rm i} \LL =
(\bar\psi_1 \gamma^A \psi_2 -  \bar\psi_2 \gamma^A \psi_1)(\phi^A +
\frac{1}{\mas_3} \partial^A \phi)\,.
\ee

To summarize, in order to obtain hermitian vertices we should supplement our
vertices by appropriate hermitian conjugated expressions. Note also that, for
the space-time dimensions when Dirac fields allow restriction to Majorana
fields, our vertices are automatically hermitian.

{\bf Vertices for massive spin-$\frac{3}{2}$ fermionic field and massive
spin-$s$ bosonic field}. For the reader convenience we worked out the
explicit expressions for cubic K- and F-vertices,
\rf{intvermmm01k},\rf{intvermmm01f} involving two massive spin-$\frac{3}{2}$
fermionic fields and one massive spin-$s$ bosonic field,
\beq
&& p_\smp3^-(\frac{3}{2},\frac{3}{2},s; 0,0,s-2) = K^\smonetwo
(L^\smthree)^{s-2} Q^\smtwothree Q^\smthreeone \,,
\\[2pt]
&& p_\smp3^-(\frac{3}{2},\frac{3}{2},s; 0,0,s) = K^\smonetwo (L^\smthree)^s
Q^\smonetwo  \,,
\\[2pt]
&& p_\smp3^-(\frac{3}{2},\frac{3}{2},s; 0,1,s-1) = K^\smonetwo L^\smtwo
(L^\smthree)^{s-1} Q^\smthreeone  \,,
\\[2pt]
&& p_\smp3^-(\frac{3}{2},\frac{3}{2},s; 1,1,s) = K^\smonetwo L^\smone
L^\smtwo (L^\smthree)^s \,,
\\[10pt]
&& p_\smp3^-(\frac{3}{2},\frac{3}{2},s; 0,0,s-3) = F (L^\smthree)^{s-3}
Q^\smtwothree Q^\smthreeone \,,
\\[2pt]
&& p_\smp3^-(\frac{3}{2},\frac{3}{2},s; 0,0,s-1) = F (L^\smthree)^{s-1}
Q^\smonetwo \,,
\\[2pt]
&& p_\smp3^-(\frac{3}{2},\frac{3}{2},s; 0,1,s-2) = F L^\smtwo
(L^\smthree)^{s-2} Q^\smthreeone  \,,
\\[2pt]
&& p_\smp3^-(\frac{3}{2},\frac{3}{2},s; 1,1,s-1) = F L^\smone L^\smtwo
(L^\smthree)^{s-1} \,.
\eeq

\medskip

{\bf Acknowledgments}. This work was supported by the INTAS project
03-51-6346, by the RFBR Grant No.05-02-17654, RFBR Grant for Leading
Scientific Schools, Grant No. LSS-4401.2006.2 and the Alexander von Humboldt
Foundation Grant PHYS0167.

\setcounter{section}{0} \setcounter{subsection}{0}
\appendix{ Notation }

{\bf Notation in basis of Lorentz algebra $so(d-1,1)$}. Our conventions are
as follows. $x^A$ denotes coordinates in $d$-dimensional flat space-time,
while $\partial_A$ denotes derivatives with respect to $x^A$, $\partial_A
\equiv \partial/\partial x^A$. Vector indices of the Lorentz algebra
$so(d-1,1)$ take the values $A,B,C,E=0,1,\ldots ,d-1$. We use the mostly
positive flat metric tensor $\eta^{AB}$. For simplicity, we drop $\eta_{AB}$
in the scalar products, i.e., we use $X^AY^A \equiv \eta_{AB}X^A Y^B$.

\medskip
\noindent {\bf $\gamma^A$ and $\Gamma_*$ -matrices for even $d$}. For
$d=2\nsf$, $\gamma^A$-matrices are $2^\nsf \times 2^\nsf$ Dirac matrices in
$d$-dimensional Minkowski space, $ \{ \gamma^A,\gamma^B\} = 2\eta^{AB}$,
while the matrix $\Gamma_*$ is defined as
\be \label{manold-11122011-01} \Gamma_* \equiv \epsilon \gamma^0 \gamma^1
\ldots \gamma^{d-1} \,, \qquad \quad
\Gamma_*^2 =1\,. \ee

\medskip
\noindent {\bf $\gamma^A$ and $\Gamma_*$ -matrices for odd $d$}. For
$d=2\nsf+1$, we use matrices $\gamma^A$, $\Gamma_*$ which are the $2^{\nsf+1}
\times 2^{\nsf+1}$ Dirac matrices in $d+1$ dimensional Minkowski space,
\be  \{ \gamma^A,\gamma^B\} = 2\eta^{AB}\,, \qquad \{\gamma^A, \Gamma_*\}=
0\,, \qquad \Gamma_*^2 =1 \,.  \ee
The use of such matrices implies that Lagrangian for $2^{\nsf+1}$th component
Dirac field $\Psi$ given by
\be \label{manold-30122011-09} \i \LL = \bar\Psi(\gamma^A \partial^A + \mas)\Psi\,,\ee
describes two massive spin-$\half$ Dirac fields, when $d$ is odd. In order to
obtain description of massive spin-$\half$ field in terms of $2^\nsf$th
component Dirac fields we introduce the decomposition
\beq
\label{manold02-l18122011-02}&& \Psi = \Psi_+ + \Psi_-\,, \qquad \Psi_\pm
\equiv \half (1\pm \sigma_*)\Psi\,,
\\
&& \sigma_* \equiv \i^{\nsf+1}\gamma^0\ldots \gamma^{d-1}\,, \qquad
[\gamma^A,\sigma_*] = 0 \,, \qquad \sigma_*^2 =1 \,,\qquad\sigma_*^\dagger =
\sigma_*\,.
\eeq
In terms of $2^\nsf$th component Dirac fields $\Psi_\pm$, Lagrangian
\rf{manold-30122011-09} takes the form
\be \i \LL = \bar\Psi_+(\gamma^A \partial^A + \mas)\Psi_+ +
\bar\Psi_-(\gamma^A
\partial^A + \mas)\Psi_-\,. \ee
Thus, for odd $d$, our description of massive fields involves reducible set
of massive fields. We use such description in order to treat the fermionic
fields in odd and even dimensions on an equal footing. For the case of odd
$d$, the description for single massive field can be obtained by using
decomposition given in \rf{manold02-l18122011-02}. We note also that, for odd
$d$, use of the matrix $\Gamma_*$ is motivated by the desire to get
constraints for the fermionic field in the form given in \rf{fer05mix} (for
details, see below).

We use matrices constructed out of $\gamma^A$ and $\Gamma_*$ matrices,
\be
\label{manold02-18122011-01} \gamma^{AB} \equiv \frac{1}{2}\gamma^A\gamma^B -
(A\leftrightarrow B)\,,
\qquad\qquad
U \equiv \frac{1}{\sqrt{2}} (1 + {\rm i}\Gamma_*)\,.
\ee
The $\gamma^A$-matrices satisfy the hermitian conjugation rule given by
$\gamma^{A\dagger} = \gamma^0 \gamma^A\gamma^0$.

\medskip
\noindent {\bf Notation in basis of $so(d-2)$ algebra}. We decompose  the
Lorentz frame space-time coordinates $x^A$
into light-cone frame space-time coordinates $x^\pm$, $x^I$ defined by%
\be x^\pm \equiv \frac{1}{\sqrt{2}}(x^{d-1}  \pm x^0)\,,\qquad x^I\,,\quad
I=1,\ldots, d-2\,, \ee
and treat $x^{+}$ as an evolution parameter. Vector indices of the $so(d-2)$
algebra take values $I,J,K=1,\ldots,d-2$. In the light-cone frame, vector of
the Lorentz algebra $X^A$ is decomposed as $X^+,X^-,X^I$ and a scalar product
of two vectors is then represented as
\be \eta_{AB}X^A Y^B = X^+Y^- + X^-Y^+ +X^IY^I\,, \ee
where the covariant and contravariant components of vectors are related as
$X^+=X_-$, $X^-=X_+$, $X^I=X_I$. This leads to the following conventions for the
derivatives:
\be \partial^I=\partial_I\equiv\partial/\partial x^I\,, \qquad
\partial^\pm=\partial_\mp \equiv
\partial/\partial x^\mp \,.
\ee
In the light-cone frame, we introduce the matrix $\gamma_*$ and projectors
$\Pi^\oplussm$, $\Pi^\ominussm$ defined by
\beq \label{manold-17122011-01} && \Gamma_* \equiv \gamma^{+-} \gamma_*\,,
\qquad \Pi^\oplussm \equiv \half \gamma^- \gamma^+\,,\qquad \Pi^\ominussm
\equiv \half \gamma^+ \gamma^-\,,
\\[5pt]
&& \Pi^\oplussm\Pi^\oplussm=\Pi^\oplussm\,,\qquad \Pi^\ominussm \Pi^\ominussm
=\Pi^\ominussm\,,\qquad \Pi^\oplussm\Pi^\ominussm=0\,,\qquad \Pi^\oplussm +
\Pi^\ominussm =1\,.
\eeq

\medskip
\noindent {\bf Bosonic oscillators}. We use a set of the bosonic creation
operators $\alpha_n^I$, $\zeta_n$, and the respective set of annihilation
operators $\bar\alpha_n^I$, $\bar\zeta_n$,  $m,n=1,\ldots \nnu$, where $\nnu$
is arbitrary integer $\nnu \geq 1$. These operators are referred to as
oscillators. Commutation relations, the vacuum, and hermitian conjugation
rules are defined as
\beq
\label{manold02-17122011-03} &&{}
[\bar\alpha_m^I,\alpha_n^J]=\eta^{IJ}\delta_{mn}\,, \qquad
[\bar\zeta_m,\zeta_n]=\delta_{mn}\,, \qquad \bar\alpha_m^I |0\rangle =
0\,,\qquad \bar\zeta_m|0\rangle = 0\,,\qquad
\nonumber\\[-10pt]
&&
\\[-10pt]
&& \alpha_n^{I \dagger} = \bar\alpha_n^I\,, \qquad\qquad \qquad
\zeta_n^\dagger = \bar\zeta_n\,,  \qquad \quad m,n=1,\ldots \nnu\,.
\nonumber
\eeq
To discuss totally symmetric field, it is sufficient to use one sort of
oscillators $\alpha^I$, $\zeta$, $\bar\alpha^I$, $\bar\zeta$. Commutation
relations, the vacuum, and hermitian conjugation rules are defined as%
\be
\label{man02-291102011-01}
[\bar\alpha^I,\alpha^J]=\eta^{IJ}\,, \qquad [\bar\zeta,\zeta]=1\,, \qquad
\bar\alpha^I|0\rangle = 0 \,,\qquad \bar\zeta|0\rangle =0\,,\qquad \alpha^{I
\dagger} =\bar\alpha^I \,, \qquad \zeta^\dagger = \bar\zeta\,.
\ee

\medskip
\noindent {\bf Fermionic oscillators}. We introduce fermionic oscillators
$\theta^\asf$, $\eta^\asf$, $p_{\theta \asf}$, $p_{\eta\asf}$ subject to the
algebraic constraints
\be \Pi^\oplussm \theta =\theta\,, \qquad \Pi^\oplussm \eta =\eta\,, \qquad
p_\theta \Pi^\oplussm = p_\theta \,,\qquad p_\eta \Pi^\oplussm = p_\eta \,,
\ee
where the projector $\Pi^\oplussm$ is defined in \rf{manold-17122011-01}.
Anticommutation relations of the fermionic oscillators, the vacuum, and
hermitian conjugation rules are defined as follows,
\be
\label{ferant} \{ \theta^\asf, p_{\theta\bsf} \} =
\Pi^\oplussm{}^\asf{}_\bsf\,, \qquad \{ \eta^\asf, p_{\eta\bsf} \} =
\Pi^\oplussm{}^\asf{}_\bsf \,,\qquad \theta |0\rangle=0\,,\quad p_\eta
|0\rangle=0\,, \quad \theta^\dagger = p_\theta\,,\quad \eta^\dagger =
p_\eta\,,
\ee
where, in \rf{ferant}, the spinor indices $\asf,\bsf$ are shown
explicitly.

\medskip
\noindent {\bf Fourier transform of fields}. Throughout this paper we use
ket-vectors depending on the light-cone time $x^+$ and momenta $p=\{\beta,
p^I\}$. For the reader convenience we now explain how our ket-vectors are
related to the one depending on space-time coordinates $x=\{x^+,x^-,x^I\}$.
This is to say that for bosonic ket-vector $\phik$ \rf{manold-07122011-10}
and fermionic ket-vector $\psik$ \rf{manold-08122011-01} we assume the
following Fourier transform
\beq
&& \label{r4ter0}  |\phi(x,\alpha)\rangle = \int \frac{d^{d-1}
p}{(2\pi)^{(d-1)/2}} e^{{\rm i}(x^+ p^- +  x^- \beta + x^I p^I) } {\rm i}^{-N_\zeta}
|\phi(x^+,p,\alpha)\rangle \,,
\\
\label{manold02-18122011-03} && |\psi(x,\alpha)\rangle  = \int \frac{d^{d-1}
p}{(2\pi)^{(d-1)/2}} e^{{\rm i}( x^+ p^- + x^- \beta + x^I p^I) }\,\,  {\rm
i}^{- N_\zeta} \widehat{U} |\psi(p,\alpha)\rangle\,,
\\
&& \hspace{2cm} \widehat{U} \equiv p_\theta U \theta +
(U^\dagger)^\asf{}_\bsf \eta^\bsf \cdot p_{\eta\asf} \,,\qquad \quad N_\zeta
\equiv \sum_{n=1}^\nnu \zeta_n\bar\zeta_n \,,
\eeq
where $p^-$ is given in \rf{intver8} , while the matrix $U$ is defined in
\rf{manold02-18122011-01}. We assume that the ket-vector
$|\psi(x,\alpha)\rangle$ is decomposed into the fermionic oscillators as in
\rf{fer01nnn},
\beq \label{manold-20122011-01}
|\psi(x,\alpha)\rangle =   ( p_\theta \psi(x,\alpha) +
\psi^\dagger(x,\alpha)\eta) |0\rangle\,,
\eeq
while fields $\psi(x,\alpha)$, $\psi^\dagger(x,\alpha)$ are decomposed into
the bosonic oscillators as in
\rf{manold-07122011-14},\rf{manold-07122011-15}. The component fields
entering $\psi(x,\alpha)$, $\psi^\dagger(x,\alpha)$ are hermitian conjugated,
\be \label{manold-20122011-02} \psi^{I_1\ldots I_n}(x)= (\psi^{\dagger
I_1\ldots I_n}(x))^\dagger\,.\ee
In terms of the Fourier modes, relation \rf{manold-20122011-02} implies
\beq
&& \hspace{-1cm} \psi_{s_1\ldots s_\nnu}^{I_1^1\ldots I_{s_1}^1\ldots
I_1^\nnu \ldots I_{s_\nnu }^\nnu }(-p) = (\psi_{s_1\ldots s_\nnu}^{\dagger
I_1^1\ldots I_{s_1}^1\ldots I_1^\nnu \ldots I_{s_\nnu }^\nnu }(p))^\dagger\,,
\hspace{2.7cm} \hbox{ for massless field},
\nonumber\\[-10pt]
&&
\\[-10pt]
&&  \hspace{-1cm} \psi_{s_1\ldots s_\nnu}^{I_1^1\ldots I_{t_1}^1\ldots
I_1^\nnu \ldots I_{t_\nnu}^\nnu }(-p) =
(-)^{\sum_{n=1}^\nu t_n} (\psi_{s_1\ldots s_\nnu}^{\dagger I_1^1\ldots
I_{t_1}^1\ldots I_1^\nnu \ldots I_{t_\nnu}^\nnu }(p))^\dagger \,,
\hspace{1cm} \hbox{ for massive field}. \qquad
\nonumber
\eeq

We now make comment on the Fourier transform in \rf{manold02-18122011-03}.
Using representation for ket-vector $|\psi(x,\alpha)\rangle$ in
\rf{manold-20122011-01}, we note that the ket-vector
$\psi(x,\alpha)|0\rangle$ satisfies the standard constraint
\be (\gamma^I \bar\alpha_n^I -\bar\zeta_n)\psi(x,\alpha)|0\rangle = 0\,, \ee
commonly used in the literature. It is the desire to get constraints in the
form given in \rf{fer05mix} that motivates us to use the Fourier transform
given in \rf{manold02-18122011-03}.

Note also that fermionic fields in \rf{manold-07122011-14},
\rf{manold-07122011-15} satisfy the algebraic constraint

\be \Pi^\oplussm  \psi_{s_1 \ldots s_\nnu } (p,\alpha)
= \psi_{s_1 \ldots s_\nnu } (p,\alpha)\,.\ee


\end{document}